\documentclass[aps,twocolumn,pra,superscriptaddress,nofootinbib,amsmath,amssymb,floats,floatfix,english,letterpaper]{quantumarticle}
\pdfoutput=1
\usepackage{polski}
\usepackage[utf8]{inputenc}
\usepackage[T1]{fontenc}
\usepackage{ifthen}
\usepackage{graphicx}
\usepackage{babel}
\usepackage{color}
\usepackage{bm}
\usepackage[normalem]{ulem}
\usepackage{float}
\usepackage{placeins}
\usepackage{chngcntr}
\usepackage{enumitem}
\usepackage{multirow}

\newcommand{\ifis}[2]{
\ifthenelse{\equal{#1}{}}{}{#2}
}

\def\be{\begin{equation}}
\def\ee{\end{equation}}
\def\bea{\begin{eqnarray}}
\def\eea{\end{eqnarray}}
\def\bsu{\begin{subequations}}
\def\esu{\end{subequations}}
\def\bi{\begin{itemize}}
\def\ei{\end{itemize}}

\newcommand{\op}[1]{\widehat{#1}}
\newcommand{\dagop}[1]{\widehat{#1}^{\dagger}}
\newcommand{\bo}[1]{{\mathbf{#1}}}
\newcommand{\mc}[1]{{\mathcal{#1}}}
\newcommand{\wt}[1]{{\widetilde{#1}}}
\newcommand{\wb}[1]{{\overline{#1}}}
\newcommand{\ul}[1]{{\underline{#1}}}
\newcommand{\nonu}{\nonumber}

\newcommand{\bra}[1]{\langle#1\vert}
\newcommand{\ket}[1]{\vert#1\rangle}

\DeclareMathOperator{\TR}{Tr}

\newcommand{\Tr}[1]{\TR\left[{#1}\right]}
\newlength{\templength}

\newenvironment{eq}{\begin{equation}}{\end{equation}}

\newcommand{\eqn}[1]{(\ref{#1})}

\renewcommand{\eq}[2]{\begin{equation}\label{#1}#2\end{equation}}
\newcommand{\eqs}[2]{\begin{subequations}\label{#1}\begin{eqnarray}#2\end{eqnarray}\end{subequations}}
\newcommand{\eqa}[2]{\begin{eqnarray}\label{#1}#2\end{eqnarray}}

\setlist[itemize]{nosep}
\setlist[enumerate]{nosep,nolistsep}

\newcommand{\ve}{\varepsilon}

\begin{document}

\title{Multi-time correlations in the positive-P, Q, and doubled phase-space representations}

\author{Piotr Deuar}
\email{deuar@ifpan.edu.pl}
\affiliation{Institute of Physics, Polish Academy of Sciences, Aleja Lotnik\'ow 32/46, 02-668 Warsaw, Poland}

\begin{abstract} 
A number of physically intuitive results for the calculation of multi-time correlations in phase-space representations of quantum mechanics are obtained. 
They relate time-dependent stochastic samples to multi-time observables, and rely on the presence of derivative-free operator identities. 
In particular, expressions for time-ordered normal-ordered observables in the positive-P distribution are derived which replace Heisenberg operators with the bare time-dependent stochastic variables, confirming extension of earlier such results for the Glauber-Sudarshan P. 
Analogous expressions are found for the anti-normal-ordered case of the doubled phase-space 
Q representation, along with conversion rules among doubled phase-space s-ordered representations. 
The latter are then shown to be readily exploited to further calculate anti-normal and mixed-ordered multi-time observables in the positive-P, Wigner, and doubled-Wigner representations. 
Which mixed-order observables are amenable and which are not is indicated, and explicit tallies are given up to 4th order. 
Overall, the theory of quantum multi-time observables in phase-space representations is extended, allowing non-perturbative treatment of many cases.
The accuracy, usability, and scalability of the results to large systems is demonstrated using stochastic simulations of the unconventional photon blockade system and a related Bose-Hubbard chain.
In addition, a robust but simple algorithm for integration of stochastic equations for phase-space samples is provided.
\end{abstract} 

\maketitle 

\section{Introduction}
\label{INTRO}
Phase-space representations of quantum mechanics such as the Wigner, P, positive-P, Q and related approaches are a powerful tool for the study and understanding of quantum mechanics \cite{QuantumNoise,Drummond80,Blakie08,Polkovnikov10,Lee95}. Their use in recent times has been directed particularly as a tool in quantum information science (see  \cite{Ferrie11} and \cite{Sperling20} for recent reviews) and for the simulation of large-scale quantum dynamics.
Negativity of Wigner, P, and other phase-space quasi-distributions is a major criterion for quantumness and closely related to contextuality and nonlocality in quantum mechanics \cite{Ferrie11,Gehrke12}. 
The inability to interpret the Glauber-Sudarshan P \cite{Glauber63,Sudarshan63} and Wigner distributions in terms of a classical probability density is the fundamental benchmark for quantum light \cite{Korbicz05,Sperling16}.
Quasi-probability representations arise naturally when looking for hidden variable descriptions and an ontological model of quantum theory \cite{Ferrie11,Spekkens08,Lee95,Bondar13}.
Wigner and P-distribution negativity are considered a resource for quantum computation \cite{Albarelli18,Ferrie11} and closely related to magic state distillation and quantum advantage \cite{Veitch12}; the Glauber-Sudarshan P distribution defines a hierarchy of nonclassicalities and nonclassicality witnesses \cite{Korbicz05,Vogel08} and can be experimentally reconstructed \cite{Agarwal94,Kiesel08}. 
Moreover, P, Q and Wigner distributions and have been used for simulations of models important for quantum information topics such as open spin-qubit systems \cite{Thapliyal15} and boson sampling \cite{Opanchuk18,Drummond16,Opanchuk19}.

For simulation of quantum mechanics, the prime advantage of the phase-space approach is that its computational cost typically grows only linearly with system size even in interacting systems of many particles. Therefore it provides a route to the non-perturbative treatment of the quantum dynamics of large systems. It has been used for solving and simulating a multitude of problems in various physical fields: e.g. quantum optics \cite{Carter87,Corney08,Drummond80b,Corney97,Lamprecht02,Olsen04,Dechoum04,Drummond20,Kiesewetter14}, ultracold atoms 
\cite{Steel98,Sinatra02,Carusotto01,Drummond99,Deuar09,Norrie05,Deuar07,Kheruntsyan12,Midgley09,Swislocki16,Lewis-Swan14,Hush13,Lewis-Swan15,Wuster17,Mathey14}, fermionic systems \cite{Corney04,Corney06,Aimi07a,Aimi07b}, spin systems \cite{Ng13,Barry08,Ng11}, nuclear physics \cite{Tsukiji16}, dissipative systems in condensed matter \cite{Wouters09,Chiocchetta14,Deuar21}, or cosmology \cite{Opanchuk13}. 
The vast majority of such calculations so far have considered only equal time correlations and observables. 

Multi-time correlations are also important for answering many physical questions which cannot necessarily be dealt with by monitoring the time dependence of equal time observables \cite{QuantumNoise,Berg09,Polkovnikov10,Lax68,Vogel08}. For example, the determination of lifetimes in an equilibrium or stationary state, response functions, out-of-time-order correlations (OTOCs), or finding the time resolution required to observe a transient phenomenon.
However their calculation in the phase-space framework has been restricted because few general results have been available. What is known is largely restricted to time- and normal-ordered correlations in the Glauber-Sudarshan P representation\cite{Agarwal70c,QuantumNoise}, time-symmetric ones in the truncated Wigner representation \cite{Berg09,Plimak14}, or the corresponding linear response corrections for closed systems. 
The Glauber-Sudarshan P representation results provide a particularly intuitive framework, simply replacing Heisenberg operators $\dagop{a}(t)$ and $\op{a}(t)$ with time-dependent phase-space variables in normal-ordered observables such as $\dagop{a}(t_1)\op{a}(t_2)$. They also apply to open systems \cite{QuantumNoise,Agarwal69b}. Such ordering corresponds to general photon counting measurements \cite{Glauber63,Glauber63a,Kelley64}. Still, while scaling with system size is excellent, both above approaches usually end up being only approximate because one is forced 
to omit part of the full quantum mechanics in their numerical implementations \cite{QuantumNoise}.
Hence, a broader application of full quantum phase-space methods to multi-time observables would be advantageous. This is especially so, since recent years have shown a lot of interest in such quantities, be it in the study of nonclassicality in the time dimension \cite{Atalaya18,Ringbauer18,Moreva17,Krumm16},  
time crystals \cite{Zhang17,Else16,Wilczek12,Syrwid17}, quantum technologies \cite{Liew10,Kal15,Moczala-Dusanowska19} where time correlations and time resolution are crucial, methods development
\cite{Plimak14,Klobas20}, and the OTOCs that have recently found to be important for the study of quantum scrambling \cite{Swingle16,Bohrdt17,Garttner18}, quantum chaos \cite{Maldacena16,Shen17} and many-body localisation \cite{Fan17,He17}.
Therefore, this paper sets out to extend the infrastructure available for multi-time observables with phase-space methods.

The positive-P representation \cite{Drummond80} does not suffer from the limitations of the Glauber-Sudarshan P (seen e.g. in Sec.~\ref{GSPBREAK}) due to its anchoring in doubled phase space. This allows it to incorporate the full quantum mechanics for most Hamiltonians, including all two-body interactions. Importantly, as we shall see, the same applies for its differently ordered analogues like the doubled-Wigner and doubled-Q. While the positive-P is known to be limited to short times for closed systems due to a noise amplification instability \cite{Gilchrist97,Deuar06a}, this abates in dissipative systems \cite{Gilchrist97,Deuar21}. Recent work has shown that a very broad range of driven  
dissipative Bose-Hubbard models can be simulated with the positive-P into and beyond the stationary state \cite{Deuar21}.
This covers systems of current interest such as micropillars, transmon qubits, and includes strong quantum effects such two-photon interference in the unconventional photon blockade \cite{Liew10,Bamba11}, making its extension quite timely.

Therefore, after basic background in Sec.~\ref{BKG}, the first order of business in this paper in Sec.~\ref{PPSEC} is to extend the long known Glauber-Sudarshan P results on time- and normal-ordered multi-time correlations to the positive-P representation.
This is sort of an obvious extension, but was missing from the literature, and allows complete quantum calculations.
It also lets us explain the ins and outs and set the stage for the less obvious and broader extensions that follow:

The first one of those in Sec.~\ref{QREP} considers general representations and reformulates the case of Q representations in an operational form that can be extended to doubled phase-space. This gives stochastic access to anti-normal ordered observables such as $\op{a}(t_1)\dagop{a}(t_2)$. 
It is then shown in Sec.~\ref{S3} how the Gaussian convolution relationship between Wigner, P, and Q distributions allows one to easily evaluate anti-normal ordered correlations in Wigner and P. This is extended also to a wide range of mixed ordered observables such as $\dagop{a}(t_1)\op{a}(t_2)\dagop{a}(t_3)\op{a}(t_4)$  in Sec.~\ref{GENORD}, and further extended to cover doubled phase-space representations in Sec.~\ref{OD}.
In Sec.~\ref{EX}, the use and accuracy of this approach is demonstrated on the numerical example of the unconventional photon blockade, a system that may have application to the creation of single-photon sources.
Finally, Appendix~\ref{NUMERIX} gives details of a robust algorithm for integrating the resulting phase-space stochastic equations, a matter that has been somewhat neglected in the literature to date.

\section{Background}
\label{BKG}

\subsection{Positive-P representation}
\label{PPDEF}

Consider an $M$-mode system (modes $1,\dots,j,\dots,M$) with bosonic annihilation operators $\op{a}_j$ for the $j$th mode.
The density operator of the system is written in terms of bosonic coherent states of complex amplitude $\alpha_j$ defined relative to the vacuum:
\eq{cohstate}{
\ket{\alpha_j}_j = e^{-|\alpha_j|^2/2}\,e^{\alpha_j\widehat{a}^{\,\dagger}_j}\ket{\rm vac},
}
using  \cite{Drummond80}: 
\eqs{PP}{
\op{\rho} &=& \int d^{4M}\bm{\lambda}\ P_+(\bm{\lambda})\,\op{\Lambda}(\bm{\lambda})\label{PPrho}\\
\op{\Lambda}(\bm{\lambda}) &=& \bigotimes_j \frac{\ket{\alpha_j}_j\bra{\beta^*_j}_j}{\bra{\beta_j^*}_j\ket{\alpha_j}_j}.\label{LambdaPP}
}
The kernel $\op{\Lambda}$ involves ``ket'' $\ket{\alpha_j}_j$ and ``bra''  $\bra{\beta^*_j}_j$ states described by separate and independent variables. 
Let us define the container variable 
\eq{lambda}{
\bm{\lambda} = \left\{\alpha_1,\dots,\alpha_M,\beta_1,\dots,\beta_M\right\} = \{ \lambda_{\mu} \} 
}
as a shorthand to hold the full configuration information. 
Individual complex variables $\lambda_{\mu}$ can be indexed by $\mu=1,\dots,2M$.
The \emph{positive-P} distribution $P_+$ is guaranteed real and non-negative \cite{Drummond80}. This is essential for the utility of the method, which lies in its ability to represent full quantum mechanics using stochastic trajectories of the \emph{samples} $\bm{\lambda}$ of the distribution $P_+$. Notably, the size of these samples scales merely linearly with $M$, allowing for the simulation of the full quantum dynamics of systems with even millions of modes \cite{Deuar07,Kheruntsyan12}.

The distribution \eqn{PP} can be compared to the simpler and more widely known \emph{Glauber-Sudarshan P representation} \cite{Glauber63,Sudarshan63}, which uses a single set of coherent state amplitudes and ties the ``bra'' and ''ket'' amplitudes to be equal. 
It is written
\eq{GSP}{
\op{\rho} = \int d^{2M}\bo{\bm{\alpha}}\ P(\bm{\alpha}) \bigotimes_j \ket{\alpha_j}_j\bra{\alpha_j}_j, 
}
where the multimode coherent amplitude is $\bm{\alpha}=\{\alpha_1,\dots,\alpha_M\}$. This representation gives a well behaved representation of only a subset of possible quantum states \cite{Cahill69b,Sperling16}, though a very useful one that includes Gaussian density operators \cite{Agarwal70b}.

For full computational utility of the positive-P method, all quantum mechanical actions should be able to be written in terms of the samples $\bm{\lambda}$. 
Central to this are the following identities: 
\eqs{PPiden}{
\op{a}_j\op{\Lambda} &=& \alpha_j\op{\Lambda},\label{5a}\\
\dagop{a}_j\op{\Lambda} &=& \left[\beta_j+\frac{\partial}{\partial\alpha_j}\right]\op{\Lambda},\label{5b}\\
\op{\Lambda}\op{a}_j &=& \left[\alpha_j+\frac{\partial}{\partial\beta_j}\right]\op{\Lambda},\label{5c}\\
\op{\Lambda}\dagop{a}_j &=& \beta_j\op{\Lambda}.\label{5d}
}

A general master equation for the density matrix written in Lindblad form with reservoir operators $\op{R}_n$ is 
\eq{master}{
\frac{\partial\op{\rho}}{\partial t} = -\frac{i}{\hbar}\left[\op{H},\op{\rho}\right] + \sum_{n}\left[2\op{R}_n\op{\rho}\dagop{R}_n-\dagop{R}_n\op{R}_n\op{\rho}-\op{\rho}\dagop{R}_n\op{R}_n\right].
}
This can be transformed by standard methods \cite{QuantumNoise}, using identities \eqn{PPiden}, to a partial differential equation (PDE) for $P_+$ of the following general form:
\eq{pdeforP}{ 
\frac{\partial P_+}{\partial t} = \left\{\sum_{n=1}^{n_{\rm max}} \sum_{\mu_1,\dots,\mu_k}^{2M}\left(\prod_{k=1}^{n}\frac{\partial}{\partial\lambda_{\mu_k}}\right) F_{\mu_1,\dots,\mu_n}(\bm{\lambda})\right\}P_+,
}
with maximum differential order $n_{\rm max}$ and coefficients $F$ that depend on the details of \eqn{master}.
Most cases have $n_{\rm max}\le2$, in which case this is a Fokker-Planck equation (FPE):
\eq{FPEgen}{
\frac{\partial P_+}{\partial t} = \left\{\sum_{\mu}\frac{\partial}{\partial\lambda_{\mu}}\left[-A_{\mu}(\bm{\lambda})\right] 
+ \sum_{\mu\nu}\frac{\partial^2}{\partial\lambda_{\mu}\partial\lambda_{\nu}}\frac{D_{\mu\nu}(\bm{\lambda})}{2}
\right\} P_+.
}
An exception occurs if irreducible higher order partial derivatives appear, e.g. due to explicit three-body interactions in the Hamiltonian. Most first principles models use only two-particle interactions, though. 

Standard methods convert an FPE like \eqn{FPEgen} to the following Ito stochastic equations of the samples \cite{QuantumNoise}:
\eq{stoch}{
\frac{d\lambda_{\mu}}{dt} = A_{\mu}(\bm{\lambda}) + \sum_{\sigma}B_{\mu\sigma}(\bm{\lambda})\xi_{\sigma}(t)
}
where 
\eq{BBT}{
D_{\mu\nu} = \sum_{\sigma}B_{\mu\sigma}B_{\nu\sigma}
}
i.e. $D=BB^T$ in matrix notation, and $\xi_{\sigma}(t)$ are independent real white noises with variances
\eq{xi}{
\langle\xi_{\mu}(t)\xi_{\nu}(t')\rangle_{\rm stoch} = \delta(t-t')\delta_{\mu\nu}.
}
The notation $\langle \cdot \rangle_{\rm stoch}$ indicates a stochastic average over the samples.
 The equations \eqn{stoch} give us quantum mechanical evolution in terms of the samples, which becomes ever more exact as the number of samples grows.
Quantum expectation values at a given time are evaluated via
\eq{obs}{
\langle \dagop{a}_{j_1}\cdots\dagop{a}_{j_{\mc{N}}} \op{a}_{k_1}\cdots\op{a}_{k_{\mc{M}}}\rangle = 
\langle \beta_{j_1}\cdots\beta_{j_{\mc{N}}} \alpha_{k_1}\cdots\alpha_{k_{\mc{M}}}\rangle_{\rm stoch}.
}
The equivalence can be written more explicitly in terms of $\mc{S}$ individual samples $\bm{\lambda}^{(u)}$ labelled by $u=1,\dots,\mc{S}$ as:
\eqa{obsu}{
\lefteqn{\langle \dagop{a}_{j_1}\cdots\dagop{a}_{j_{\mc{N}}} \op{a}_{k_1}\cdots\op{a}_{k_{\mc{M}}}\rangle}\qquad& \\
&\qquad=\lim_{\mc{S}\to\infty}\frac{1}{\mc{S}}\sum_u
\beta^{(u)}_{j_1}\cdots\beta^{(u)}_{j_{\mc{N}}} \alpha^{(u)}_{k_1}\cdots\alpha^{(u)}_{k_{\mc{M}}}.\nonu
}
The mode labels in the above can be in any combination, provided all annihilation operators are to the right of all creation operators (which is called ``normal ordering''). 
Any single-time operator can be expressed as a sum of normally ordered terms like \eqn{obs}.

\subsection{Multi-time averages in open systems}
\label{S2aiii}

The situation is much more complicated when the operators in the expectation value are not evaluated all at the same time. 
The root of the difficulty is that multi-time commutation relations usually depend non-trivially on the full system dynamics, and a reduction of arbitrary multi-time operators to a normal-ordered form is not generally possible.

To describe what is meant at the operator level it is first helpful to introduce the two-sided evolution operator $\breve{V}(t_1,t_2)$ such that evolution by the master equation \eqn{master} can be summarised as 
\eq{Vtt}{
\op{\rho}(t_2) = \breve{V}(t_2,t_1)\op{\rho}(t_1).
}
Notably, the evolution operator has the semigroup property \cite{Spohn80}
$\breve{V}(t_3,t_2)\breve{V}(t_2,t_1)=\breve{V}(t_3,t_1)$. 
We will use the convention that two-sided operators, indicated by a breve~$\breve{}$\ , act on everything to their right. Hence
\eq{ordering}{
\breve{V}AB = \breve{V}\left\{AB\right\} \neq \left\{\breve{V}A\right\}B.
}

Let us also define the \emph{time-ordered form} via
\eq{timeordered}{
\langle \op{A}_1(t_1)\op{A}_2(t_2)\cdots\op{A}_{{\mc{N}}}(t_{\mc{N}}) \op{B}_1(s_1)\op{B}_2(s_2)\cdots\op{B}_{\mc{M}}(s_{\mc{M}})\rangle
}
where the $\op{A}_j(t)$ and $\op{B}_j(t)$ are Heisenberg picture operators, and the times obey 
\eqs{torder}{
t_1\le t_2 \le &\dots& \le t_{\mc{N}} \\
s_1\ge s_2 \ge &\dots& \ge s_{\mc{M}}.
}
There are $\mc{N}$ operators $\op{A}$ with times labelled $t_1,\dots,t_{\mc{N}}$ increasing to the right, inward, and $\mc{M}$ operators $\op{B}$ with times labelled $s_1,\dots,s_{\mc{M}}$ increasing to the left, also inward.
The location of the inner ``meeting point'' is arbitrary, and the number of $\op{A}$ or $\op{B}$ operators can also be zero. No particular constraints are imposed on the $\op{A}$ and $\op{B}$ operators, except that they should be single time quantities.
These operators refer to measurements made at the respective times $\tau_r$, when the density operator was $\op{\rho}(\tau_r)$. 
Between measurements the state evolves according to the master equation \eqn{master} (i.e. \eqn{Vtt}).

It has been shown \cite{QuantumNoise} that multi-time correlations that correspond to sequences of measurements can always be written in the above form \eqn{timeordered}.
Therefore this time ordering is not an arbitrary one, and not particularly restrictive in itself.
It has also been shown that a general time-ordered correlation function \eqn{timeordered} obeying \eqn{torder} can be written in a form that uses the $\breve{V}$. To do so it is necessary to be careful about operator ordering.
Following  \cite{QuantumNoise},
let us order all the times $t_p$ and $s_q$ in the correlation in sequence from earliest to latest. Let us then rename them $\tau_r$ so that
\eq{taur}{
  \tau_1 \le \tau_2 \le \dots \le \tau_{\mc{R}-1} \le \tau_{\mc{R}}
}
with $\mc{R}=\mc{N}+\mc{M}$. We also define the corresponding two-sided operators $\breve{F}_r$
\eq{Fdef}{
\breve{F}_r\op{\rho} = \left\{\begin{array}{l@{\quad}l}
\op{\rho}\op{A}_p & \text{if $\tau_r=t_p$}\\
\op{B}_q\op{\rho}   & \text{if $\tau_r=s_q$}
\end{array}
\right.
}

Then, 
\eqa{basicOp}{
\lefteqn{\langle \op{A}_1(t_1)\op{A}_2(t_2)\cdots\op{A}_{\mc{N}}(t_{\mc{N}}) \op{B}_1(s_1)\op{B}_2(s_2)\cdots\op{B}_{\mc{M}}(s_{\mc{M}})\rangle}&\nonu\\
&= {\rm Tr}\Big[
\breve{F}_{\mc{R}} \breve{V}(\tau_{\mc{R}},\tau_{\mc{R}-1}) \breve{F}_{\mc{R}-1} \breve{V}(\tau_{\mc{R}-1},\tau_{\mc{R}-2})\cdots \nonu\\
&\qquad\qquad\qquad\qquad\qquad\cdots\breve{F}_2 \breve{V}(\tau_2,\tau_1)\breve{F}_1\op{\rho}(\tau_1))
\Big].\qquad
}

\subsection{Multi-time correlations in the Glauber-Sudarshan~P representation}
\label{S2aii}

Time-ordered correlations for the Glauber-Sudarshan P representation and a range of other single-phase space representations like the Husimi Q were first studied by Agarwal and Wolf in a Greens function framework \cite{Agarwal69a} extended to the case of open systems \cite{Agarwal69b} and for Hamiltonian systems in a formal integral form \cite{Agarwal70c}. 
Later, Gardiner \cite{QuantumNoise} used a different more operational approach to derive equivalent expressions for 
normally and time-ordered operator averages in the Glauber-Sudarshan P representation. These have an intuitive form similar to the single-time stochastic expression \eqn{obs}, as follows: 
\eqa{Pmultitime}{
\langle \dagop{a}_{p_1}(t_1)\cdots\dagop{a}_{p_{\mc{N}}}(t_{\mc{N}}) \op{a}_{q_1}(s_1)\cdots\op{a}_{q_{\mc{M}}}(s_{\mc{M}})\rangle = \qquad\qquad\\
\qquad\qquad\langle \alpha^*_{p_1}(t_1)\cdots\alpha^*_{p_{\mc{N}}}(t_{\mc{N}}) \alpha_{q_1}(s_1)\cdots\alpha_{q_{\mc{M}}}(s_{\mc{M}})\rangle_{\rm stoch}\nonu
}
provided the times are ordered according to \eqn{torder}. 
The requirement that the operators in \eqn{Pmultitime} be normally ordered is an additional constraint on top of time-ordering \eqn{torder}, but one that leads to all operators sorted as they occur in photo-counting
theory \cite{Kelley64}. It covers a very large subset of the potentially physically interesting correlations.
The Gardiner approach is more amenable to extension to doubled phase space and will be used in what follows.

This ordering can be contrasted with the \emph{time-symmetric} ordering for which straightforward  truncated Wigner correspondences for closed system evolution were found in \cite{Berg09,Plimak14,Polkovnikov10}. Examples of time-symmetric ordered quantities are $\tfrac{1}{2}\left[\dagop{a}(t_2)\op{a}(t_1)+\op{a}(t_1)\dagop{a}(t_2)\right]$ and 
$\tfrac{1}{4}\left[\op{a}(t_1)\op{a}(t_2)\dagop{a}(t_3)+\op{a}(t_2)\dagop{a}(t_3)\op{a}(t_1)+\op{a}(t_1)\dagop{a}(t_3)\op{a}(t_2)\right.$ $\left.+\dagop{a}(t_3)\op{a}(t_2)\op{a}(t_1)
\right]$.

\section{Time ordered moments in the positive-P representation}
\label{PPSEC}
\subsection{Normally ordered observables}
\label{S2c}

To derive an expression like \eqn{Pmultitime} for the \emph{positive}-P representation, we will follow Gardiner's approach \cite{QuantumNoise} that was previously used for the Glauber-Sudarshan~P. 
Smaller steps will, however, be taken here to draw attention to a few subtleties that will be necessary later.

\subsubsection{First order correlation function}

First consider the correlation 
\eq{basic}{
G^{(1)}(t',t)=\langle\dagop{a}_j(t')\op{a}_k(t)\rangle. 
}
Comparing to \eqn{basicOp}, we can identify $\op{A}_1 = \dagop{a}_j$, $\op{B}_1=\op{a}_k$, and the times $t_1=t'$, $s_1=t$. For this low order correlation, the time ordering \eqn{torder} sets no additional conditions. 
There are two possibilities for $\tau_1$, depending on whether time $t$ or $t'$ is later. Consider first $t'\ge t$, so that $\tau_1=t$, $\tau_2=t'$. 
Using \eqn{basicOp} and \eqn{Fdef} we have that 
\eqa{basicVa}{
\langle\dagop{a}_j(t')\op{a}_k(t)\rangle &=& \Tr{ \breve{V}(t',t) \left\{\op{a}_k(t)\op{\rho}(t)\right\}\dagop{a}_j(t') }\\
&=& \Tr{ \dagop{a}_j(t') \breve{V}(t',t) \left\{\op{a}_k(t)\op{\rho}(t)\right\} \label{basicV}}
}
The 2nd line follows from the cyclic property of traces. The $\{\cdot\}$ is kept for now for clarity. 
We can see that the evolution operator from $t$ to $t'$ acts to the right on all the quantities at time $t$, while the later-time operator $\dagop{a}_j(t')$ acts only on the evolved quantities to its right. This makes intuitive physical sense. In the second case of $t'<t$, we have $\tau_1=t'$, $\tau_2=t$, and get that
\eq{basicV2}{
\langle\dagop{a}_j(t')\op{a}_k(t)\rangle = 
 \Tr{ \op{a}_k(t) \breve{V}(t,t') \left\{\op{\rho}(t')\dagop{a}_j(t')\right\} }.
}
This again has the intuitive form of the operator $\breve{V}$ acting to the right on all the earlier-time quantities.

Take the first case with $t'\ge t$. 
Expressing the density matrix in \eqn{basicV} in the positive-P representation \eqn{PPrho}, 
\eqa{corr1}{
G^{(1)}&=&\Tr{ \dagop{a}_j(t')\breve{V}(t',t)\left\{\!\int d^{4M}\bm{\lambda}\,P_+(\bm{\lambda},t)\,\op{a}_k(t) \op{\Lambda}(\bm{\lambda})\right\}  }\nonu\\
&&\hspace*{-2.5em}=\Tr{ \dagop{a}_j(t')\breve{V}(t',t)\left\{\!\int d^{4M}\bm{\lambda}\,\alpha_k P_+(\bm{\lambda},t)\,\op{\Lambda}(\bm{\lambda})\right\}  }.\qquad\label{corr1b}
}
The 2nd line follows from application of \eqn{5a}. 
We cannot do the same for $\dagop{a}_j(t')$ yet, because the kernel $\op{\Lambda}$ finds itself inside the prior action of the $\breve{V}$ operator.

To deal with this, consider now the action of the evolution operator $\breve{V}$ on distributions $P_+$.  i.e. the action of the PDE \eqn{pdeforP}. If we define the conditional distribution
$\mc{P}(\bm{\lambda},t'|\ul{\bm{\lambda}},t)$ as the solution of this PDE at time $t'\ge t$ starting from the initial condition $\delta^{4M}(\bm{\lambda}-\ul{\bm{\lambda}})$, i.e. the ``propagator'', then it can be used to formally write 
\eqa{VP}{
\lefteqn{\breve{V}(t',t) \left\{\!\int d^{4M}\bm{\lambda}\,P_+(\bm{\lambda},t)\op{\Lambda}(\bm{\lambda})\right\} }&& \nonu\\
&=&\breve{V}(t',t) \left\{\!\int d^{4M}\bm{\lambda}\int d^{4M}\ul{\bm{\lambda}}\,\delta^{4M}(\bm{\lambda}-\ul{\bm{\lambda}})\,P_+(\ul{\bm{\lambda}},t)\op{\Lambda}(\bm{\lambda})\right\} \hspace*{-1em} \nonu\\
&=& \int d^{4M}\bm{\lambda}\int d^{4M}\ul{\bm{\lambda}}\, \mc{P}(\bm{\lambda},t'|\ul{\bm{\lambda}},t) P_+(\ul{\bm{\lambda}},t)\op{\Lambda}(\bm{\lambda}).
}
This now contains no more two-sided operators. 
Through this convolution, $P_+$ is expressed in $\ul{\bm{\lambda}}$ variables which accompany the earliest time $t$ to aid for later interpretation as part of a joint probability, while the kernel $\op{\Lambda}$ is expressed in the variables $\bm{\lambda}$ that accompany later times, ready for application of the next operator identity.

Notice that there are no particularly stringent assumptions about $P_+$ for \eqn{VP} itself to apply. For example, \eqn{VP} applies equally well if one replaces $P_+$ with some complex distribution function $\wt{P}$. This point will soon be useful. However, there \emph{was} an assumption that the propagator $\mc{P}$ is well behaved. This is certainly true if the PDE was of a Fokker-Planck form \eqn{FPEgen}, and therefore is always justified if we have an exact mapping of the master equation \eqn{master} to stochastic equations \eqn{stoch}. However, in some other cases of third/higher order terms in the PDE, it might not. We will not be concerned with such cases here. 

Now in \eqn{corr1b}, $\breve{V}$ is acting on a distribution $\wt{P}(\bm{\lambda},t)=\alpha_k P_+(\bm{\lambda},t)$. Using \eqn{VP} we get
\eqa{corr2}{
&G^{(1)}=\Tr{ \dagop{a}_j(t')\!\int d^{4M}\bm{\lambda}\,\int d^{4M}\ul{\bm{\lambda}} \mc{P}(\bm{\lambda},t'|\ul{\bm{\lambda}},t) \wt{P}(\ul{\bm{\lambda}},t)\,\op{\Lambda}(\bm{\lambda})  }\hspace*{-3em}\nonu\\
&=\Tr{ \dagop{a}_j(t')\!\int d^{4M}\bm{\lambda}\,\int d^{4M}\ul{\bm{\lambda}}\,\ul{\alpha}_k \mc{P}(\bm{\lambda},t'|\ul{\bm{\lambda}},t) P_+(\ul{\bm{\lambda}},t)\,\op{\Lambda}(\bm{\lambda})  }.\hspace*{-3em}\nonu\\
}
The two-way operator acting on the right that was $\breve{V}$, has now been gotten rid of, by virtue of being incorporated in the propagator $\mc{P}$. Therefore, the remaining operator $\dagop{a}_j$ can now be shifted to the right due to the cyclic property of the trace, and then processed via \eqn{5d} as so:
\eqs{corr3}{
&G^{(1)}\!=\!\int d^{4M}\bm{\lambda}\int d^{4M}\ul{\bm{\lambda}} \mc{P}(\bm{\lambda},t'|\ul{\bm{\lambda}},t) \wt{P}(\ul{\bm{\lambda}},t)\Tr{\op{\Lambda}(\bm{\lambda})\dagop{a}_j(t')  }\hspace*{-1em}\nonu\\
&= \int d^{4M}\bm{\lambda}\,\beta_j\int d^{4M}\ul{\bm{\lambda}}\, \ul{\alpha}_k \mc{P}(\bm{\lambda},t'|\ul{\bm{\lambda}},t) P_+(\ul{\bm{\lambda}},t)\Tr{\op{\Lambda}(\bm{\lambda})  }\hspace*{-1em}\nonu\\
&\hspace*{-3.2em}= \int d^{4M}\bm{\lambda}\,\beta_j\int d^{4M}\ul{\bm{\lambda}}\, \ul{\alpha}_k \mc{P}(\bm{\lambda},t'|\ul{\bm{\lambda}},t) P_+(\ul{\bm{\lambda}},t). 
}
The last line follows from $\Tr{\op{\Lambda}}=1$, which is pre-set by the definition \eqn{LambdaPP}.

The quantity $\mc{P}P_+$ is the just the joint probability 
\eq{jointP}{
P(\bm{\lambda},t';\ul{\bm{\lambda}},t) = \mc{P}(\bm{\lambda},t'|\ul{\bm{\lambda}},t) P_+(\ul{\bm{\lambda}},t) 
}
of having configuration $\ul{\bm{\lambda}}$ at time $t$ and configuration $\bm{\lambda}$ at time $t'$. 
(Provided $\mc{P}$ is well behaved, positive, real, as mentioned before, which is the case for any model fully described by an FPE \eqn{FPEgen}). 
Therefore, 
\eq{corr4}{
G^{(1)}=\int d^{4M}\bm{\lambda}\,\int d^{4M}\ul{\bm{\lambda}}\ \beta_j\ul{\alpha}_k\ P(\bm{\lambda},t';\ul{\bm{\lambda}},t).
}
At this stage we can identify probability with stochastic realisations. The noises $\xi_{\mu}(t)$ introduced during time evolution are independent from each other, independent at each time step, independent for each sample's trajectory. They are also independent of any other random variables used to produce the initial ensemble $\{\bm{\lambda}^{(1)},\dots,\bm{\lambda}^{(\mc{S})}\}(t)$ that samples $P_+(\bm{\lambda},t)$. 
Therefore, the evolved configuration $\bm{\lambda}^{(u)}(t')$ at time $t'$ depends only on mutually independent random variables that consist of $\bm{\lambda}^{(u)}(t)$ and the noise history of the $u$th trajectory. 
As a result, the combination of initial configuration $\bm{\lambda}^{(u)}(t)$ and the evolved configuration $\bm{\lambda}^{(u)}(t')$, together form an unbiased sample of the joint distribution  $P(\bm{\lambda},t';\ul{\bm{\lambda}},t)$. 
We arrive then at the final result that 
\eq{G1result}{
G^{(1)}(t',t)=\langle\dagop{a}_j(t')\op{a}_k(t)\rangle = \langle\beta_j(t')\alpha_k(t)\rangle_{\rm stoch}.
}
The procedure for the case $t'<t$ gives a result analogous to \eqn{corr4}:
\eq{corr4b}{
G^{(1)}=\int d^{4M}\bm{\lambda}\,\int d^{4M}\ul{\bm{\lambda}}\ \ul{\beta}_j\alpha_k\ P(\bm{\lambda},t;\ul{\bm{\lambda}},t'),
}
which once again leads to \eqn{G1result}.

It is especially important to note that the quantity to be averaged comes from the \emph{time evolution of individual sample trajectories}. Explicitly:
\eq{G1resultu}{
\langle\dagop{a}_j(t')\op{a}_k(t)\rangle = \lim_{\mc{S}\to\infty} \frac{1}{\mc{S}}\sum_{u=1}^{\mc{S}} \beta^{(u)}_j(t')\alpha^{(u)}_k(t).
}
This allows for very efficient calculations.

\subsubsection{Higher order correlations}

Other time and normal ordered correlations follow a similar pattern. For example, when $t''\ge t'\ge t$,
\eqa{corraaa1}{
\lefteqn{\langle\op{a}_j(t'')\op{a}_k(t')\op{a}_l(t)\rangle}&&\\
&&\qquad= \Tr{ \op{a}_j(t'')\breve{V}(t'',t')\left\{\op{a}_k(t')\breve{V}(t',t) \left\{\op{a}_l(t)\op{\rho}(t)\right\}\right\} }.\nonu
}
In this case, following a similar procedure to before, one can act alternately with \eqn{5a} on the kernel to extract a factor of $\alpha$, and \eqn{VP} to convert the evolution operators to propagators. 
One finds
\eqa{corraaa2}{
\langle\op{a}_j(t'')\op{a}_k(t')\op{a}_l(t)\rangle&=& \int d^{4M}\bm{\lambda} d^{4M}\ul{\bm{\lambda}}d^{4M}\ul{\ul{\bm{\lambda}}}\ \,\alpha_j\ul{\alpha}_k \ul{\ul{\alpha}}_l\nonu\\
&&\hspace*{-5em}\times \mc{P}(\bm{\lambda},t''|\ul{\bm{\lambda}},t')\mc{P}(\ul{\bm{\lambda}},t'|\ul{\ul{\bm{\lambda}}},t) P_+(\ul{\ul{\bm{\lambda}}},t).
}
Since the times are ordered $t''\ge t'\ge t$, and the conditional probabilities follow from the evolution of the FPE, $\ul{\ul{\bm{\lambda}}}$ are parent variables of the $\ul{\bm{\lambda}}$ and so on, and the product of conditional probabilities is just the joint probability. Hence  \eqn{corraaa2} is 
\eq{corraaa3}{
\int d^{4M}\bm{\lambda} d^{4M}\ul{\bm{\lambda}}d^{4M}\ul{\ul{\bm{\lambda}}}\ \,\alpha_j\ul{\alpha}_k \ul{\ul{\alpha}}_l\ P(\bm{\lambda},t'';\ul{\bm{\lambda}},t';\ul{\ul{\bm{\lambda}}},t),
}
and
\eq{corraaa4}{
\langle\op{a}_j(t'')\op{a}_k(t')\op{a}_l(t)\rangle = \langle\alpha_j(t'')\alpha_k(t')\alpha_l(t)\rangle_{\rm stoch}.
}
Working similarly, using just the 1st and 4th identities in \eqn{PPiden}, one readily but somewhat cumbersomely finds that the stochastic estimator for the general time-and-normal-ordered correlation function is
\eqa{PPmultitime}{
\lefteqn{\langle \dagop{a}_{p_1}(t_1)\cdots\dagop{a}_{p_{\mc{N}}}(t_{\mc{N}}) \op{a}_{q_1}(s_1)\cdots\op{a}_{q_{\mc{M}}}(s_{\mc{M}})\rangle }\qquad&&\nonu\\
&&\hspace*{-1em}=\langle \beta_{p_1}(t_1)\cdots\beta_{p_{\mc{N}}}(t_{\mc{N}}) \alpha_{q_1}(s_1)\cdots\alpha_{q_{\mc{M}}}(s_{\mc{M}})\rangle_{\rm stoch}.\qquad
}
Here, of course \eqn{torder} must hold, and the stochastic averaging is over products constructed using values from the evolution of a single sample, as in \eqn{G1resultu}.
It confirms the suspicion and intuition that the behaviour of the positive-P representation in this regard should be similar to the earlier expression \eqn{Pmultitime}, for the Glauber-Sudarshan~P.

\subsection{Other ordering in the positive-P}
\label{S2d}
Now to see the limitations of this scheme, consider the anti-normally (but time-ordered) ordered correlation
\eq{anti}{
\mc{A}=\langle\op{a}_j(t')\dagop{a}_k(t)\rangle
}
with $t'>t$. The first point to make is that we cannot rearrange this to a normal-ordered form like $\langle\dagop{a}_k(t)\op{a}_j(t')+\delta_{jk}\rangle$ and then use \eqn{PPmultitime} because generally $\left[\op{a}_j(t'),\dagop{a}_k(t)\right]\neq\delta_{jk}$ when $t\neq t'$. Instead it is some time-dependent operator. 
Now applying \eqn{basicOp}, \eqn{anti} can be written
\eq{antiV}{
\hspace*{-0.5em}\mc{A}=\langle\op{a}_j(t')\dagop{a}_k(t)\rangle = \Tr{ \op{a}_j(t')\breve{V}(t',t) \left\{\dagop{a}_k(t)\op{\rho}(t)\right\} }. \hspace*{-0.5em}
}
Upon expansion, we will need to act on $\op{\Lambda}$ using the 2nd identity in \eqn{PPiden}, \eqn{5b},  to convert $\dagop{a}_k$ to variable form. Thus
\eqa{anti2}{
\mc{A}&=&{\rm Tr}\Bigg[ \op{a}_j(t')\breve{V}(t',t)\Bigg\{\nonu\\
&&\int d^{4M}\bm{\lambda}\, P_+(\bm{\lambda},t)\ \left[\beta_k+\frac{\partial}{\partial\alpha_k}\right]\op{\Lambda}(\bm{\lambda})\Bigg\}\Bigg].\qquad
}
This is not of a form amenable to \eqn{VP}. We can soldier on applying integration by parts and assuming negligible boundary terms to obtain
\eqa{anti3}{
\mc{A}&=&{\rm Tr}\Bigg[ \op{a}_j(t')\breve{V}(t',t)\Bigg\{\\
&&\!\int d^{4M}\bm{\lambda}\, \left[\beta_k P_+(\bm{\lambda},t)-\frac{\partial P_+(\bm{\lambda},t)}{\partial\alpha_k}\right]\op{\Lambda}(\bm{\lambda})\Bigg\}\Bigg].\nonu
}
The matter of whether boundary terms can be discarded has been studied in depth \cite{Gilchrist97,Deuar02,Kinsler91,Smith89,Carusotto03b,DeuarPhD}. The summary is that one can determine operationally in a stochastic simulation whether boundary terms are negligible or not. If deemed negligible, then integration by parts is justified. 
In \eqn{anti3} we can now identify a distribution $\wt{P}(\bm{\lambda},t)=[\beta_k-\frac{\partial}{\partial\alpha_k}]P_+(\bm{\lambda},t)$ to act on with \eqn{VP}. Doing so gives:
\eqa{anti4}{
\mc{A}&=& {\rm Tr}\Bigg[ \op{a}_j(t') \!\int\int d^{4M}\bm{\lambda}\, d^{4M}\ul{\bm{\lambda}}\\
&&\, \mc{P}(\bm{\lambda},t'|\ul{\bm{\lambda}},t)\left[ \ul{\beta}_kP_+(\ul{\bm{\lambda}},t)-\frac{\partial P_+(\ul{\bm{\lambda}},t)}{\partial\ul{\alpha}_k}\,\right]\op{\Lambda}(\bm{\lambda})  \Bigg].\nonu
}
The second operator $\op{a}_j(t')$ can now act on $\op{\Lambda}$ from the left.
One obtains 
\eqa{anti5}{
\mc{A}&=& \int\int d^{4M}\bm{\lambda}\, d^{4M}\ul{\bm{\lambda}}\,\\
&& \mc{P}(\bm{\lambda},t'|\ul{\bm{\lambda}},t)\alpha_j\left[ \ul{\beta}_kP_+(\ul{\bm{\lambda}},t)-\frac{\partial P_+(\ul{\bm{\lambda}},t)}{\partial\ul{\alpha}_k}\,\right].\nonu
}
While this is formally acceptable (given those negligible boundary terms), and could be used for some analytic work in small systems such as demonstrated in \cite{QuantumNoise}, 
unfortunately the derivative of $P_+$ is not amenable to interpretation in terms of stochastic samples. 
At least not the direct samples we are investigating here. 
It may be partially treatable using the quantum jump and response theory approach previously applied to truncated Wigner \cite{Berg09,Polkovnikov10,Plimak14}, which is a topic for another time.

However -- in Sections~\ref{S3} and~\ref{OD}, a different direct way to evaluate anti-normal ordered observables such as $\mc{A}=\langle\op{a}_j(t')\dagop{a}_k(t)\rangle$ will be demonstrated.

\section{Other phase-space representations}
\label{OTH}

In the derivations of Sec.~\ref{S2c}, one can see that the crucial aspect for obtaining a stochastically useful expression is to use only those identities which do not contain derivatives%
\footnote{Strictly speaking, a small exception to this appears if the derivative appears only at the final time when the only remaining operator is $\op{\Lambda}$ since Tr$[\frac{\partial}{\partial\lambda_{\mu}}\op{\Lambda}]=\frac{\partial}{\partial\lambda_{\mu}}{\rm Tr}[\op{\Lambda}]=0$ removes any awkward terms. 
Such a case can, however, also be treated by an identity without any derivatives after using the cyclic property of the trace at the right step.}.
This suggests that convenient expressions for multi-time correlations similar to \eqn{PPmultitime} will be obtainable whenever the operators in the correlation can be converted to phase-space variables without resorting to identities with derivatives. 

However, it happens that such identities without derivatives are not particularly abundant in other phase-space representations. 
For example, the Wigner representation \cite{Wigner32,Moyal49,Polkovnikov10} has derivatives in all identities, as does its dimension-doubled analogue \cite{Plimak01,Hoffmann08}. 
Notably, while the trace of the symmetric form in the Wigner representation corresponds to $|\alpha|^2$ with no corrections: ${\rm Tr}\left[\tfrac{1}{2}\left(\dagop{a}\op{a}+\op{a}\dagop{a}\right)\op{\Lambda}_{\rm Wig}\right] = |\alpha|^2$, this does not remove derivatives in the corresponding identity%
\footnote{Here, $\op{\Lambda}_{\rm Wig}=\op{\Lambda}_s$ from \eqn{Psdef} with $s=0$.}. The best that appears to be achievable in this way is
\eq{Wigid}{
\frac{\dagop{a}\op{a}+\op{a}\dagop{a}}{2}\op{\Lambda}_{\rm Wig} = \left[|\alpha|^2+\frac{1}{4}\frac{\partial^2}{\partial\alpha\partial\alpha^*}\right]\op{\Lambda}_{\rm Wig}.
}	
Hence, the path-integral and time-symmetric approach \cite{Berg09} is more suited to the Wigner representation.
Also the phase-space representations developed for spin systems \cite{Barry08,Mandt15,Ng13}, contain derivatives for all identities.  

One notable exception is the Q representation, which admits derivative-free identities similar to \eqn{PPiden} for \emph{anti}-normally ordered operators, and so is a good candidate for convenient phase-space expressions.
Time-ordered anti-normal operators occur for example in the theory of photon detectors that operate via emission rather than absorption of photons \cite{Mandel66}. Formal integral expressions for this kind of correlation were also provided in \cite{Agarwal70c}. Below, the operational stochastic expressions are found using the Gardiner approach.

\subsection{The case of the Q representation}
\label{QREP}

The Husimi Q representation \cite{Husimi40} is defined as 
\eq{Q}{
Q(\bm{\alpha}) = \frac{1}{\pi^M}\bra{\bm{\alpha}}\,\op{\rho}\,\ket{\bm{\alpha}} 
}
and is positive for any $\op{\rho}$. 
Due to the Q distribution being defined in this explicit way, rather than the implicit form \eqn{PP},
observable expressions in the Q distribution have traditionally been found by simply expanding the trace: 
\eq{tralpha}{
\Tr{\op{O}\,\op{\rho}\,} = \frac{1}{\pi^M}\int d^{2M}\bm{\alpha}\ \bra{\bm{\alpha}}\op{O}\,\op{\rho}\,\ket{\bm{\alpha}},
}
and applying the eigenvalue equation for coherent states 
\eq{eigalpha}{
\op{a}_j\ket{\bm{\alpha}}=\alpha_j\ket{\bm{\alpha}}.
}
For anti-normal ordered correlations at equal times, the cyclic property of traces gives
$\langle \op{a}_{j_1}\cdots\op{a}_{j_{\mc{N}}} \dagop{a}_{k_1}\cdots\dagop{a}_{k_{\mc{M}}}\rangle = 
\Tr{\dagop{a}_{k_1}\cdots\dagop{a}_{k_{\mc{M}}}\op{\rho}\ \op{a}_{j_1}\cdots\op{a}_{j_{\mc{N}}}
}$
which immediately leads to
\eq{obsQ}{\langle \op{a}_{j_1}\cdots\op{a}_{j_{\mc{N}}} \dagop{a}_{k_1}\cdots\dagop{a}_{k_{\mc{M}}}\rangle 
= 
\langle \alpha_{j_1}\cdots\alpha_{j_{\mc{N}}} \alpha^*_{k_1}\cdots\alpha^*_{k_{\mc{M}}}\rangle_{\rm stoch}.
}
However, this traditional approach fails with time-ordered correlations. 
Whatever way one orders the operators, working this way on expression \eqn{basicOp} will lead to a form
\eqa{QbasicOptrad}{
\lefteqn{\frac{1}{\pi^M}\int d^{2M}\bm{\alpha}\bra{\bm{\alpha}}\Bigg[\breve{F}_R }&&\\
&&
\breve{V}(\tau_R,\tau_{R-1})\left\{ 
\breve{F}_{R-1} \breve{V}(\tau_{R-1},\tau_{R-2}) \big\{  
\cdots \breve{F}_1\op{\rho}(\tau_1) \big\}
\right\}\Bigg]\ket{\bm{\alpha}}\nonu
}
in which it is the \emph{latest time} operator $\breve{F}_R(\tau_R)$  that acts on the outer coherent states. One can convert $\breve{F}_R$ to phase-space variables using \eqn{eigalpha}, and move the state vectors $\ket{\bm{\alpha}}$ or $\bra{\bm{\alpha}}$ closer to the density matrix and the form \eqn{Q}. However, in the next putative step, there is no clear way to convert the evolution operator $\breve{V}(\tau_R,\tau_{R-1})$ to phase space form. Moreover, when there is phase-space diffusion, the propagator $\mc{P}$ is well-behaved only in the forward time direction, so there is no way to act with  $\mc{P}$ on the outer state vectors and variables which correspond to later times than the inner ones.

Therefore, we proceed in a non-traditional way, using an implicit form similar to what was done for the positive-P distribution in Sec.~\ref{S2c}.
The Q representation is the \mbox{$s\to-1$} limiting case of the family of s-ordered representations $W_s(\bm{\alpha})$ studied by Cahill and Glauber \cite{Cahill69a,Cahill69b} (the Glauber-Sudarshan~P and Wigner correspond to $s=1$ and $s=0$, respectively).
All these distributions can be written using coherent displacement operators
\eq{Dalpha}{
\op{D}_j(\alpha_j) = e^{\alpha_j\widehat{a}^{\,\dagger}_j-\alpha_j^*\op{a}_j},\ ;\ 
\op{D}(\bm{\alpha}) = \prod_{j} \op{D}_j(\alpha_j),
}
in an implicit form similar to \eqn{PP}:
\eqs{Psdef}{
\op{\rho} &=& \int d^{2M}\bm{\alpha}\,W_s(\bm{\alpha}) \op{\Lambda}_s(\bm{\alpha})\label{Psdefrho}\\
\op{\Lambda}_s(\bm{\alpha}) &=& \prod_j \op{D}_j(\alpha_j) \op{T}_j(0,-s) \op{D}_j(-\alpha_j),\label{PsdefLambda}
}
with the base operator \cite{Cahill69b} 
\eq{Ts}{
 \op{T}_j(0,-s) = \frac{2}{1+s}\left(\frac{s-1}{1+s}\right)^{\widehat{a}^{\,\dagger}_j\op{a}_j}.
}
One has $\Tr{\op{\Lambda}_s}=1$ and the operator identities 
\eqs{siden}{
\op{a}_j\op{\Lambda}_s &=& \left[\alpha_j-\frac{1-s}{2}\frac{\partial}{\partial\alpha^*_j}\right]\op{\Lambda}_s,\\
\dagop{a}_j\op{\Lambda}_s &=& \left[\alpha^*_j+\frac{1+s}{2}\frac{\partial}{\partial\alpha_j}\right]\op{\Lambda}_s,\\
\op{\Lambda}_s\op{a}_j &=& \left[\alpha_j+\frac{1+s}{2}\frac{\partial}{\partial\alpha^*_j}\right]\op{\Lambda}_s,\\
\op{\Lambda}_s\dagop{a}_j &=& \left[\alpha^*_j-\frac{1-s}{2}\frac{\partial}{\partial\alpha_j}\right]\op{\Lambda}_s,
}
which can be verified by equating the LHS and RHS when $\op{T}(0,-s)$ is expanded in number states.
Therefore, in the limit $s\to-1$ corresponding to the Q representation with $\op{\Lambda}_{-1}=\op{\Lambda}_Q$, 
\eqs{Qiden}{
\op{a}_j\op{\Lambda}_Q &=& \left[\alpha_j-\frac{\partial}{\partial\alpha^*_j}\right]\op{\Lambda}_Q,\\
\dagop{a}_j\op{\Lambda}_Q &=& \alpha^*_j\op{\Lambda}_Q,\label{idQb}\\
\op{\Lambda}_Q\op{a}_j &=& \alpha_j\op{\Lambda}_Q,\label{idQc}\\
\op{\Lambda}_Q\dagop{a}_j &=& \left[\alpha^*_j-\frac{\partial}{\partial\alpha_j}\right]\op{\Lambda}_Q.
}
We see then that multi-time correlations in which only the orderings $\dagop{a}_j\op{\Lambda}_Q$ and $\op{\Lambda}_Q\op{a}_j$ appear could have the capacity to correspond to simple stochastic expressions. 
This implies anti-normal ordered moments since by the cyclic property of traces
\eq{anmom}{
\Tr{\dagop{a}_{q_1}\cdots\dagop{a}_{q_{\mc{M}}}\op{\rho}\,\op{a}_{p_1}\cdots\op{a}_{p_{\mc{N}}}}
=\langle \op{a}_{p_1}\cdots\op{a}_{p_{\mc{N}}}\dagop{a}_{q_1}\cdots\dagop{a}_{q_{\mc{M}}}\rangle.
}

The implicit form \eqn{Psdef} for $\op{\rho}$ allows us to proceed in a similar fashion to Sec.~\ref{S2c}.
Consider then the anti-normal and time-ordered correlation
\eq{defAd}{
\mc{A}' =\langle \op{a}_{p_1}(t_1)\cdots\op{a}_{p_{\mc{N}}}(t_{\mc{N}})\dagop{a}_{q_1}(s_1)\cdots\dagop{a}_{q_{\mc{M}}}(s_{\mc{M}})\rangle
}
with times obeying \eqn{torder}.
A potentially awkward issue is that the kernel $\op{\Lambda}_s$ is not very well bounded in the \mbox{$s\to-1$} limit, with
$\bra{\bm{\beta}}\op{\Lambda}_s(\bm{\alpha})\ket{\bm{\beta}} = \frac{2}{\ve}\exp\left[-2|\bm{\alpha}-\bm{\beta}|^2/\ve\right]$ where $s=\ve-1$. 
To gauge whether this is a problem, we will work using infinitesimal $\ve>0$ in which case 
\eq{LambdaQlim}{
\op{\Lambda}_s \to \op{\Lambda}_Q + \mc{O}(\ve).
}
Using the form \eqn{basicOp} on \eqn{anmom} with now $B=\dagop{a}$ and $A=\op{a}$, one obtains
\eqa{QbasicOp}{
\lefteqn{\mc{A}' = \mc{O}(\ve) + \int d^{2M}\bm{\alpha}\,
{\rm Tr}\Bigg[\breve{F}_R\breve{V}(\tau_R,\tau_{R-1})\Bigg\{ 
}&&\\
&&\cdots\breve{F}_3
\breve{V}(\tau_3,\tau_2)\Big\{\breve{F}_2
\breve{V}(\tau_2,\tau_1)\big\{W_s(\bm{\alpha},\tau_1) \breve{F}_1 \op{\Lambda}_Q(\bm{\alpha}) \big\}
\Big\}
\Bigg\}\Bigg].\nonu
}
The part in the inner $\{\ \}$ brackets will be either 
\eq{eior1}{
\begin{array}{l@{\qquad\qquad}l}
W_s(\bm{\alpha},t_1)\op{\Lambda}_Q\op{a}_{p_1} & \text{if $t_1\le s_{\mc{M}}$}\\
W_s(\bm{\alpha},s_{\mc{M}})\dagop{a}_{q_{\mc{M}}}\op{\Lambda}_Q   & \text{otherwise.}
\end{array}
}
Both cases allow us to use derivative-free identities \eqn{idQc} or \eqn{idQb}, to replace \eqn{eior1} with $\wt{P}_s(\bm{\alpha},\tau_1)\op{\Lambda}_Q$ where
\eq{eior2}{
\begin{array}{l@{\quad\qquad}l}
\wt{P}_s = \alpha_{p_1}W_s(\bm{\alpha},s_{\mc{M}}) & \text{if $t_1\le s_{\mc{M}}$}\\
\wt{P}_s = \alpha^*_{q_{\mc{M}}}W_s(\bm{\alpha},t_1) & \text{otherwise.}
\end{array}
}
Now assuming an acceptable propagator $\mc{P}_s(\bm{\alpha},\tau_2|\ul{\bm{\alpha}},\tau_1)$ exists for the evolution of the s-ordered distribution $W_s$, use of \eqn{VP} 
leads to 
\eqa{QbasicOp2}{
\lefteqn{\mc{A}' = \mc{O}(\ve) + \int d^{2M}\bm{\alpha}\,\int d^{2M}\ul{\bm{\alpha}}\,
{\rm Tr}\Bigg[\breve{F}_R\breve{V}(\tau_R,\tau_{R-1})\Bigg\{ 
}&&\nonu\\
&&\cdots\breve{F}_3
\breve{V}(\tau_3,\tau_2)\Big\{
\mc{P}_s(\bm{\alpha},\tau_2|\ul{\bm{\alpha}},\tau_1)\wt{P}_s(\ul{\bm{\alpha}},\tau_1)\breve{F}_2\op{\Lambda}_Q
\Big\}
\Bigg\}\Bigg],\qquad
}
in which the objects related to the first time interval have been fully converted to phase-space quantities. 
Proceeding in this fashion with increasing time for the remaining $\op{a}$ and $\dagop{a}$ operators, one arrives at
\eqa{QbasicOp3}{
\lefteqn{\mc{A}' = \bigg[\int d^{2M}\bm{\alpha}(\tau_1)\,\int d^{2M}\bm{\alpha}(\tau_2)\,\cdots\,\int d^{2M}\bm{\alpha}(\tau_R) }&&\nonu\\
&&\qquad\qquad\qquad\alpha_{p_1}(t_1)\cdots\alpha_{p_{\mc{N}}}(t_{\mc{N}})\alpha^*_{q_1}(s_1)\cdots\alpha^*_{q_{\mc{M}}}(s_{\mc{M}}) \times \nonu\\
&&\mc{P}_s(\bm{\alpha}(\tau_R),\tau_R|\bm{\alpha}(\tau_{R-1}),\tau_{R-1})\cdots
\mc{P}_s(\bm{\alpha}(\tau_2),\tau_2|\bm{\alpha}(\tau_1),\tau_1)\nonu\\
&&\qquad\qquad\qquad\qquad\qquad\times W_s(\bm{\alpha}(\tau_1),\tau_1)\bigg]
+\mc{O}(\ve).
}
Variables $\bm{\alpha}, \ul{\bm{\alpha}}$ etc were relabelled to $\bm{\alpha}(\tau_R), \bm{\alpha}(\tau_{R-1})$.

 In the limit $\ve\to 0$ that we are considering, $W_s(\bm{\alpha})\to Q(\bm{\alpha})$, which is real non-negative. To be consistent, the propagator $\mc{P}_{-1}$ must also be real nonnegative. 
When the master equation \eqn{master} is faithfully reproduced by a Fokker-Planck equation for $Q(\bm{\alpha})$, this will be the case. 
Then,  the 
$\mc{P}_s\dots\mc{P}_sW_s$ factors can be interpreted similarly to \eqn{jointP} as the joint probability of samples $\bm{\alpha}(\tau_1), \bm{\alpha}(\tau_2), \dots$ at successive times. 
With that, we arrive at the hoped for result that a time ordered (as \eqn{torder}) and \emph{anti}-normally ordered correlation is evaluated as
\eqa{Qmultitime}{
\lefteqn{\langle \op{a}_{p_1}(t_1)\cdots\op{a}_{p_{\mc{N}}}(t_{\mc{N}}) \dagop{a}_{q_1}(s_1)\cdots\dagop{a}_{q_{\mc{M}}}(s_{\mc{M}})\rangle }\qquad&&\nonu\\
&&\hspace*{-0.5em}=\langle \alpha_{p_1}(t_1)\cdots\alpha_{p_{\mc{N}}}(t_{\mc{N}}) \alpha^*_{q_1}(s_1)\cdots\alpha^*_{q_{\mc{M}}}(s_{\mc{M}})\rangle_{\rm stoch}\qquad
}
using \emph{Q representation} samples $\alpha_j$.
Stochastic averaging is over products constructed using values from the evolution of a single sample, like in \eqn{G1resultu}.

\section{Evaluation by conversion to the Q representation}
\label{S3}

\subsection{Conversion between samples of s-ordered distributions}
\label{CONVs}
The s-ordered distributions $W_s$ introduced in \eqn{Psdef} are mutually related by \cite{Cahill69b} 
\eqa{sschange}{
\lefteqn{W_s(\bm{\alpha}')}&& \\&&= \left(\frac{2}{s_0-s}\right)^M \int \frac{d^{2M}\bm{\alpha}}{\pi^M} \exp\left[-\frac{2|\bm{\alpha}'-\bm{\alpha}|^2}{s_0-s}\right] W_{s_0}(\bm{\alpha}).\nonu
}
in the sense that $W_s$ and $W_{s_0}$ represent the same quantum density matrix $\op{\rho}$. 
When $s_0>s$, this is a Gaussian convolution of the more normally-ordered distribution $W_{s_0}$, 
and reflects the well known property that Q distributions ($s=-1$) are broader and more smoothed than Wigner ($s=0$), which are in turn broader than Glauber-Sudarshan P distributions ($s=1$) for the same state. 
Importantly for us here, this means that if we have samples $\bm{\alpha}$ of a more normally ordered distribution, we can easily also obtain samples $\bm{\alpha}'$ of the less normally ordered distributions simply by adding Gaussian noise.
The prescription is
\eq{converts}{
\alpha'_j = \alpha_j+\sqrt{\frac{s_0-s}{2}}\ \zeta_j
}
for each mode $j$, with each $\zeta_j$ a complex random variable of variance 1:
\eq{zetadef}{
\langle\zeta_j\rangle_{\rm stoch} = 0\ ;\ 
\langle\zeta_j\zeta_k\rangle_{\rm stoch} = 0\ ;\ 
\langle\zeta^*_j\zeta_k\rangle_{\rm stoch} = 1.
}
In particular, 
converting samples of a P distribution to samples of Q requires
\eq{convPQ}{
\alpha'_j = \alpha_j+\zeta_j.
}
Converting samples of a Wigner distribution (if it is nonnegative to begin with) to samples of Q can be done with
\eq{convWQ}{
\alpha'_j = \alpha^{\rm Wig}_j+\frac{\zeta_j}{\sqrt{2}}.
}

\subsection{Evaluation of anti-normal ordered moments starting from P and Wigner representations}
\label{ANP}

Therefore, if one has samples $\bm{\alpha}(t_0)$ of a P or Wigner representation up to a time $t_0$ (e.g. from a prior stochastic evolution), 
the prescription \eqn{converts} can be used to convert them to samples $\bm{\alpha}'(t_0)$ of the Q representation at that time. 
The noises $\zeta_j$ are generated just once at this time.  
Subsequent evolution according to the Q stochastic equations then leads to Q samples $\bm{\alpha}'(t)$ at later times $t>t_0$. These can be directly used in expression \eqn{Qmultitime} to evaluate anti-normal ordered multi-time observables.

This is potentially a little less convenient that remaining in one distribution throughout because the conversion time $t_0$ has to be chosen before starting a simulation, and the anti-normal ordered operators cannot extend to times before $t_0$. It does preserve the principal advantages of phase-space simulation, though: intuitive and computationally tractable expressions for observables, stochasticity, gentle scaling with system size. The evolution equations \eqn{stoch} are usually of similar form in all s-ordered representations, apart from simplification at special $s$ values.  

One cannot convert the other way with this procedure (e.g. from Wigner to P), so normally ordered multi-time correlations cannot be extracted this way from samples of Wigner or Q distributions.

\subsection{Mixed-order moments}
\label{NNP}

The above discussion suggests a way that some mixed-order multi-time correlations that contain both normal and anti-normal factors could be evaluated. Suppose early time operators  ($t\le t_0$) are normally ordered (can be evaluated using the P distribution), while later time operators ($t>t_0$) are anti-normally ordered (can be evaluated using the Q representation). A switch according to \eqn{convPQ} could be made at $t_0$ after evaluating any normally-ordered factors, and the Q distribution evolved and used at later times to obtain the remaining inner factors with $t>t_0$. Let us check in detail whether this is feasible.

Consider the time-ordered but neither anti- or normally ordered correlation 
\eqa{mixcor}{
&&\lefteqn{ \mc{A}'' = \langle \dagop{a}_{p_1}(t_1)\op{a}_{p_2}(t_2) \dagop{a}_{q_1}(s_1)\op{a}_{q_2}(s_2)\rangle }\nonu\\
&&=
 \int d^{2M}\bm{\alpha}\,{\rm Tr}\bigg[\,
\dagop{a}_{q_1}\breve{V}(s_1,t_2)\bigg\{ \\
&&
\breve{V}(t_2,s_2)\Big\{
\op{a}_{q_2}\breve{V}(s_2,t_1)\big\{
P(\bm{\alpha},t_1) \op{\Lambda}_1(\bm{\alpha})\dagop{a}_{p_1}\big\}
\Big\}\op{a}_{p_2}
\bigg\}\bigg]\nonu
}
whose times satisfy \eqn{torder}, and $t_1\le s_2\le t_2< s_1$ (for instance). 
The initial expansion of $\op{\rho}$ is made in the $P$ representation, where $P(\bm{\alpha})=W_1(\bm{\alpha})$ and $\op{\Lambda}_1(\bm{\alpha})=\ket{\bm{\alpha}}\bra{\bm{\alpha}}$.
The first two  operators $\dagop{a}_{p_1}$, $\op{a}_{q_2}$, and first two $\breve{V}$ convert easily to phase-space variables via \eqn{5a} and \eqn{5d}, giving

\eqa{mixeq1}{
&&\lefteqn{
\mc{A}'' = \iiint d^{2M}\bm{\alpha}\,d^{2M}\ul{\bm{\alpha}}\,d^{2M}\ul{\ul{\bm{\alpha}}}\,{\rm Tr}\Bigg[\ \dagop{a}_{q_1}\breve{V}(s_1,t_2)\Bigg\{ }	\\
&&
\mc{P}_1(\bm{\alpha},t_2|\ul{\ul{\bm{\alpha}}},s_2)
\mc{P}_1(\ul{\ul{\bm{\alpha}}},s_2|\ul{\bm{\alpha}},t_1)
P(\ul{\bm{\alpha}},t_1)
\,\ul{\ul{\alpha}}_{q_2}\,\ul{\alpha}^*_{p_1}\op{\Lambda}_1(\bm{\alpha})\op{a}_{p_2}\Bigg\}\Bigg].\nonu\hspace*{-1em}
}
Further work by this route is now closed because $\op{\Lambda}_1\op{a}_{p_2}$ involves \eqn{5c} and derivatives. 
Instead, another result that follows from from \cite{Cahill69a} 
 allows us to convert kernels:
\eqa{kernelchange}{
\lefteqn{\op{\Lambda}_s(\bm{\alpha})}&& \\&&= \left(\frac{2}{s-s_0}\right)^M \int \frac{d^{2M}\bm{\alpha}'}{\pi^M} \exp\left[-\frac{2|\bm{\alpha}'-\bm{\alpha}|^2}{s-s_0}\right] \op{\Lambda}_{s_0}(\bm{\alpha}').\nonu
}
Taking $s=1$ and $s_0=-1$, to move to a Q representation kernel, one finds
\eqa{kernelPQ}{
\op{\Lambda}_1(\bm{\alpha}) = \int \frac{d^{2M}\bm{\zeta}}{\pi^M} \exp\left[-|\bm{\zeta}|^2\right] \op{\Lambda}_Q(\bm{\alpha}').
}
where $\bm{\zeta}=\bm{\alpha}'-\bm{\alpha}=\{\zeta_j\}$ has exactly the properties of the noise in \eqn{zetadef}, and $\bm{\alpha}'=\bm{\alpha}+\bm{\zeta}$ is given by \eqn{convPQ}.
After substituting \eqn{kernelPQ} into \eqn{mixeq1}, applying \eqn{idQc} to $\op{\Lambda}_Q(\bm{\alpha}')\op{a}_{p_2}$, and defining the distribution
\eqa{PtildeQ}{
\wt{P}(\bm{\alpha}',t_2) &=& \alpha'_{p_2}\!\!\int \frac{d^{2M}\bm{\alpha}}{\pi^M} e^{-|\bm{\alpha}'-\bm{\alpha}|^2} \!\!\!\iint\!\!d^{2M}\ul{\bm{\alpha}}\,d^{2M}\ul{\ul{\bm{\alpha}}} \qquad\\
&&\hspace*{-2em}
\mc{P}_1(\bm{\alpha},t_2|\ul{\ul{\bm{\alpha}}},s_2)
\mc{P}_1(\ul{\ul{\bm{\alpha}}},s_2|\ul{\bm{\alpha}},t_1)
P(\ul{\bm{\alpha}},t_1)
\,\ul{\ul{\alpha}}_{q_2}\,\ul{\alpha}^*_{p_1}  .\nonu
}
we have
\eq{mixeq1a}{
\mc{A}'' = \int d^{2M}\bm{\alpha}'\,{\rm Tr}\Bigg[\ \dagop{a}_{q_1}\breve{V}(s_1,t_2)\Bigg\{ 
\wt{P}(\bm{\alpha}',t_2)\, \op{\Lambda}_Q(\bm{\alpha}')\Bigg\}\Bigg].
}
Now using \eqn{VP}, and taking care with variable labels:
\eqa{mixeq1b}{
\mc{A}'' &=& \iint d^{2M}\bm{\alpha}'\,d^{2M}\ul{\bm{\alpha}}'\    
\mc{P}_{-1}(\bm{\alpha}',s_1|\ul{\bm{\alpha}}',t_2) 
\wt{P}(\ul{\bm{\alpha}}	',t_2)\,\nonu\\
&&\qquad\qquad\qquad\qquad\times {\rm Tr}\left[\dagop{a}_{q_1}\op{\Lambda}_Q(\bm{\alpha}')\right].
}
After the variable change caused by \eqn{VP}, now
\eq{alfui}{
\ul{\bm{\alpha}}' = \bm{\alpha}+\bm{\zeta}.
}
Notice also the $-1$ label on the latest propagator in \eqn{mixeq1b}, indicating that it is according to the Q representation equations in the time interval $(t_2,s_2]$. Finally, applying \eqn{idQb}, one arrives at:
\eqa{mixeq2}{
&&\lefteqn{
\mc{A}'' = \int d^{2M}\ul{\bm{\alpha}}\,d^{2M}\ul{\ul{\bm{\alpha}}}\, d^{2M}\bm{\alpha}\,d^{2M}\bm{\zeta}\,\left(\frac{e^{-|\bm{\zeta}|^2}}{\pi^M}\right)\,d^{2M}\bm{\alpha}' }\nonu\\
&&
\qquad\times\ \ul{\ul{\alpha}}_{q_2}\,\ul{\alpha}^*_{p_1}  (\alpha_{p_2}+\zeta_{p_2}) \alpha^{\prime *}_{q_1}\mc{P}_{-1}(\bm{\alpha}',s_1|\bm{\alpha}+\zeta,t_2)\nonu\\
&&
\qquad\times
\mc{P}_1(\bm{\alpha},t_2|\ul{\ul{\bm{\alpha}}},s_2)
\mc{P}_1(\ul{\ul{\bm{\alpha}}},s_2|\ul{\bm{\alpha}},t_1)
P(\ul{\bm{\alpha}},t_1).
}
This encodes the following sequence of operations:
\begin{enumerate}
\item Start with initial samples $\ul{\bm{\alpha}}$ at $t_1$.
\item Propagate P representation equations to $s_2$ obtaining samples $\ul{\ul{\bm{\alpha}}}$.
\item Propagate P representation equations to $t_2$ obtaining samples $\bm{\alpha}$.
\item Add Gaussian noise as per \eqn{alfui}, to get samples $\ul{\bm{\alpha}}'$.
\item Propagate Q representation equations to $s_1$ obtaining samples $\bm{\alpha}'$.
\end{enumerate}
Along the way, samples are collected to use in the final stochastic average.
The $\mc{P}_s$ factors in \eqn{mixeq2} together with the Gaussian factor in the top line form the joint probability 
$P(\bm{\alpha}',s_1;\ul{\bm{\alpha}}',t_2;\ul{\ul{\bm{\alpha}}},s_2;\ul{\bm{\alpha}},t_1)$. Therefore, the final estimator for the observable $\mc{A}''$ is 
\eqa{estmixcor}{
\lefteqn{\mc{A}''=\langle \dagop{a}_{p_1}(t_1)\op{a}_{p_2}(t_2) \dagop{a}_{q_1}(s_1)\op{a}_{q_2}(s_2)\rangle  = }\qquad\qquad\qquad&&\\
&&\langle \alpha_{p_1}^*(t_1)
\alpha'_{p_2}(t_2)
\alpha^{\prime *}_{q_1}(s_1)
\alpha_{q_2}(s_2)\rangle_{\rm stoch}.\nonu
}
Primed variables $\alpha'$ are samples of the Q distribution, while un-primed ones $\alpha$ are samples of the P distribution.
This all matches intuitively with the evolution and the expectation that anti-normally ordered elements will use Q distribution samples and normally-ordered elements samples of the P distribution.

\subsection{Most general ordering case}
\label{GENORD}

The widest generalisation of this procedure to other (time-ordered) cases is as follows:
If there is an early time set of normally ordered operators, on either side of the correlation, it can be dealt with by sampling the P distribution according to the replacements 
\eq{replP}{
\op{a}_j\to\alpha_j\qquad;\qquad \dagop{a}_j\to\alpha_j^*.
}
 Once this avenue becomes exhausted, one adds noise via \eqn{convPQ} to convert $\alpha$'s to Q distribution samples $\alpha'$. If the remaining later time inner factors are anti-normally ordered, they can then be dealt with using the replacement
\eq{replQ}{
\op{a}_j\to\alpha'_j\qquad;\qquad \dagop{a}_j\to\alpha_j^{\prime *}.
}
The above covers both fully normal and fully anti-normal ordered products as special cases.
On the other hand, the case of an early time anti-normally ordered block and a later time normally ordered block containing several times is not amenable to this approach because one cannot stochastically convert Q samples to P samples. 

A large number of cases can also be reduced to a form amenable to this procedure  by the use of the commutator $[\op{a},\dagop{a}]=1$ for equal time factors.
Also, simulations starting in a Wigner representation can be used for evaluation of outer symmetric ordered parts, and then a switch can be made to the Q representation via \eqn{convWQ} to evaluate any remaining inner anti-normal ordered parts.

\subsection{Correlation function coverage}
\label{TABS}

Table~\ref{tab:combos} counts the number of distinct 1, 2, and 3 annihilation/creation operator products that can  be evaluated using the various methodology described above. All one and two-factor combinations can be evaluated (though 4/12 of the latter require use of the Q representation). For third order correlations, almost all time-ordered cases can be evaluated (74 out of 80). The majority require use of the Q representation or a shift from a P to Q representation as described in Sec.~\ref{GENORD} to work. There are 6 exceptions that cannot be evaluated: 
$\langle\dagop{a}(t_2)\dagop{a}(t_3)\dagop{a}(t_1)\rangle$, 
$\langle\op{a}(t_2)\dagop{a}(t_3)\op{a}(t_1)\rangle$, 
$\langle\op{a}(t_1)\op{a}(t_3)\op{a}(t_2)\rangle$, 
$\langle\op{a}(t_1)\dagop{a}(t_3)\op{a}(t_2)\rangle$, 
$\langle\op{a}(t_3)\op{a}(t_2)\dagop{a}(t_1)\rangle$, 
$\langle\dagop{a}(t_3)\op{a}(t_2)\dagop{a}(t_1)\rangle$, 
where $t_1<t_2<t_3$ is assumed. 
These all share the feature that the earliest factor already requires the Q representation, while the next operator in time requires the P representation to which one cannot return.
There are also a number of correlations that are not time ordered, for which the earliest time $t_1$ is on the middle operator, and these are not possible to evaluate according to the schemes presented here.

\begin{table}
\begin{center}
\begin{tabular}{|l|c|c|c|}
\hline
Order 					& 1st			& 2nd			& 3rd	\\
(number of operators)		& order		& order		& order	\\
\hline\hline
\bf Total permutations		&	\bf2		&	\bf12		& 	\bf104	\\
\hline					
single time correlations		&	2		&	4		&	8	\\
\hline
multi-time accessible	&&&\\
 with P representation 		&	--		&	4		&	22	\\
\hline
additional accessible	&&&\\
 with Q representation		&	--		&	4		&	22	\\
\hline
additional accessible	&&&\\
 with mixed order (Sec.~\ref{GENORD}) &	--		&	--		&	22	\\
\hline 
\bf Total doable			&	\bf2		&	\bf12		&	\bf74	\\
\hline
time ordered not doable		&	--		&	--		&	6	\\
\hline
Not time ordered, not doable	&	--		&	--		&	24	\\
\hline
\end{tabular}
\end{center}\vspace*{-0.5cm}
\caption{A tally of $\op{a}$, $\dagop{a}$ product permutations that can/cannot be evaluated with the various approaches discussed. The general form considered is $\langle \op{A}(t_a)\op{B}(t_b)\op{C}(t_c)\rangle$, where $\op{A}, \op{B}, \op{C}$ can be either of $\op{a}$ or $\dagop{a}$ (same mode), and the time arguments can take up to three distinct times $t_1<t_2<t_3$. Permutations with the same time topology (e.g. $\op{A}(t_1)\op{B}(t_1)\op{C}(t_2)$ and $\op{A}(t_2)\op{B}(t_2)\op{C}(t_3)$) are counted only once. 
\label{tab:combos}}
\end{table}

The greatest interest in multi-time correlations usually concerns those involving two times (say $t=0$ and $t=\tau>0$). Examples are counting correlations like $\langle\dagop{a}(0)\dagop{a}(\tau)\op{a}(\tau)\op{a}(0)\rangle$, and pair correlations such as $\langle\dagop{a}(0)\dagop{a}(0)\op{a}(\tau)\op{a}(\tau)\rangle$. 
The case count for these is summarised by Table~\ref{tab:2time}. 
A we can see, all 160 kinds of time ordered four-operator products of this form can be evaluated, including atypical combinations such as  $\langle\op{a}(0)\dagop{a}(\tau)\dagop{a}(\tau)\op{a}(0)\rangle$, but very many require the Q representation. (72 out of the 160 accessible ones require the use of the mid-simulation switching to the Q representation described in Sec.~\ref{GENORD}).
The only correlations that are inaccessible are the non time ordered ones such as e.g. $\langle\dagop{a}(\tau)\dagop{a}(0)\op{a}(0)\op{a}(\tau)\rangle$.

\begin{table}
\begin{center}
\begin{tabular}{|l|c|c|c|}
\hline
Order 					& 2nd			& 3rd			& 4th	\\
(number of operators)		& order		& order		& order	\\
\hline\hline
\bf Total permutations		&	\bf12		&	\bf56		& 	\bf240	\\
\hline					
single time correlations		&	4		&	8		&	16	\\
\hline
multi-time accessible	&&&\\
 with P representation 		&	4		&	14		&	36	\\
\hline
additional accessible	&&&\\
 with Q representation		&	4		&	14		&	36	\\
\hline
additional accessible	&&&\\
 with mixed order (Sec.~\ref{GENORD}) &	--		&	12		&	72	\\
\hline 
\bf Total doable			&	\bf12		&	\bf48		&	\bf160	\\
\hline
time ordered not doable		&	--		&	--		&	--	\\
\hline
Not time ordered, not doable	&	--		&	8		&	80	\\
\hline
\end{tabular}
\end{center}\vspace*{-0.5cm}
\caption{A tally of $\op{a}$, $\dagop{a}$ products involving up to four operators, evaluated at one of two times. The general form considered is $\langle \op{A}(t_a)\op{B}(t_b)\op{C}(t_c)\op{D}(t_d)\rangle$, where $\op{A}, \op{B}, \op{C}, \op{D}$ can be either of $\op{a}$ or $\dagop{a}$ (same mode), and the time arguments can take up to two distinct times $t=0$ and $t=\tau>0$.
\label{tab:2time}}
\end{table}

The change of distribution can introduce additional restrictions. In particular, Wigner and Q distributions have a tendency to involve higher order derivatives in the PDE \eqn{pdeforP} than P distributions, so that a standard diffusion process no longer captures the full quantum dynamics. An example are two-photon losses with operators $\op{R}=\op{a}^2$ for which P distributions produce only 2nd order derivatives but Q distributions 4th order terms, and Wigner distributions 3rd order ones. This precludes fully accurate calculation of correlation functions such as $\langle\dagop{a}(0)\op{a}(t)\op{a}(t)\dagop{a}(0)\rangle=\langle\alpha^*(0)\alpha'(t)\alpha'(t)\alpha^{\prime *}(0)\rangle_{\rm stoch}$ requiring Q evolution after the changeover, but not cases where the change to a Q distribution is only needed at the final time such as $\langle\dagop{a}(0)\op{a}(t)\dagop{a}(t)\op{a}(0)\rangle=\langle\alpha^*(0)\alpha'(t)\alpha^{\prime *}(t)\alpha(0)\rangle_{\rm stoch}$.
Stochastic techniques for dealing with higher order PDE terms have been investigated \cite{Jumarie99,Olsen02,Jumarie05,Plimak03} particularly in \cite{Drummond14} for doubled phase space, though attempts to date have shown strong time and stability limitations.

\section{Q representations and s-ordering in doubled phase space}
\label{OD}
To use the mechanisms described in Sec.~\ref{S3} in the doubled phase-space representations that give more complete coverage of quantum mechanics, we need also to consider the doubled phase space Q representation and how to switch to it from the positive-P.

\subsection{Doubled phase space s-ordered representations}
\label{DOUBLE}
The s-ordered representations were first generalised to doubled phase space by de Oliveira \cite{deOliveira92c}. Later studies  \cite{Plimak01,Hoffmann08,Deuar09b} used a different normalisation that is closer to the original single-phase space formulation \eqn{Dalpha} of Cahill and Glauber \cite{Cahill69a}. It will also be used here.
The explicit expansion of the density matrix was given in \cite{Deuar09b} as\footnote{In the supplemental material therein.}:
\eqs{PPsdef}{
\op{\rho} &=& \int d^{4M}\bm{\lambda}\,W^+_s(\bm{\lambda}) \op{\Lambda}^+_s(\bm{\lambda})\label{PPsdefrho}\\
\op{\Lambda}^+_s(\bm{\lambda}) &=& \prod_j \op{d}_j(\alpha_j,\beta_j) \op{T}_j(0,-s) \op{d}_j(-\alpha_j,-\beta_j),\qquad
}
where the displacement-like operator $\op{d}$ is
\eq{dalpha}{
\op{d}_j(\alpha_j,\beta_j) = e^{\alpha_j\dagop{a}_j-\beta_j\op{a}_j}\quad;\quad 
\op{d}_j(\alpha,\alpha^*)=\op{D}_j(\alpha),
}
obtained by the replacement $\alpha^*\to\beta$ in \eqn{Dalpha}. To distinguish from single phase space, the superscript ${}^+$ is used where necessary.
The usual properties $\Tr{\op{\Lambda}^+_s}=1$ and $\op{d}^{-1}(\alpha,\beta)=\op{d}(-\alpha,-\beta)$ continue to apply. 
The operator identities are now 

\eqs{sidenPP}{
\op{a}_j\op{\Lambda}^+_s &=& \left[\alpha_j-\frac{1-s}{2}\frac{\partial}{\partial\beta_j}\right]\op{\Lambda}^+_s,\\
\dagop{a}_j\op{\Lambda}^+_s &=& \left[\beta_j+\frac{1+s}{2}\frac{\partial}{\partial\alpha_j}\right]\op{\Lambda}^+_s,\\
\op{\Lambda}^+_s\op{a}_j &=& \left[\alpha_j+\frac{1+s}{2}\frac{\partial}{\partial\beta_j}\right]\op{\Lambda}^+_s,\\
\op{\Lambda}^+_s\dagop{a}_j &=& \left[\beta_j-\frac{1-s}{2}\frac{\partial}{\partial\alpha_j}\right]\op{\Lambda}^+_s.
}
The $s\to1$ limit of all the above gives the positive-P representation,
$s=0$ the  doubled-Wigner representation of \cite{Hoffmann08}, and the limit $s\to-1$ a doubled phase space analogue to the Q representation (``doubled-Q'').  Eqs. \eqn{sidenPP} are equivalent to the correspondences found in \cite{deOliveira92c}. 
An advantage of the doubled-Q and doubled Wigner representations relative to their single-phase space analogues is that all 2nd order derivative terms in the PDE can be made positive-definite and converted fully to stochastic equations via the same analytic kernel trick as for the positive-P representation. For example, spontaneous emission in a Schwinger boson system need not be amputated like in \cite{HUber20b} for the Wigner representation.

The kernel transform between different orderings in doubled phase-space is found to be
\eqa{kernelchangePP}{
\lefteqn{\op{\Lambda}^+_s(\bm{\alpha},\bm{\beta})}&& \\&&= \left(\frac{2}{s-s_0}\right)^M \!\!\!\!\int \frac{d^{2M}\bm{\zeta}}{\pi^M} \exp\left[-\frac{2|\bm{\zeta}|^2}{s-s_0}\right] \op{\Lambda}^+_{s_0}(\bm{\alpha}+\bm{\zeta},\bm{\beta}+\bm{\zeta}^*),\nonu\hspace*{-2em}
}
which is easily verified with the help of 
\eq{from69a}{
\op{T}_j(0,-s)=\frac{1}{\pi}\int d^2\gamma e^{-\tfrac{1}{2}s|\gamma|^2}\op{D}_j(\gamma)  \\ 
}
from \cite{Cahill69a}, the easy to show identities
\eqs{dd}{
\op{D}_j(\alpha)\op{D}_j(\gamma)&=&\op{D}_j(\alpha+\gamma)e^{\tfrac{1}{2}(\alpha\gamma^*-\alpha^*\gamma)} \\  
\op{d}_j(\alpha,\beta)\op{d}_j(\alpha',\beta')&=&\op{d}_j(\alpha+\alpha',\beta+\beta')e^{\tfrac{1}{2}(\alpha\beta'-\beta\alpha')}\!,\qquad 
}
and Gaussian integrals.
 The distribution transform
\eqa{sschangePP}{
\lefteqn{W^+_s(\bm{\alpha},\bm{\beta})}&& \\&&= \left(\frac{2}{s_0-s}\right)^M \!\!\!\!\int \frac{d^{2M}\bm{\zeta}}{\pi^M} \exp\left[-\frac{2|\bm{\zeta}|^2}{s_0-s}\right] W^+_{s_0}(\bm{\alpha}+\bm{\zeta},\bm{\beta}+\bm{\zeta}^*).\nonu\hspace*{-1em}
}
  is readily found by equating two \eqn{PPsdefrho} expansions of $\op{\rho}$ which have different $s$, and applying \eqn{kernelchangePP} to the one with higher $s$ ($=s_0$).  

\subsection{Non normally-ordered correlations in the positive-P representation}

The ideas from Sec.~\ref{S3} can be used to treat non normally-ordered correlations in the positive-P representation, which -- unlike the Glauber-Sudarshan P --  is applicable for all quantum states and systems. 
With the tools for the doubled s-ordered phase space described in Sec.~\ref{DOUBLE},  derivation of the expressions for mixed-time expectation values follows the same path as in Secs.~\ref{CONVs} and~\ref{NNP},  but some care needs to be taken to incorporate the doubled phase-space.
We obtain that: 

(1) To shift samples from a more to a less normally ordered  doubled-phase space representation, one adds the following noise:
\eq{convertsPP}{
\alpha'_j = \alpha_j+\sqrt{\frac{s_0-s}{2}}\ \zeta_j\ ;\ 
\beta'_j = \beta_j+\sqrt{\frac{s_0-s}{2}}\ \zeta^*_j.
}
Notably -- the same noise for $\alpha$ and $\beta^*$, which was not obvious \emph{a priori}. 
The prefactor is $1$ for positive-P to ``doubled''-Q, and $1/\sqrt{2}$ for positive-P to doubled-Wigner.

(2) Anti-normal ordered mixed-time correlations are evaluated as
\eqa{QmultitimePP}{
\lefteqn{\langle \op{a}_{p_1}(t_1)\cdots\op{a}_{p_{\mc{N}}}(t_{\mc{N}}) \dagop{a}_{q_1}(s_1)\cdots\dagop{a}_{q_{\mc{M}}}(s_{\mc{M}})\rangle }\qquad&&\nonu\\
&&\hspace*{-1em}=\langle \alpha'_{p_1}(t_1)\cdots\alpha'_{p_{\mc{N}}}(t_{\mc{N}}) \beta'_{q_1}(s_1)\cdots\beta'_{q_{\mc{M}}}(s_{\mc{M}})\rangle_{\rm stoch}.\qquad
}
where $\alpha'$ and $\beta'$ are doubled-Q representation samples, possibly created via \eqn{convertsPP} (with $s_0-s=2$) from initial positive-P samples, and later evolved via the appropriate doubled-Q representation evolution equations.

(3) For mixed ordering, the procedure in Sec.~\ref{GENORD} follows with the same structure, except that the correspondences are
\eq{replPP}{
\op{a}_j\to\alpha_j\qquad;\qquad \dagop{a}_j\to\beta_j.
}
in the positive-P and 
\eq{replQQ}{
\op{a}_j\to\alpha'_j\qquad;\qquad \dagop{a}_j\to\beta'_j.
}
in the doubled-Q. For example, the expression for the correlation $\mc{A}''$ \eqn{estmixcor} using samples starting in the positive-P is
\eq{estmixcorPP}{
\mc{A}''=
\langle \beta_{p_1}(t_1)
\alpha'_{p_2}(t_2)
\beta'_{q_1}(s_1)
\alpha_{q_2}(s_2)\rangle_{\rm stoch}
}

(4) The correlation tallies of Sec.~\ref{TABS} apply without change to the doubled phase space representations.

\section{Test example: unconventional photon blockade}
\label{EX}

\subsection{The system}
Consider the two-site Bose-Hubbard Hamiltonian
\eqa{Ham}{
\op{H} &=& \sum_{j=1,2} \dagop{a}_j\left[-\Delta+\frac{U}{2}\dagop{a}_j\op{a}_j\right]\op{a}_j\nonu\\
&&+ F\left[\dagop{a}_1 +\op{a}_1\right] - J \left[\dagop{a}_1\op{a}_2+\dagop{a}_2\op{a}_1\right]
}
with standard annihilation operators $\op{a}_j$ for modes $j$ using units $\hbar=m=1$. This describes two sites with (real) tunnelling $J$, local on-site interaction constant $U$, detuning $\Delta$, and a coherent drive $F$ (real) only on the first site.
There is also a decay process with rate $\gamma$ that is treated by describing the evolution of the system with the master equation
\eqa{masterH2}{
\frac{\partial\op{\rho}}{\partial t} &=& -i\left[\op{H},\op{\rho}\right] + \frac{\gamma\wb{N}}{2}\sum_j\left[2\dagop{a}_j\op{\rho}\op{a}_j-\op{a}_j\dagop{a}_j\op{\rho} -\dagop{\rho}\op{a}_j\op{a}_j\right]\nonu\\
&&+ \frac{\gamma(\wb{N}+1)}{2}\sum_j\left[2\op{a}_j\op{\rho}\dagop{a}_j-\dagop{a}_j\op{a}_j\op{\rho} -\op{\rho}\dagop{a}_j\op{a}_j\right].
}
To test the performance a bit more beyond the standard model, we have also addded a thermal bath with mean occupation $\wb{N}$.
 Such a system can be realised using e.g. micropillars \cite{Goblot19,Schneider16} or transmon qubits \cite{Schmidt13,Houck12}. 
The nontrivial feature here is a two-boson destructive interference effect between photons injected by the drive, and other photons that have been previously injected, tunnelled to site 2 and then back, returning with a relative phase of $\pi$ \cite{Bamba11}. The result of this is that in the steady state only single photons can be present at the driven site 1, providing possibly an avenue to create single-photon sources \cite{Liew10}. 
The lack of double occupation is evidenced in a single-time two body correlation function
$g_{11} = \langle\op{a}_1^{\dagger 2}\op{a}_1^2\rangle / \langle\dagop{a}_1\op{a}_1\rangle^2$,
which is very close to zero.
However, for practical application, it is of particular interest to find out how large a time mismatch between measured photons can be accommodated without significantly increasing $g^{(2)}_{11}$ from this low level. If it is too short, then the single-photon source will have a too short active time for practical applications. Therefore the correlation function of particular interest is
\eq{g2t}{
g_{11}(\tau) = \frac{\langle\dagop{a}_1(t)\dagop{a}_1(t+\tau)\op{a}_1(t+\tau)\op{a}_1(t)\rangle}{n_1^2}
}
in the stationary state, with delay time $\tau$. The mean occupation is $n_1=\langle\dagop{a}_1\op{a}_1\rangle$.

The Ito stochastic evolution equations in the positive-P ($s=1$) and doubled-Q ($s=-1$) representations are 
\eqs{ppeq}{
\frac{d\alpha_j}{dt} &=& \left[-iU(\alpha_j\beta_j+s-1)-\frac{\gamma}{2}+\sqrt{-isU}\,\xi_j(t)\right]\alpha_j -iF_j \hspace*{-1em}\nonu\\ 
&&-i\sum_kH^{\rm sp}_{jk}\alpha_k+ \sqrt{\gamma\left(\wb{N}+\tfrac{1-s}{2}\right)}\,\eta_j(t)\label{ppeqa}\\
\frac{d\beta_j}{dt} &=& \left[iU(\alpha_j\beta_j+s-1)-\frac{\gamma}{2}+\sqrt{isU}\,\wt{\xi}_j(t)\right]\beta_j +iF_j \hspace*{-1em}\nonu\\ 
&&+i\sum_kH^{\rm sp}_{jk}\beta_k + \sqrt{\gamma\left(\wb{N}+\tfrac{1-s}{2}\right)}\,\eta^*_j(t).
}
These were derived in \cite{Deuar21} for the positive-P. Here the matrix elements of the one-particle Hamiltonian are $H^{\rm sp}_{jj}=-\Delta$, $H^{\rm sp}_{12}=H_{21}=-J$, the drives are $F_1=F$, $F_2=0$,  
while  $\xi_j$ and $\wt{\xi}_j$ are delta-correlated independent white real noises with variance
\eq{xieg}{
\langle\xi_j(t)\xi_{j'}(t')\rangle = \delta_{jj'}\delta(t-t') 
}
and $\eta_j$ are similarly correlated independent complex noises:
\eq{eta}{
\langle\eta_j(t)\eta^*_{j'}(t')\rangle = \delta_{jj'}\delta(t-t');\qquad \langle\eta_j(t)\eta_{j'}(t')\rangle=0.
}
Appendix~\ref{N-NOISE} gives some detail on generation of the noise.
In some cases, calculations with these stochastic equations can already be faster than brute force calculations directly with the density matrix $\op{\rho}$ in a suitably truncated number state basis. For large systems, they are the only tractable way to access full quantum mechanics.
Appendix~\ref{NUMERIX} gives details of an integration algorithm that is robust to the multiplicative noise appearing in \eqn{ppeq}, and was used for the simulations reported here and elsewhere \cite{nocut,Deuar21,Ross21,Swislocki16,Deuar16,Kheruntsyan12,Deuar14,Deuar11,Deuar13,Ng13,Deuar09b}.

\subsection{Two-time photon-photon correlations}

Consider first the strong antibunching parameters studied in \cite{Bamba11} and \cite{Deuar21}: $U=0.0856$, $J=3$, $\Delta=-0.275$, $\gamma=1$, $F=0.01$. 
We will study the correlations primarily in the stationary state. To this end, a positive-P simulation is initialised in the vacuum 	 $\bm{\alpha}=\bm{\beta}=0$, and evolved up to $t=30$. The stationary state is attained after $t\gtrsim 15$, and we calculate multi time correlation functions from times $t_0=20$ to $t_0+\tau$ using the available samples. ($\mc{S}=2^{16}$ in all cases). 
Uncertainty in predictions is calculated using sub-ensemble averaging, as explained in Appendix~\ref{N-ERR}.
These errors are shown as triple lines in all the plots.

The two-time photon-photon correlations between one photon measured at site 1 at time $t_0$ and the other at site $j$ after a delay time of $\tau$  are \eq{g11tau}{
g_{1,j}(\tau) = \frac{\langle\dagop{a}_1(t_0)\dagop{a}_j(t_0+\tau)\op{a}_j(t_0+\tau)\op{a}_1(t_0)\rangle}{n_1(t_0)n_j(t_0+\tau)},
}
where $n_j(t)=\langle\dagop{a}(t)\op{a}(t)\rangle={\rm Re}\langle\alpha_j(t)\beta_j(t)\rangle_{\rm stoch}$ is the mean occupation of site $j$. In the 
positive-P representation the correlation \eqn{g11tau} is calculated via
\eq{g11tauPP}{
g_{1,j}(\tau) = \frac{{\rm Re}\left\langle\alpha_1(t_0)\beta_1(t_0)\alpha_j(t_0+\tau)\beta_j(t_0+\tau)\right\rangle_{\rm stoch}}{n_1(t_0)n_j(t_0+\tau)}.
}
We can take the real parts above, because the imaginary parts must converge to zero in the $\mc{S}\to\infty$ limit. This behaviour is verified in the simulations.

\begin{figure}[htb]
\begin{center}
\includegraphics[width=0.8\columnwidth]{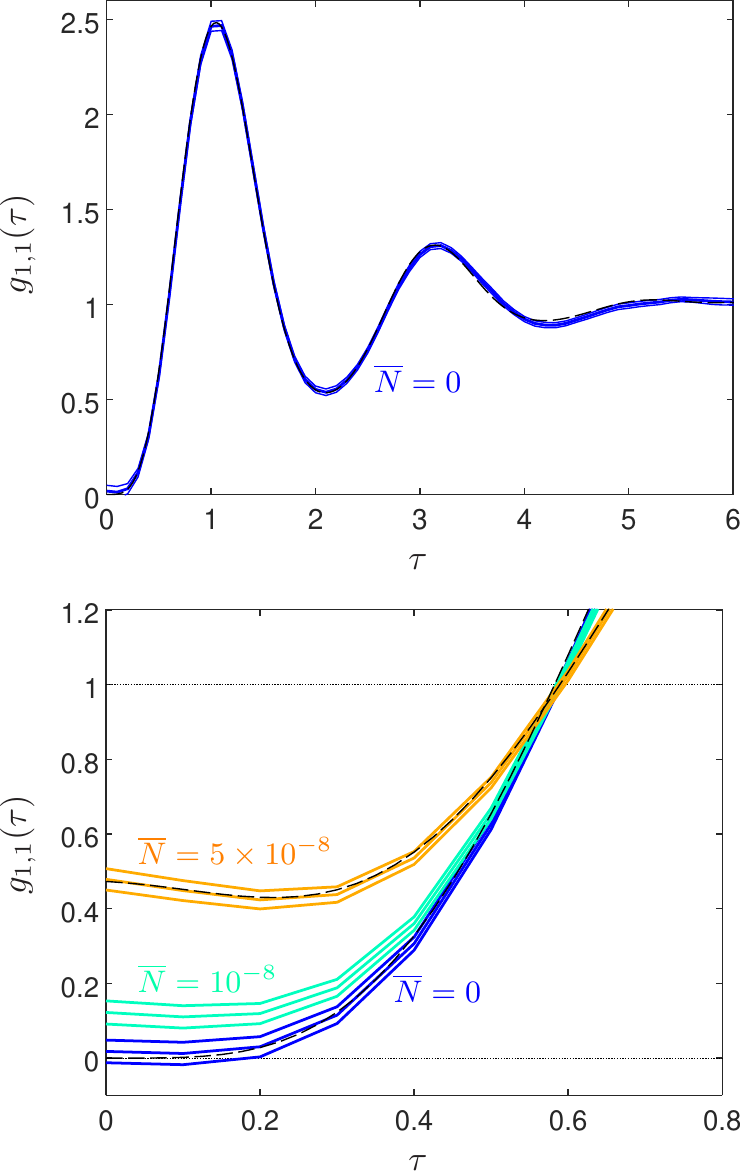}
\end{center}\vspace*{-0.5cm}
\caption{Time duration of antibunching in the unconventional photon blockade system, $F=0.01$.
Top: clean system \mbox{($\wb{N}=0$),} strong antibunching. 
Bottom: degradation with background thermal intensity $\wb{N}>0$.
Black dashed lines: direct solution of the master equation. 
Triple lines show $1\sigma$ statistical uncertainty.
\label{fig:Nbar}}
\end{figure}

Fig.~\ref{fig:Nbar} (top) shows the multi-time local photon correlation \eqn{g11tau} at site 1 (i.e. $j=1$) for the clean case \mbox{($\wb{N}=0$),} and verifies that the result perfectly matches the exact brute force solution. The desired anti-correlation dip is seen around $\tau=0$, along with characteristic oscillations out to delay times of about $\tau=5$. The bottom panel shows how the anti-correlation dip degrades when the system is linked to a particle reservoir (growing $\wb{N}$). Notably the dip timescale does not change appreciably as the minimum correlation rises. A 10\% remnant correlation which might still be acceptable for applications is found when $\wb{N}=10^{-8}=0.026n_1$. This sets a limit on how much background photon reservoir is acceptable.

\begin{figure}[htb]
\begin{center}
\includegraphics[width=\columnwidth]{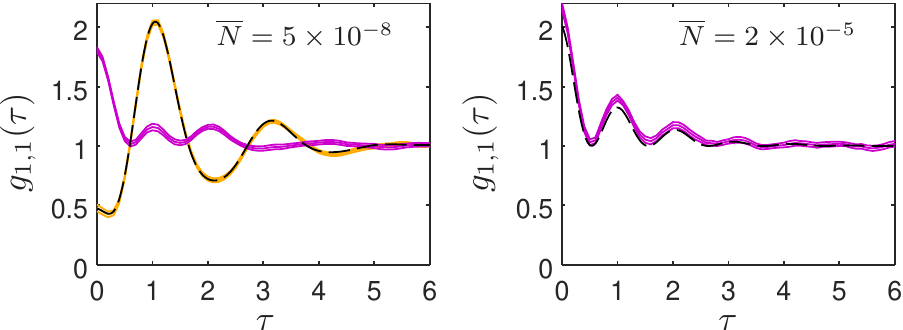}
\end{center}\vspace*{-0.5cm}
\caption{Breakdown of the Glauber-Sudarshan~P representation: 
shown in pink; 
Positive-P simulation: yellow; 
exact: black dashed.
$F=0.01$.
\label{fig:Glauber}}
\end{figure}

\subsection{Breakdown of the Glauber-Sudarshan~P}
\label{GSPBREAK}
This system also shows the breakdown of the single-phase-space P representation very clearly. 
The evolution equation is \eqn{ppeqa} with $s=1$, but it is a correct representation of the quantum FPE \eqn{FPEgen} only when the diffusion matrix $D_{\mu\nu}$ in the FPE has all nonnegative eigenvalues \cite{QuantumNoise}.
Here, the diffusion matrix for real, imaginary parts of $\alpha_j=\alpha_{jr}+i\alpha_{ji}$ has elements 
$D_{jr,jr}=\tfrac{U}{2}{\rm Im}(\alpha_j^2)+\tfrac{\gamma\wb{N}}{2}$, 
$D_{ji,ji}=-\tfrac{U}{2}{\rm Im}(\alpha_j^2)+\tfrac{\gamma\wb{N}}{2}$,
$D_{jr,ji}=D_{ji,jr}=-\tfrac{U}{2}{\rm Re}(\alpha_j^2)$. 
with eigenvalues $\lambda_{j,\pm} = \tfrac{\gamma\wb{N}}{2} \pm \tfrac{U}{2}|\alpha_j|^2$. These only become non-negative once 
$\gamma\wb{N}>U|\alpha_j|^2$, i.e.  $\gamma\wb{N}\gtrsim Un_j$. 
The question then is: for what parameters does the evolution remain well described?
When $\wb{N}=0$ The antibunched mode has mean occupation $n_1=3.87\times10^{-7}$, naively suggesting $\wb{N}\sim 5\times10^{-8}=1.5U n_1/\gamma$ to already be a value for which the description is good.
However, we can see in Fig.~\ref{fig:Glauber} that it does not give correct results at all. This is because the problem lies in the nr. 2 mode with $n_2=1.07\times10^{-5}$. Taking a far larger $\wb{N}=2\times 10^{-5}$ (in which case  $\gamma\wb{N}\sim 10U n_1, 7Un_2$)  gives almost correct results with the Glauber-Sudarshan~P (though still not fully), but of course antibunching in mode 1 is long gone for such a relatively high thermal noise level.

This is an indication that skimping on full quantum effects by trying approximate semiclassical methods is not a good strategy for this kind of system.

\subsection{Differently ordered correlations}

\begin{figure}[htb]
\begin{center}
\includegraphics[width=0.8\columnwidth]{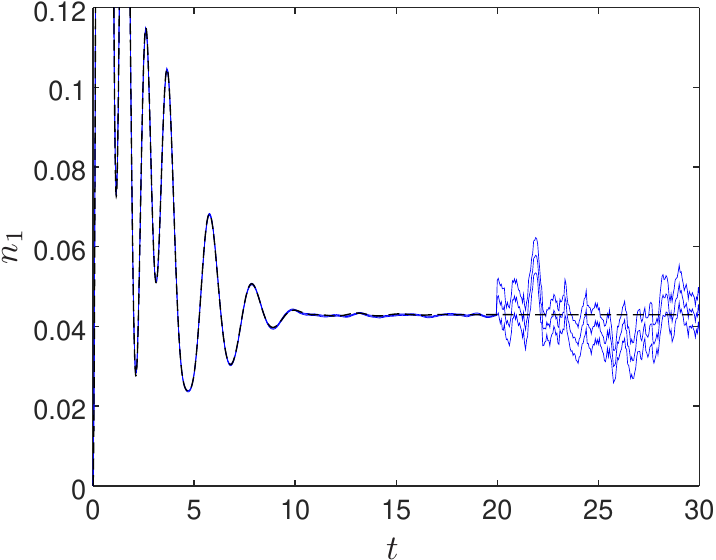}
\end{center}\vspace*{-0.5cm}
\caption{Estimation in positive-P and doubled-Q simulations: occupation of mode 1, when $F=3$, after switching from the positive-P to the doubled-Q representation at $t=t_0=20$.
\label{fig:PQnoise}}
\end{figure}

\begin{figure}[htb]
\begin{center}
\includegraphics[width=\columnwidth]{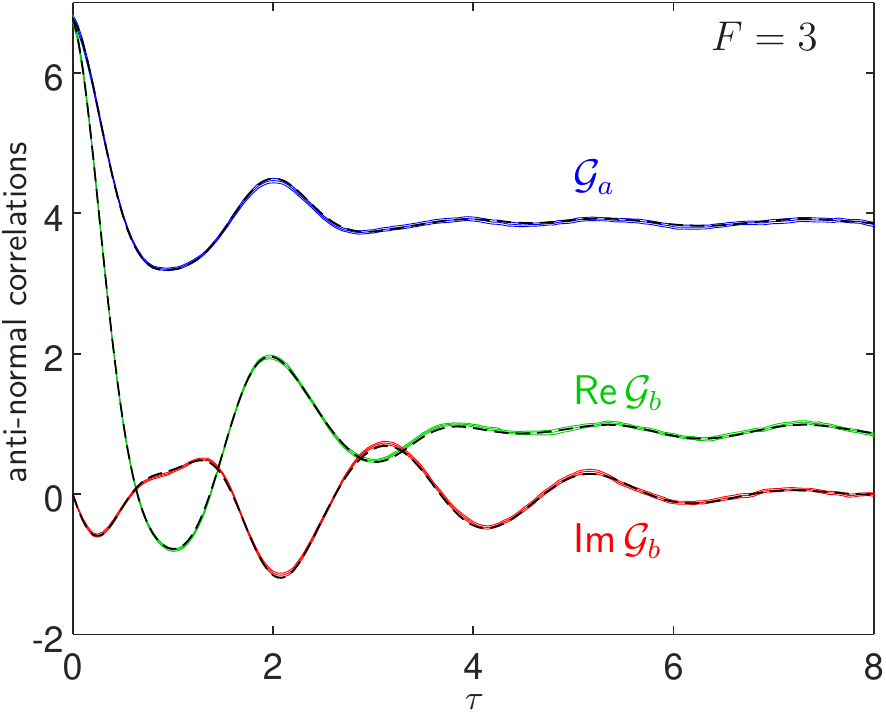}
\end{center}\vspace*{-0.5cm}
\caption{Anti-normal ordered moments. Strongly pumped case ($F_1=3$), mode 2. Shown are $\mc{G}_a$ and the real and imaginary parts of $\mc{G}_b$ as per \eqn{QQcorr}. 
Exact results: dashed black.
\label{fig:QQ}}
\end{figure}

\begin{figure}[htb]
\begin{center}
\includegraphics[width=\columnwidth]{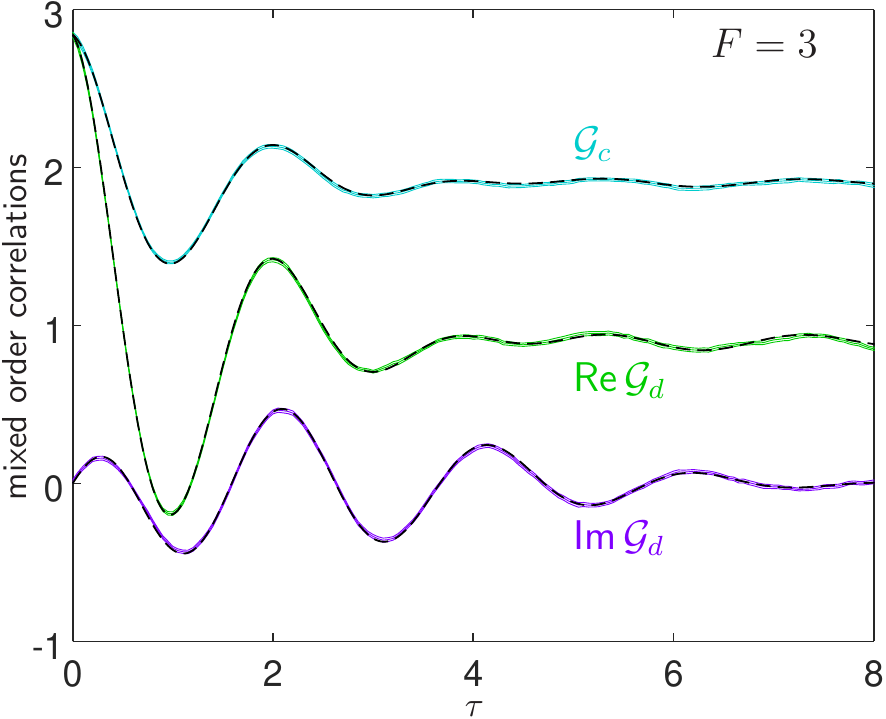}
\end{center}\vspace*{-0.5cm}
\caption{Moments that are neither normal nor anti-normal. $F=3$ like in Fig.~\ref{fig:QQ}. Shown are $\mc{G}_c$ and the real and imaginary parts of $\mc{G}_d$ as per \eqn{PQcorr}. 
Exact results: dashed black lines.
\label{fig:PQ}}
\end{figure}

To test how the prescriptions developed in Secs.~\ref{QREP} and~\ref{S3} work, we use a different driving, $F=3$, which generates larger occupations (in the stationary state $n_1\approx 0.043$ and $n_2\approx 0.98$) and as a result more interesting anti-normal and mixed-order correlations. 
The doubled-Q simulations are also too noisy to get useful predictions for the parameters in Fig.~\ref{fig:Nbar} because Q distributions have a width of $\mc{O}(1)$ even in vacuum. This is a certain limitation to the Q distribution approach. The difference in sampling accuracy can be nicely seen in Fig.~\ref{fig:PQnoise} which shows the mean and uncertainty of $n_1$ during the simulation used to generate Figs.~\ref{fig:QQ}-\ref{fig:PQ}. The samples are switched from positive-P to doubled-Q at $t=t_0=20$.

Arguably the primary practical interest lies in cases with four factors and two times -- corresponding roughly to either detection of single particles at two times, or creation and destruction of pairs. 
Fig.~\ref{fig:QQ} shows the simulation of some anti-normally ordered correlations of this kind (not normalised), evaluated using the doubled-Q representation:

\eqa{QQcorr}{
\mc{G}_a &=& \langle\op{a}_2(t_0)\op{a}_2(t_0+\tau)\dagop{a}_2(t_0+\tau)\dagop{a}_2(t_0)\rangle\nonu\\
 &=& {\rm Re}\langle\alpha'_2(t_0)\alpha'_2(t_0+\tau)\beta'_2(t_0+\tau)\beta'_2(t_0)\rangle_{\rm stoch}\\
\mc{G}_b &=& \langle\op{a}_2(t_0+\tau)^2[\dagop{a}_2(t_0)]^2\rangle = \langle\alpha'_2(t_0+\tau)^2\beta'_2(t_0)^2\rangle_{\rm stoch}\nonu
}
The latter $\mc{G}_b$ is an anomalous correlation that has both real and imaginary components. 
Similarly, Fig.~\ref{fig:PQ} shows simulations of a number of mixed-order correlations which absolutely require a swapping from positive-P to doubled-Q representation using the procedure of Sec.~\ref{GENORD}:

\eqa{PQcorr}{
\mc{G}_c &=& \langle\op{a}_2(t_0+\tau)\dagop{a}_2(t_0+\tau)\dagop{a}_2(t_0)\op{a}_2(t_0)\rangle \nonu\\
&=& {\rm Re}\langle\alpha'_2(t_0+\tau)\beta'_2(t_0+\tau)\beta'_2(t_0)\alpha_2(t_0)\rangle_{\rm stoch}\nonu\\
\mc{G}_d &=& \langle\op{a}_2(t_0)[\dagop{a}_2(t_0+\tau)]^2\op{a}_2(t_0)\rangle \nonu\\
&=& \langle\alpha'_2(t_0)\beta'_2(t_0+\tau)^2\alpha_2(t_0)\rangle_{\rm stoch}
}
In all cases, primed variables are evaluated in the doubled-Q representation, un-primed in the positive-P.

Notably, in both figures the stochastic simulations perfectly and very accurately agree with brute force calculations using the density matrix. This is despite the regime being one which is very poorly treated by approximate semiclassical methods (occupations are $\mc{O}(1)$ or smaller). This is strong evidence that the intuitive expressions and approach laid out in the previous sections is appropriate.

\subsection{Correlation dynamics in the stationary state}

As a larger-sized example, consider the same Hamiltonian and master equation as \eqn{Ham} and \eqn{masterH2} (take $\wb{N}=0$), but with a longer chain of 32 sites, which is already beyond or at the limit of the capabilities of alternative methods such as brute force or corner space renormalisation \cite{Finazzi15,Casteels16}. 
The tunnelling term in \eqn{Ham} now becomes 
$-J\sum_{j=1}^{31}\left[\dagop{a}_j\op{a}_{j+1}+\dagop{a}_{j+1}\op{a}_j\right],$
and a driving of $F=3$ remains only at the $j=1$ site.
While the stationary state does not show any time-dependent change in expectation values, we should expect to still see signatures of transport in its multi-time correlations as a function of distance and delay time if the method is good.

\begin{figure}[htb]
\begin{center}
\includegraphics[width=\columnwidth]{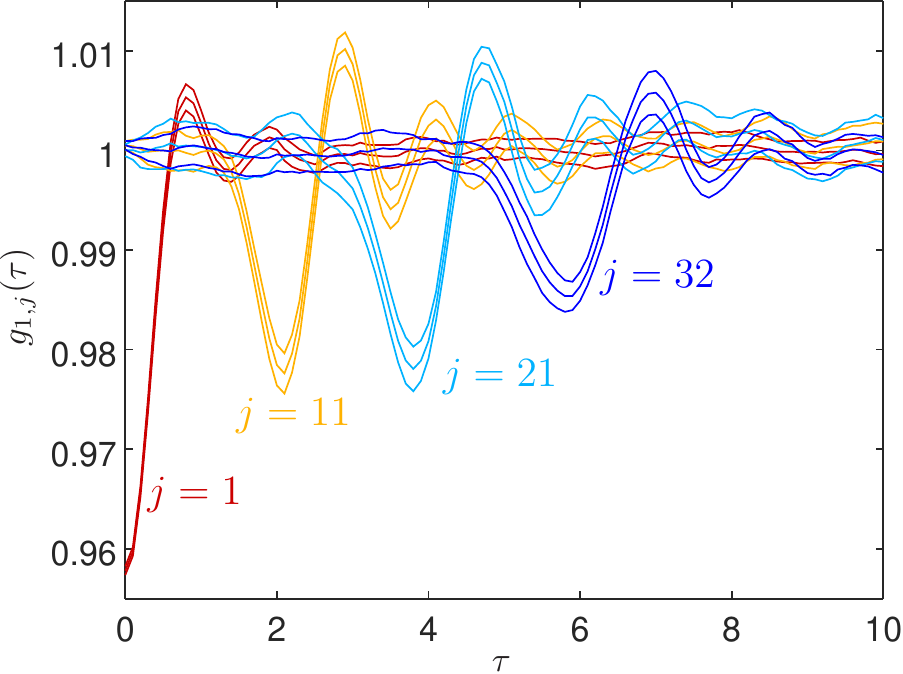}
\end{center}\vspace*{-0.5cm}
\caption{Spreading of correlations in a 32 site chain, in the stationary state. $F=3$. The correlation is given by \eqn{g11tau}. 
\label{fig:sound}}
\end{figure}
	
Fig.~\ref{fig:sound} shows such a calculation in the 32 site system, plotting the normalised photon-photon correlation $g_{1,j}(\tau)$ with spatial separation $j-1$ and delay time $\tau$, as defined in \eqn{g11tau}. The correlations are evaluated from $t_0=20$. Despite this being the stationary state, a very clear anti-correlation signal can be seen moving steadily with delay time. 
Its speed is approximately $2J$, twice the tunnelling rate. Correlation waves often travel at twice the characteristic speed for single particles \cite{Cheneau12,Lauchli08,stob}, so this is not unexpected. However, the speed is clearly not related to the superfluid speed of sound, here $\sim\sqrt{Un_j}$, which is far lower. 

\section{Summary}
\label{CONC}

The known framework for evaluation of multi-time observables from the work of Gardiner in the P representation \cite{QuantumNoise} has been extended here to include
firstly: the much more widely applicable positive-P representation \eqn{PPmultitime}; Secondly the Q \eqn{Qmultitime}, doubled-Q and other single and double phase space s-ordered representations; Thirdly: other orderings such as anti-normal and mixed-ordered observable products (Sec.~\ref{S3}, and especially \eqn{estmixcor} and the algorithm of Sec.~\ref{GENORD}). 
These results allow the evaluation of a very wide range of quantum multi-time observables in bosonic systems (classified in Tables.~\ref{tab:combos}-\ref{tab:2time})
in a way that contains the full quantum mechanics, and is scalable to large systems. Systems with localised interactions require computational effort proportional to $M$ or $M\log M$ with the number of modes or sites, $M$.

While out-of-time correlations (OTOCs) are not directly accessible through the mechanism outlined here, time reversal schemes like those used in experiment \cite{Garttner17} or theory \cite{Dowling05,Swingle16,Bohrdt17} could be attempted by changing the signs of constants in the equation of motion. When combined with the techniques developed here, this would give access to a wide range of information about quantum chaos \cite{Bohrdt17,Maldacena16,Foini19}, many body localisation \cite{Fan17,He17}, and quantum phase transitions \cite{Shen17} in larger systems than were accessible to date \cite{Bohrdt17,Garttner17}.

Along the way, a number of additional results were obtained regarding conversion formulae between different orderings in the doubled-phase space representations (\eqn{kernelchangePP} and \eqn{sschangePP}) and their stochastic samples \eqn{convertsPP}. A clear case of the breakdown of the Glauber-Sudarshan P representation was seen in Fig.~\ref{fig:Glauber}. Also, in Sec.~\ref{EX} some results regarding the correlation functions (Figs.~\ref{fig:QQ} and~\ref{fig:PQ}), susceptibility to background thermal density (Fig.~\ref{fig:Nbar}) and signal speed (Fig.~\ref{fig:sound}) in the unconventional photon blockade system were obtained. 

Finally, Appendix~\ref{NUMERIX} details a convenient algorithm for simulation of phase-space stochastic equations, an item that has been hard to find in the literature in the past.

\acknowledgments
I am grateful to Micha{\l} Matuszewski, Marzena Szyma\'nska, and Alex Ferrier 
for discussions on topics leading to this paper, and to Wouter Vestraelen for discussions of numerics. 
This research was supported by the National Science Centre (Poland) grant No. 2018/31/B/ST2/01871. 

\appendix

\section{Numerical techniques for stochastic simulation of phase-space trajectories}
\label{NUMERIX}

This appendix details how the stochastic equations for the examples in Sec.~\ref{EX} were integrated, and describes a robust and general algorithm that is particularly applicable to stochastic trajectories generated by phase-space descriptions of quantum systems. The bulk of the method is based on the semi-implicit midstep algorithm described in \cite{Drummond91,Werner97}, adding some modified propagators as per \cite{nocut} which allow for more efficient treatment of exponential and dominant deterministic processes. Some practical elements that are usually skipped over in the literature are also pointed out. Example textbooks for the broad topic of stochastic integration include \cite{Kloeden92,Ikeda88}.

\subsection{Integration algorithm}
\label{N-INT}
Let us introduce the following notation. The Ito stochastic equations for a set of variables $\vec{v}$ with elements $v_j$ indexed by $j$ will be written
\eq{eqs}{
\frac{d v_j}{dt} = \mc{D}_j(\vec{v},t) = A_j(\vec{v},t) + X_j(\vec{v},t)
}
where $X_j$ are noise terms with zero mean $\langle X_j \rangle_{\rm stoch}=0$, whereas $A_j$ are deterministic ``drift terms'' that contain no explicit noise contribution, apart from the statistical spread of the variables $v_j$ themselves. At the level of the Ito equations \eqn{eqs} we also require that the noise is not correlated in time $\langle X_j(t) X_{j'}(t') \rangle_{\rm stoch} \propto \delta(t-t')$, i.e. proportional to Wiener increments. For example, in \eqn{ppeq}:
\eqs{Xj}{
X_{\alpha_j} &=& \alpha_j\sqrt{-isU}\,\xi_j + \sqrt{\gamma\left(\wb{N}+\frac{1-s}{2}\right)}\,\eta_j,\qquad\\
X_{\beta_j} &=& \beta_j\sqrt{isU}\,\wt{\xi}_j + \sqrt{\gamma\left(\wb{N}+\frac{1-s}{2}\right)}\,\eta^*_j.
}

For various reasons, the standard mathematical environments and numerical packages
tend not to perform well when integrating stochastic equations\footnote{A notable exception is the XMDS package \cite{XMDS2}.}, especially ones that contain variable-dependent diffusion coefficients such as in \eqn{Xj} or the prototypical  Kubo oscillator \cite{Kubo85} with multiplicative noise. Phase space methods that treat the full quantum dynamics also typically contain such terms.
Once one leaves the simplest Ito-Euler algorithm which requires very small step sizes, two of the recurring issues are that 
advanced integration algorithms usually assume sufficiently continuous derivatives (completely violated by white noise) or
try to adapt the timestep in real time (which can introduce systematic errors when noise is involved). A third difficulty is the need for autocorrelation corrections (seen Sec.~\ref{N-STRAT})  which become much more difficult to derive as the complexity of the algorithm grows.

The semi-implicit midpoint approach \cite{Drummond83,Smith89} has been shown to be robust against spurious numerical instability as carefully analysed in \cite{Drummond91,Werner97},
but remains simple enough for any autocorrelation corrections to be readily calculated. 

Suppose we are to advance time by $\Delta t$, starting from variables $v_j^{(0)}$ at time $t$ to $v'_j$ at $t+\Delta t$. 
Define a constituent substep or propagator that advances times by $\tau$ and variables to
$v_j^{\rm step}\left[\vec{v}_j^{(0)},\mc{D}(\vec{v}^{\,(0)},t),\tau\right]$ which depends in order on the starting variables, the form of the derivative, and the timestep $\tau$.
The simplest choice for this propagator is the Ito-Euler form
\eq{euler}{
v_j^{\rm step}\left[v_j^{(0)},\mc{D}(\vec{v}^{\,(0)},t),\tau\right]= v_j^{(0)} + \mc{D}(\vec{v}^{\,(0)},t)\,\tau
}
which just uses the values at the beginning of the step for evaluating the derivative, as per the definition of Ito stochastic calculus.

The semi-implicit midpoint algorithm can be viewed as a container procedure that calculates the final variables $v'_j$ via a series of iterations \cite{Werner97}
\eqa{midstep}{
v^{(1)}_j &=& v^{\rm step}_j\left[v_j^{(0)},\mc{D}(\vec{v}^{\,(0)},t),\Delta t/2\right],\nonu\\
v^{(2)}_j &=& v^{\rm step}_j\left[v_j^{(0)},\mc{D}(\vec{v}^{\,(1)},t),\Delta t/2\right],\nonu\\
&\cdots&\\
v^{(n)}_j &=& v^{\rm step}_j\left[v_j^{(0)},\mc{D}(\vec{v}^{\,(n-1)},t),\Delta t/2\right],\nonu\\
v'_j(t+\Delta t) &=& v^{\rm step}_j\left[v_j^{(0)},\mc{D}(\vec{v}^{\,(n)},t+\Delta t/2),\Delta t\right].\label{vij}\qquad
}
The numbered iterations always start from the initial point, but iteratively find a self-consistent estimate of the midpoint value of the derivatives. The last iteration \eqn{vij}  uses this midpoint value and midstep time to make the final advancement of the variables. 
Usually, a single midpoint iteration (i.e. $n=1$) is sufficient.
When using the basic Ito-Euler propagator \eqn{euler} at time argument $t+\Delta t/2$ in \eqn{vij}, the final step is accurate to $\mc{O}(\Delta t^2)$ in the deterministic parts $A_j$, 
and the step's mean to order $\mc{O}(\Delta t)$ in the noise terms $X_j$ \cite{Drummond91}, provided the autocorrelation correction is incorporated into 
$v^{\rm step}_j$, as explained in Sec.~\ref{N-STRAT}. This is so-called ``weak'' stochastic convergence to $\mc{O}(\Delta t)$ in the terminology of \cite{Drummond91} because the accuracy for single trajectories is $\mc{O}(\sqrt{\Delta t})$. ``Strong'' stochastic convergence to $\mc{O}(\Delta t)$ for single trajectories requires a more involved and more inconvenient autocorrelation correction than that presented in Sec.~\ref{N-STRAT}.
Nevertheless, the ``weak'' level of accuracy in the noise term is actually quite high already. For example, even 4th or 9th order Runge-Kutta implementations made available are generally only accurate to $\mc{O}(\sqrt{\Delta t})$ in this regard \cite{XMDS2}. 

Moreover, in practice, one can often easily but substantially improve on the Euler step \eqn{euler}. For example, if $\Delta$, $\gamma$, or $U$ are large in \eqn{ppeq}, the evolution is primarily exponential. Another typical case is in continuum models when the kinetic energy or external potential energy contributions are dominant and trivial to integrate \cite{nocut}. 
A pragmatic and flexible propagator choice is to separate the evolution into parts that will be treated exponentially (E), strictly linearly (L) and the rest (R):
\eq{explin}{
\mc{D}_j(\vec{v}) = \mc{D}_j^{E}(\vec{v})\, v_j + \mc{D}_j^{R}(\vec{v}) + \mc{D}_j^{L}(\vec{v}),
}
and solve the equation for the E and R parts 
\eq{KC}{
\frac{dv_j}{d\tau} = \mc{D}_j^{E}\,v_j + \mc{D}_j^{R}
}
as if the coefficients were constant \cite{nocut}. 
This gives the full propagator
\eqa{dtsoln}{
\lefteqn{v_j^{\rm step}\left[v_j^{(0)},\mc{D}(\vec{v}^{\,(0)},t),\tau\right] =}&&\nonu\\
&\qquad& v_j^{(0)}\,e^{\tau\,\mc{D}_j^E(\vec{v}^{\,(0)},t)} +\left(e^{\tau\,\mc{D}_j^E(\vec{v}^{\,(0)},t)}-1\right)\ \frac{\mc{D}^R_j(\vec{v}^{\,(0)},t)}{\mc{D}^E_j(\vec{v}^{\,(0)},t)} \nonu\\
&&+ \mc{D}^L_j(\vec{v}^{\,(0)},t)\ \tau.
}
Then, the leading processes are often integrated exactly, with only higher order corrections needed to be worked on by the midpoint algorithm. This is particularly efficient when the stochastic terms are a perturbation on the dominant exponential deterministic evolution.
It is usually optimal to place all non-exponential parts into the ``R'' part, but some special cases can require avoidance of the nonlinear solution \cite{nocut}. 
Both $A$  drift and $X$ noise parts can enter the coefficients as one chooses. 

In the case of the calculations in this paper using \eqn{ppeq}, 
\eqa{coefs}{
\mc{D}_{\alpha_j}^E &=& -iU(\alpha_j\beta_j+s-1)-\frac{\gamma}{2} +\sqrt{-isU}\,\xi_j -iH^{\rm sp}_{jj} + C^E_{\alpha_j},\nonu\hspace*{-1em}\\
\mc{D}_{\alpha_j}^R &=& -iF_j -i\sum_{k\neq j}H^{\rm sp}_{jk}\alpha_k+\sqrt{\gamma\left(\wb{N}+\frac{1-s}{2}\right)}\,\eta_j,\nonu\\
\mc{D}_{\beta_j}^E &=& iU(\alpha_j\beta_j+s-1)-\frac{\gamma}{2} +\sqrt{isU}\,\wt{\xi}_j +iH^{\rm sp}_{jj} + C^E_{\beta_j},\nonu\hspace*{-1em}\\
\mc{D}_{\beta_j}^R &=& iF_j +i\sum_{k\neq j}H^{\rm sp}_{jk}\beta_k+\sqrt{\gamma\left(\wb{N}+\frac{1-s}{2}\right)}\,\eta^*_j,\nonu\\
&&
}
and $\mc{D}^L_j=0$ were used for the propagator \eqn{dtsoln} and combined with a single iteration ($n=1$) of the midpoint algorithm \eqn{midstep}.
Note the presence of the autocorrelation corrections 
\eqs{strat}{
C^E_{\alpha_j}&=&  \frac{isU}{2}\\
C^E_{\beta_j}&=&  -\frac{isU}{2}
}
as explained in Sec.~\ref{N-STRAT}.

Importantly -- the form of the corrections \eqn{strat} above arises when the noises $\xi$, $\eta$ etc. appearing in the $X_j$ are calculated just once at the beginning of the entire procedure for making the full $\Delta t$ step. Other noise input can change the expressions \eqn{strat}.

The actual timestep $\Delta t$ that must be used is constrained by the need for the coefficients $\mc{D}_j$ to change little over a single timestep $\Delta t$.
A good practical rule of thumb is to require
\eq{rot}{
\frac{\left|\mc{D}^{(\nu)}(\vec{v},t)\,\Delta t\right|}{\left|v_j^{(0)}\right|} \lesssim 0.1,
}
for each term $\mc{D}^{(\nu)}$ in $\mc{D} = \sum_{\nu}\mc{D}^{(\nu)}$. One caveat is that one should take care not to estimate $\mc{D}(\vec{v}^{\,(0)},t)$ on the run using noise or even variable values from individual trajectories. Doing so can, and often does, introduce autocorrelations between noise values and time step lengths, leading to systematic errors. For treating noise terms, it is useful to use the RMS estimate $X_j\approx\sqrt{\langle|X_j|^2\rangle} \propto \sqrt{\Delta t}$ based on expected mean variable values $\langle v_j\rangle$ for timestep estimates.

Finally, for completeness and general applicability beyond the examples of Sec.~\ref{EX}, it can be advantageous to treat widely differing processes using a split-step approach \cite{Werner97}. 
In particular, the treatment of kinetic energy and tunnelling is much bettered for many-mode systems.
A split step allows to take advantage of the fact that kinetic energy and tunnelling are diagonal in k-space and can be done exactly there. k-space is reached using a discrete Fast Fourier Transform that has computational cost of only $M\log M$ for an M-site system \cite{FFTW05}. 
The split-step is applied in three parts:\\
(i) First evolve terms diagonal in k-space for $\Delta t/2$;\\
(ii) then change variables to x-space and evolve the remaining terms there for $\Delta t$; \\
(iii) finally return to k-space for another k-space evolution of $\Delta t/2$. 

For cases like \eqn{ppeq} in which the main mean-field evolution has a Gross-Pitaevskii form, the split step method is symplectic (conserves energy) and has been shown to have an accuracy of $\mc{O}(\Delta t)^2$ for the long time solution \cite{Javanainen06} provided that the latest available copy of the field is input as $\vec{v}^{\,(0)}$ after each Fourier transform \cite{Feit82}. 
Within each k- or x-space split step, the midpoint method \eqn{midstep} can employed if the coefficients $\mc{D}$ of the substep are variable-dependent, or otherwise if the coefficients are constant just a plain propagator substep \eqn{dtsoln} is done. 
Thus, one has the split-step algorithm as a container for midpoint iterations, which themselves are containers for the base propagators.  
Long-range interactions (eg. dipole-dipole) can also often be treated efficiently through Fourier transforms \cite{Wuster17}.

Sumarizing, the algorithm used for the simulations of Sec.~\ref{EX} consists of \eqn{midstep}, \eqn{vij}, \eqn{dtsoln}--\eqn{strat}.

\subsection{Timestep autocorrelation correction}
\label{N-STRAT}

Realisations of Ito stochastic equations \eqn{eqs} must at the least produce the required average and noise variance to lowest order. That is, writing
\eq{deltavj}{
\Delta v_j = v'_j(t+\Delta t)-v_j^{(0)},
}
one needs
\eqa{agreem}{
\langle \Delta v_j\rangle_{\rm stoch} &=& \left\langle A_j(\vec{v}^{\,(0)})\Delta t \right\rangle_{\rm stoch} + \mc{O}(\Delta t)^{3/2}\nonu\\
\!\!\!\langle \Delta v_j \Delta v_k\rangle_{\rm stoch} &=& \left\langle X_j(\vec{v}^{\,(0)})X_k(\vec{v}^{\,(0)})\Delta t^2 \right\rangle_{\rm stoch}\!\!\!\!\!\!+\mc{O}(\Delta t)^{3/2}.\nonu\hspace*{-2em}\\
}
In orders of $\Delta t$, the drift is $A_j\Delta t\sim\mc{O}(\Delta t)$ and the noise $X_j\Delta t\sim\mc{O}(\sqrt{\Delta t})$. The time-dependence of coefficients $A$ and $X$ does not enter at this order of $\Delta t$.
A simple Ito-Euler timestep $v_j^{(0)}\to v_j^{(0)}+A_j(\vec{v}^{\,(0)})\,\Delta t+X_j(\vec{v}^{\,(0)})\,\Delta t$ 
meets both conditions \eqn{agreem}.

However, the midstep algorithm with an Ito-Euler propagator \eqn{euler} can introduce an extra correlation of order $\Delta t$ by virtue of using the same noise for both the halfstep of $\Delta t/2$ and then for the final full step. 
The first halfstep advance using \eqn{euler} produces
\eqa{deltav}{
\Delta v_j^{(1)}&=&v_j^{(1)}-v_j^{(0)} = A_j(\vec{v}^{\,(0)})\frac{\Delta t}{2}+X_j(\vec{v}^{\,(0)})\,\frac{\Delta t}{2},\nonu\\
&\sim& \mc{O}(\sqrt{\Delta t})
}
Note that the $X_j$ themselves are $\mc{O}(\frac{1}{\sqrt{\Delta t}})$, which is important below.
The next step in \eqn{midstep} involves coefficients $A_j(\vec{v}^{\,(1)})$ and $X_j(\vec{v}^{\,(1)})$. Taylor expanding around $\vec{v}^{\,(0)}$,
\eq{taylor1}{
X_j(\vec{v}^{\,(1)}) = X_j(\vec{v}^{\,(0)}) + \sum_k\frac{dX_j}{dv_k}(\vec{v}^{\,(0)})\Delta v_k^{(1)} + \mc{O} (\sqrt{\Delta t}), 
}
and analogously for $A_j$. Using \eqn{deltav} and discarding terms of subleading order $\mc{O}(\sqrt{\Delta t})$ in the new coefficients one finds
\eqa{taylor2}{
X_j(\vec{v}^{\,(1)}) &=& X_j(\vec{v}^{\,(0)}) \nonu\\
			&&+ \sum_k\frac{dX_j}{dv_k}(\vec{v}^{\,(0)})\,X_j(\vec{v}^{\,(0)})\frac{\Delta t}{2} + \mc{O}(\sqrt{\Delta t}),\nonu\\
A_j(\vec{v}^{\,(1)}) &=& A_j(\vec{v}^{\,(0)}) + \mc{O}(\sqrt{\Delta t}).
}
Notice the appearance of a term in $X_j(\vec{v}^{\,(1)})$ of the same order as the original coefficients $X_j(\vec{v}^{\,(0)})$. 
Assuming first just one $n=1$ iteration, the final step \eqn{vij} becomes
\eqa{finalv}{
v' &=& v^{(0)}_j + A_j(\vec{v}^{\,(0)})\Delta t+X_j(\vec{v}^{\,(0)})\,\Delta t\\
 &&+ \sum_k\frac{dX_j}{dv_k}(\vec{v}^{\,(0)})\,X_j(\vec{v}^{\,(0)})\frac{(\Delta t)^2}{2} + \mc{O}(\Delta t)^{3/2}.\nonu
}
It turns out that iterations of $n>1$ do not affect the expression \eqn{finalv} at $\mc{O}(\Delta t)$. 
The new term in \eqn{finalv} breaks the equivalence of the algorithm \eqn{agreem} because it is of the same order as the drift $A_j\Delta t$. 

To restore equivalence, one can add a correction $C_j$
to the drift used in the algorithm as per
\eq{correct}{
A_j \to A_j + C_j.
} 
in either the ``strongly convergent'' form 
\eq{Cjstrong}{
C_j^{\rm strong} = -\frac{\Delta t}{2} \sum_k\frac{dX_j}{dv_k}(\vec{v})\,X_j(\vec{v})
}
or 
the ``weakly convergent'' averaged form $C_j =\left\langle C_j^{\rm strong}\right\rangle_{\rm stoch}$. For the case of Ito-Euler propagator in the midstep algorithm,
\eq{Cj}{
C_j =  C_j^{\rm Strat} = -\frac{\Delta t}{2} \left\langle \sum_k\frac{dX_j}{dv_k}(\vec{v})\,X_j(\vec{v}) \right\rangle_{\rm stoch}.
}
The strong forms \eqn{Cjstrong} have different values for each stochastic realization \cite{Drummond91} and obtain $\mc{O}(\Delta t)$ timestep accuracy for the noise terms, 
while the weak forms use a pre-calculated mean \eqn{Cj}. 
The weak variant is more commonly used in practice and sufficient to obtain overall stable and accurate integration. It is what is applied in \eqn{strat} and the calculations of Sec.~\ref{EX}.

The weak autocorrelation correction $C_j$ has often been dubbed a ``Stratonovich correction'' because the form \eqn{Cj} is identical to the correction used to move between Ito and Stratonovich stochastic calculus.
However, this is a misnomer because the match is merely a coincidence that occurs when an Ito-Euler propagator is used with the midpoint algorithm. 
For the case of the partly exponential propagator \eqn{dtsoln}, the same procedure as above leads to the following, different, form of the autocorrelation corrections. When $n\ge 1$ in the midstep algorithm:
\eq{correction}{
C_j = C_j^{\rm Strat} +\frac{\Delta t}{2} \left\langle X^E_j(\vec{v}) X^L_j(\vec{v}) \right\rangle_{\rm stoch},
}
or when one uses no midstepping ($n=0$):
\eq{itocorrection}{
C_j =  -\frac{\Delta t}{2} \left\langle X^E_j(\vec{v})X^E_j(\vec{v})v_j + X^E_j(\vec{v})X^R_j(\vec{v})\right\rangle_{\rm stoch}.
} 
As it happens, in the case of the equations \eqn{ppeq}, the corrections are the same whether using an Ito-Euler propagator (\eqn{strat}) or the partially exponential one \eqn{correction}-\eqn{itocorrection}, but that is not always the case.

\subsection{Noise implementation}
\label{N-NOISE}
Since the underlying stochastic equations are defined in the infinitesimal limit $\Delta t\to0$, in principle any implementation of the noise that has zero mean and satisfies the variance conditions \eqn{xieg} and \eqn{eta} in the $\Delta t\to0$ limit is correct. By the central limit theorem, after several timesteps the effective distribution of the sum of the noises will always converge to a Gaussian one.
In practice, however, it is usually desirable to use explicitly Gaussian distributed noises in the implementation. Doing so already accurately depicts the limiting distribution for each $\Delta t$ step and avoids a potential need to reduce timestep further to capture the right noise distribution. For real noises $\xi$, one generates fresh Gaussian random variables of variance $1/\Delta t$ for each $j$ at the beginning of each timestep. Generating new ones for each time step ensures the $\delta(t-t')$ limit as $\Delta t\to0$. The Box-Muller algorithm \cite{Box58} is a simple way to obtain two such independent Gaussian variables from two uniformly distributed noises, whereas a variety of more efficient though less transparent algorithms are also known \cite{Bell68,Brent74}.
 In our case Box-Muller was used, while  the underlying uniformly distributed noise was generated using the SFMT fast Mersenne twister method \cite{SFMT} \footnote{\href{http://www.math.sci.hiroshima-u.ac.jp/m-mat/MT/SFMT/}{http://www.math.sci.hiroshima-u.ac.jp/m-mat/MT/SFMT/}}.
Complex noise $\eta$ is simply constructed as $\eta = (\xi_r+i\xi_i)/\sqrt{2}$ using two real Gaussian noises $\xi_{i,r}$. 

When using efficient random number generation such as \cite{SFMT} it turns out that the creation of Gaussian noises from uniform ones is the most computationally costly step. The Brent algorithm \cite{Brent74} alleviates this appreciably, while a simple alternative route applicable at least for stochastic simulations -- due to their central limit properties -- is to produce binomially distributed noise instead.
One adds $n_B$ uniform noises $r_n$ on $[0,1]$ as per $\xi\approx \sqrt{\frac{12}{n_B\Delta t}}\sum_{n=1}^{n_B} (r_n-\tfrac{1}{2})$. In practice $n_B=4$ or $n_B=3$ is already sufficient.

\subsection{Error estimation}
\label{N-ERR}

Statistical uncertainty in quantities $\mc{Q}$ calculated from the simulations can be robustly estimated via sub-ensemble averaging \cite{Werner97} even when the distribution of samples is unknown. 
The full ensemble of, say $\mc{S}$, realisations is divided into  a smaller number $u$ of sub-ensembles.
Auxiliary subensemble means $\wb{\mc{Q}}_{(i=1,\dots,u)}$ are calculated for each sub-ensemble individually. Then by the central limit theorem, the $1\sigma$ uncertainty in the full ensemble prediction is 
\eq{uncert}{
\Delta\mc{Q} = \sqrt{\frac{{\rm var}\left[\wb{\mc{Q}}_{(i)}\right]}{u-1}}.
}
In practice $u\lesssim\sqrt{\mc{S}}$ is useful. In the simulations reported in this paper, $u=32$ was used.

A separate issue is testing the time discretization accuracy in stochastic equations. In a simple implementation where noise is generated sequentially, changing timestep changes also the noise history, making comparison of single trajectories with different $\Delta t$ useless for determination of accuracy. Yet, comparing the ensemble means for different timesteps can be onerous in large systems. 
A solution is to compare two runs of a single realisation with identical underlying noise sources \cite{Drummond91}: One with time step $\Delta t$ and noises $\xi_j^{(i)}$ generated sequentially for time steps numbered $i=1,\dots$. The other with integration timestep $2\Delta t$ but noises generated by adding pairs of noises from the first simulation: $\xi_j^{(j)\prime} = \xi_j^{(2j-1)}+\xi_j^{(2j)}$. This allows separation of the time discretization error from the stochastic randomness.

\bibliography{artnew_tcorr}

\begin{thebibliography}{145}
\providecommand{\natexlab}[1]{#1}
\providecommand{\url}[1]{\texttt{#1}}
\expandafter\ifx\csname urlstyle\endcsname\relax
  \providecommand{\doi}[1]{doi: #1}\else
  \providecommand{\doi}{doi: \begingroup \urlstyle{rm}\Url}\fi

\bibitem[Agarwal(1969{\natexlab{a}})]{Agarwal69a}
G.~S. Agarwal.
\newblock Phase-space analysis of time-correlation functions.
\newblock \emph{Phys. Rev.}, 177:\penalty0 400--407, 1969{\natexlab{a}}.
\newblock \doi{https://doi.org/10.1103/PhysRev.177.400}.

\bibitem[Agarwal(1969{\natexlab{b}})]{Agarwal69b}
G.~S. Agarwal.
\newblock Master equations in phase-space formulation of quantum optics.
\newblock \emph{Phys. Rev.}, 178:\penalty0 2025--2035, 1969{\natexlab{b}}.
\newblock \doi{https://doi.org/10.1103/PhysRev.178.2025}.

\bibitem[Agarwal and Chaturvedi(1994)]{Agarwal94}
G.~S. Agarwal and S.~Chaturvedi.
\newblock {Scheme to measure the positive {$P$} distribution}.
\newblock \emph{Phys. Rev. A}, 49:\penalty0 R665--R667, 1994.
\newblock \doi{https://doi.org/10.1103/PhysRevA.49.R665}.

\bibitem[Agarwal and Wolf(1970{\natexlab{a}})]{Agarwal70b}
G.~S. Agarwal and E.~Wolf.
\newblock Calculus for functions of noncommuting operators and general
  phase-space methods in quantum mechanics. {II}. quantum mechanics in phase
  space.
\newblock \emph{Phys. Rev. D}, 2:\penalty0 2187--2205, 1970{\natexlab{a}}.
\newblock \doi{https://doi.org/10.1103/PhysRevD.2.2187}.

\bibitem[Agarwal and Wolf(1970{\natexlab{b}})]{Agarwal70c}
G.~S. Agarwal and E.~Wolf.
\newblock Calculus for functions of noncommuting operators and general
  phase-space methods in quantum mechanics. {III}. a generalized {Wick} theorem
  and multitime mapping.
\newblock \emph{Phys. Rev. D}, 2:\penalty0 2206--2225, 1970{\natexlab{b}}.
\newblock \doi{https://doi.org/10.1103/PhysRevD.2.2206}.

\bibitem[Aimi and Imada(2007{\natexlab{a}})]{Aimi07a}
Takeshi Aimi and Masatoshi Imada.
\newblock {Gaussian-Basis {Monte Carlo} Method for Numerical Study on Ground
  States of Itinerant and Strongly Correlated Electron Systems}.
\newblock \emph{J. Phys. Soc. Jpn.}, 76:\penalty0 084709, 2007{\natexlab{a}}.
\newblock \doi{https://doi.org/10.1143/JPSJ.76.084709}.

\bibitem[Aimi and Imada(2007{\natexlab{b}})]{Aimi07b}
Takeshi Aimi and Masatoshi Imada.
\newblock {Does Simple Two-Dimensional {Hubbard} Model Account for High-$T_c$
  Superconductivity in Copper Oxides?}
\newblock \emph{J. Phys. Soc. Jpn.}, 76:\penalty0 113708, 2007{\natexlab{b}}.
\newblock \doi{https://doi.org/10.1143/JPSJ.76.113708}.

\bibitem[Albarelli et~al.(2018)Albarelli, Genoni, Paris, and
  Ferraro]{Albarelli18}
Francesco Albarelli, Marco~G. Genoni, Matteo G.~A. Paris, and Alessandro
  Ferraro.
\newblock Resource theory of quantum non-{Gaussianity} and {Wigner} negativity.
\newblock \emph{Phys. Rev. A}, 98:\penalty0 052350, 2018.
\newblock \doi{https://doi.org/10.1103/PhysRevA.98.052350}.

\bibitem[Atalaya et~al.(2018)Atalaya, Hacohen-Gourgy, Martin, Siddiqi, and
  Korotkov]{Atalaya18}
Juan Atalaya, Shay Hacohen-Gourgy, Leigh~S. Martin, Irfan Siddiqi, and
  Alexander~N. Korotkov.
\newblock Multitime correlators in continuous measurement of qubit observables.
\newblock \emph{Phys. Rev. A}, 97:\penalty0 020104, 2018.
\newblock \doi{https://doi.org/10.1103/PhysRevA.97.020104}.

\bibitem[Bamba et~al.(2011)Bamba, Imamo\ifmmode~\breve{g}\else \u{g}\fi{}lu,
  Carusotto, and Ciuti]{Bamba11}
Motoaki Bamba, Atac Imamo\ifmmode~\breve{g}\else \u{g}\fi{}lu, Iacopo
  Carusotto, and Cristiano Ciuti.
\newblock Origin of strong photon antibunching in weakly nonlinear photonic
  molecules.
\newblock \emph{Phys. Rev. A}, 83:\penalty0 021802, 2011.
\newblock \doi{https://doi.org/10.1103/PhysRevA.83.021802}.

\bibitem[Barry and Drummond(2008)]{Barry08}
D.~W. Barry and P.~D. Drummond.
\newblock Qubit phase space: {SU$(n)$} coherent-state {$P$} representations.
\newblock \emph{Phys. Rev. A}, 78:\penalty0 052108, 2008.
\newblock \doi{https://doi.org/10.1103/PhysRevA.78.052108}.

\bibitem[Bell(1968)]{Bell68}
J.~R. Bell.
\newblock Algorithm 334, normal random deviates.
\newblock \emph{Communications of the ACM}, 11\penalty0 (7):\penalty0 498,
  1968.
\newblock \doi{https://doi.org/10.1145/363397.363547}.

\bibitem[Berg et~al.(2009)Berg, Plimak, Polkovnikov, Olsen, Fleischhauer, and
  Schleich]{Berg09}
B.~Berg, L.~I. Plimak, A.~Polkovnikov, M.~K. Olsen, M.~Fleischhauer, and W.~P.
  Schleich.
\newblock Commuting heisenberg operators as the quantum response problem:
  Time-normal averages in the truncated {Wigner} representation.
\newblock \emph{Phys. Rev. A}, 80:\penalty0 033624, 2009.
\newblock \doi{https://doi.org/10.1103/PhysRevA.80.033624}.

\bibitem[Blakie et~al.(2008)Blakie, Bradley, Davis, Ballagh, and
  Gardiner]{Blakie08}
P.~B. Blakie, A.~S. Bradley, M.~J. Davis, R.~J. Ballagh, and C.~W. Gardiner.
\newblock Dynamics and statistical mechanics of ultra-cold {Bose} gases using
  c-field techniques.
\newblock \emph{Advances in Physics}, 57\penalty0 (5):\penalty0 363--455, 2008.
\newblock \doi{https://doi.org/10.1080/00018730802564254}.

\bibitem[Bohrdt et~al.(2017)Bohrdt, Mendl, Endres, and Knap]{Bohrdt17}
A.~Bohrdt, C.~B. Mendl, M.~Endres, and M.~Knap.
\newblock Scrambling and thermalization in a diffusive quantum many-body
  system.
\newblock \emph{New Journal of Physics}, 19\penalty0 (6):\penalty0 063001,
  2017.
\newblock \doi{https://doi.org/10.1088/1367-2630/aa719b}.

\bibitem[Bondar et~al.(2013)Bondar, Cabrera, Zhdanov, and Rabitz]{Bondar13}
Denys~I. Bondar, Renan Cabrera, Dmitry~V. Zhdanov, and Herschel~A. Rabitz.
\newblock {Wigner} phase-space distribution as a wave function.
\newblock \emph{Phys. Rev. A}, 88:\penalty0 052108, 2013.
\newblock \doi{https://doi.org/10.1103/PhysRevA.88.052108}.

\bibitem[Box and Muller(1958)]{Box58}
G.~E.~P. Box and Mervin~E. Muller.
\newblock {A Note on the Generation of Random Normal Deviates}.
\newblock \emph{The Annals of Mathematical Statistics}, 29\penalty0
  (2):\penalty0 610 -- 611, 1958.
\newblock \doi{https://doi.org/10.1214/aoms/1177706645}.

\bibitem[Brent(1974)]{Brent74}
Richard~P. Brent.
\newblock Algorithm 488, a {Gaussian} pseudo-random number generator.
\newblock \emph{Communications of the ACM}, 17\penalty0 (12):\penalty0 704,
  1974.
\newblock \doi{https://doi.org/10.1145/361604.361629}.

\bibitem[Cahill and Glauber(1969{\natexlab{a}})]{Cahill69a}
K.~E. Cahill and R.~J. Glauber.
\newblock Ordered expansions in boson amplitude operators.
\newblock \emph{Phys. Rev.}, 177:\penalty0 1857--1881, 1969{\natexlab{a}}.
\newblock \doi{https://doi.org/10.1103/PhysRev.177.1857}.

\bibitem[Cahill and Glauber(1969{\natexlab{b}})]{Cahill69b}
K.~E. Cahill and R.~J. Glauber.
\newblock Density operators and quasiprobability distributions.
\newblock \emph{Phys. Rev.}, 177:\penalty0 1882--1902, 1969{\natexlab{b}}.
\newblock \doi{https://doi.org/10.1103/PhysRev.177.1882}.

\bibitem[Carter et~al.(1987)Carter, Drummond, Reid, and Shelby]{Carter87}
S.~J. Carter, P.~D. Drummond, M.~D. Reid, and R.~M. Shelby.
\newblock Squeezing of quantum solitons.
\newblock \emph{Phys. Rev. Lett.}, 58:\penalty0 1841--1844, 1987.
\newblock \doi{https://doi.org/10.1103/PhysRevLett.58.1841}.

\bibitem[Carusotto and Castin(2001)]{Carusotto01}
I.~Carusotto and Y.~Castin.
\newblock An exact stochastic field method for the interacting {Bose} gas at
  thermal equilibrium.
\newblock \emph{Journal of Physics B: Atomic, Molecular and Optical Physics},
  34\penalty0 (23):\penalty0 4589, 2001.
\newblock \doi{https://doi.org/10.1088/0953-4075/34/23/305}.

\bibitem[Carusotto and Castin(2003)]{Carusotto03b}
Iacopo Carusotto and Yvan Castin.
\newblock Exact reformulation of the bosonic many-body problem in terms of
  stochastic wave functions: an elementary derivation.
\newblock \emph{Ann. Henri Poincar{\'e}}, 4\penalty0 (2):\penalty0 783--792,
  2003.
\newblock \doi{https://doi.org/10.1007/s00023-003-0961-7}.

\bibitem[Casteels et~al.(2016)Casteels, Rota, Storme, and Ciuti]{Casteels16}
W.~Casteels, R.~Rota, F.~Storme, and C.~Ciuti.
\newblock Probing photon correlations in the dark sites of geometrically
  frustrated cavity lattices.
\newblock \emph{Phys. Rev. A}, 93:\penalty0 043833, 2016.
\newblock \doi{https://doi.org/10.1103/PhysRevA.93.043833}.

\bibitem[Cheneau et~al.(2012)Cheneau, Barmettler, Poletti, Endres, Schau\ss,
  Fukuhara, Gross, Bloch, Kollath, and Kuhr]{Cheneau12}
M.~Cheneau, P.~Barmettler, D.~Poletti, M.~Endres, P.~Schau\ss, T.~Fukuhara,
  C.~Gross, I.~Bloch, C.~Kollath, and S.~Kuhr.
\newblock Light-cone-like spreading of correlations in a quantum many-body
  system.
\newblock \emph{Nature}, 484:\penalty0 484--487, 2012.
\newblock \doi{https://doi.org/10.1038/nature10748}.

\bibitem[Chiocchetta and Carusotto(2014)]{Chiocchetta14}
Alessio Chiocchetta and Iacopo Carusotto.
\newblock Quantum langevin model for nonequilibrium condensation.
\newblock \emph{Phys. Rev. A}, 90:\penalty0 023633, 2014.
\newblock \doi{https://doi.org/10.1103/PhysRevA.90.023633}.

\bibitem[Corney and Drummond(2004)]{Corney04}
J.~F. Corney and P.~D. Drummond.
\newblock {Gaussian Quantum {Monte Carlo} Methods for Fermions and Bosons}.
\newblock \emph{Phys. Rev. Lett.}, 93:\penalty0 260401, 2004.
\newblock \doi{https://doi.org/10.1103/PhysRevLett.93.260401}.

\bibitem[Corney et~al.(1997)Corney, Drummond, and Liebman]{Corney97}
J.~F. Corney, P.~D. Drummond, and A.~Liebman.
\newblock Quantum noise limits to terabaud communications.
\newblock \emph{Opt. Commun.}, 140:\penalty0 211--215, 1997.
\newblock \doi{https://doi.org/10.1016/S0030-4018(97)00191-0}.

\bibitem[Corney et~al.(2006)Corney, Drummond, Heersink, Josse, Leuchs, and
  Andersen]{Corney06}
Joel~F. Corney, Peter~D. Drummond, Joel Heersink, Vincent Josse, Gerd Leuchs,
  and Ulrik~L. Andersen.
\newblock Many-body quantum dynamics of polarization squeezing in optical
  fibers.
\newblock \emph{Phys. Rev. Lett.}, 97:\penalty0 023606, 2006.
\newblock \doi{https://doi.org/10.1103/PhysRevLett.97.023606}.

\bibitem[Corney et~al.(2008)Corney, Heersink, Dong, Josse, Drummond, Leuchs,
  and Andersen]{Corney08}
Joel~F. Corney, Joel Heersink, Ruifang Dong, Vincent Josse, Peter~D. Drummond,
  Gerd Leuchs, and Ulrik~L. Andersen.
\newblock Simulations and experiments on polarization squeezing in optical
  fiber.
\newblock \emph{Phys. Rev. A}, 78:\penalty0 023831, 2008.
\newblock \doi{https://doi.org/10.1103/PhysRevA.78.023831}.

\bibitem[de~Oliveira(1992)]{deOliveira92c}
Fernando A.~M. de~Oliveira.
\newblock s-order nondiagonal quasiprobabilities.
\newblock \emph{Phys. Rev. A}, 45:\penalty0 5104--5112, 1992.
\newblock \doi{https://doi.org/10.1103/PhysRevA.45.5104}.

\bibitem[Dechoum et~al.(2004)Dechoum, Drummond, Chaturvedi, and
  Reid]{Dechoum04}
K.~Dechoum, P.~D. Drummond, S.~Chaturvedi, and M.~D. Reid.
\newblock Critical fluctuations and entanglement in the nondegenerate
  parametric oscillator.
\newblock \emph{Phys. Rev. A}, 70:\penalty0 053807, 2004.
\newblock \doi{https://doi.org/10.1103/PhysRevA.70.053807}.

\bibitem[Dennis et~al.(2013)Dennis, Hope, and Johnsson]{XMDS2}
Graham~R. Dennis, Joseph~J. Hope, and Mattias~T. Johnsson.
\newblock {XMDS2}: Fast, scalable simulation of coupled stochastic partial
  differential equations.
\newblock \emph{Computer Physics Communications}, 184\penalty0 (1):\penalty0
  201--208, 2013.
\newblock \doi{https://doi.org/10.1016/j.cpc.2012.08.016}.

\bibitem[Deuar(2005)]{DeuarPhD}
P.~Deuar.
\newblock \emph{First-principles quantum simulations of many-mode open
  interacting {Bose} gases using stochastic gauge methods}.
\newblock PhD thesis, University of Queensland, arXiv:cond-mat/0507023, 2005.
\newblock URL \url{https://arxiv.org/abs/cond-mat/0507023}.

\bibitem[Deuar(2009)]{Deuar09b}
P.~Deuar.
\newblock Simulation of complete many-body quantum dynamics using controlled
  quantum-semiclassical hybrids.
\newblock \emph{Phys. Rev. Lett.}, 103:\penalty0 130402, 2009.
\newblock \doi{https://doi.org/10.1103/PhysRevLett.103.130402}.

\bibitem[Deuar(2016)]{Deuar16}
P.~Deuar.
\newblock A tractable prescription for large-scale free flight expansion of
  wavefunctions.
\newblock \emph{Computer Physics Communications}, 208:\penalty0 92 -- 102,
  2016.
\newblock \doi{http://dx.doi.org/10.1016/j.cpc.2016.08.004}.

\bibitem[Deuar and Drummond(2002)]{Deuar02}
P.~Deuar and P.~D. Drummond.
\newblock Gauge {$P$} representations for quantum-dynamical problems: Removal
  of boundary terms.
\newblock \emph{Phys. Rev. A}, 66:\penalty0 033812, 2002.
\newblock \doi{https://doi.org/10.1103/PhysRevA.66.033812}.

\bibitem[Deuar and Drummond(2006)]{Deuar06a}
P.~Deuar and P.~D. Drummond.
\newblock First-principles quantum dynamics in interacting {Bose} gases: {I}.
  the positive {P} representation.
\newblock \emph{Journal of Physics A: Mathematical and General}, 39\penalty0
  (5):\penalty0 1163, 2006.
\newblock \doi{https://doi.org/10.1088/0305-4470/39/5/010}.

\bibitem[Deuar and Drummond(2007)]{Deuar07}
P.~Deuar and P.~D. Drummond.
\newblock Correlations in a {BEC} collision: First-principles quantum dynamics
  with 150\,000 atoms.
\newblock \emph{Phys. Rev. Lett.}, 98:\penalty0 120402, 2007.
\newblock \doi{https://doi.org/10.1103/PhysRevLett.98.120402}.

\bibitem[Deuar et~al.(2009)Deuar, Sykes, Gangardt, Davis, Drummond, and
  Kheruntsyan]{Deuar09}
P.~Deuar, A.~G. Sykes, D.~M. Gangardt, M.~J. Davis, P.~D. Drummond, and K.~V.
  Kheruntsyan.
\newblock Nonlocal pair correlations in the one-dimensional {Bose} gas at
  finite temperature.
\newblock \emph{Phys. Rev. A}, 79:\penalty0 043619, 2009.
\newblock \doi{https://doi.org/10.1103/PhysRevA.79.043619}.

\bibitem[Deuar et~al.(2011)Deuar, Chwede\'{n}czuk, Trippenbach, and
  Zi\'{n}]{Deuar11}
P.~Deuar, J.~Chwede\'{n}czuk, M.~Trippenbach, and P.~Zi\'{n}.
\newblock {Bogoliubov} dynamics of condensate collisions using the positive-{P}
  representation.
\newblock \emph{Phys. Rev. A}, 83:\penalty0 063625, 2011.
\newblock \doi{https://doi.org/10.1103/PhysRevA.83.063625}.

\bibitem[Deuar et~al.(2013)Deuar, Wasak, Zi\'{n}, Chwede\'{n}czuk, and
  Trippenbach]{Deuar13}
P.~Deuar, T.~Wasak, P.~Zi\'{n}, J.~Chwede\'{n}czuk, and M.~Trippenbach.
\newblock Tradeoffs for number squeezing in collisions of {Bose-Einstein}
  condensates.
\newblock \emph{Phys. Rev. A}, 88:\penalty0 013617, 2013.
\newblock \doi{https://doi.org/10.1103/PhysRevA.88.013617}.

\bibitem[Deuar et~al.(2014)Deuar, Jaskula, Bonneau, Krachmalnicoff, Boiron,
  Westbrook, and Kheruntsyan]{Deuar14}
P.~Deuar, J.-C. Jaskula, M.~Bonneau, V.~Krachmalnicoff, D.~Boiron, C.~I.
  Westbrook, and K.~V. Kheruntsyan.
\newblock Anisotropy in $s$-wave {Bose}-{Einstein} condensate collisions and
  its relationship to superradiance.
\newblock \emph{Phys. Rev. A}, 90:\penalty0 033613, 2014.
\newblock \doi{https://doi.org/10.1103/PhysRevA.90.033613}.

\bibitem[Deuar and Pietraszewicz(2019)]{nocut}
Piotr Deuar and Joanna Pietraszewicz.
\newblock A semiclassical field theory that is freed of the ultraviolet
  catastrophe, 2019.
\newblock URL \url{https://arxiv.org/abs/1904.06266}.
\newblock arXiv:1904.06266.

\bibitem[Deuar et~al.(2021)Deuar, Ferrier, Matuszewski, Orso, and
  Szyma\'{n}ska]{Deuar21}
Piotr Deuar, Alex Ferrier, Micha\l{} Matuszewski, Giuliano Orso, and Marzena~H.
  Szyma\'{n}ska.
\newblock Fully quantum scalable description of driven-dissipative lattice
  models.
\newblock \emph{PRX Quantum}, 2:\penalty0 010319, 2021.
\newblock \doi{https://doi.org/10.1103/PRXQuantum.2.010319}.

\bibitem[Dowling et~al.(2005)Dowling, Drummond, Davis, and Deuar]{Dowling05}
Mark~R. Dowling, Peter~D. Drummond, Matthew~J. Davis, and Piotr Deuar.
\newblock Time-reversal test for stochastic quantum dynamics.
\newblock \emph{Phys. Rev. Lett.}, 94:\penalty0 130401, 2005.
\newblock \doi{https://doi.org/10.1103/PhysRevLett.94.130401}.

\bibitem[Drummond(1983)]{Drummond83}
P.~D. Drummond.
\newblock Central partial difference propagation algorithms.
\newblock \emph{Computer Physics Communications}, 29\penalty0 (3):\penalty0
  211--225, 1983.
\newblock \doi{https://doi.org/10.1016/0010-4655(83)90001-2}.

\bibitem[Drummond(2014)]{Drummond14}
P.~D. Drummond.
\newblock Fundamentals of higher order stochastic equations.
\newblock \emph{J. Phys. A}, 47\penalty0 (33):\penalty0 335001, 2014.
\newblock \doi{https://doi.org/10.1088/1751-8113/47/33/335001}.

\bibitem[Drummond and Corney(1999)]{Drummond99}
P.~D. Drummond and J.~F. Corney.
\newblock Quantum dynamics of evaporatively cooled {Bose-Einstein} condensates.
\newblock \emph{Phys. Rev. A}, 60:\penalty0 R2661--R2664, 1999.
\newblock \doi{https://doi.org/10.1103/PhysRevA.60.R2661}.

\bibitem[Drummond and Gardiner(1980)]{Drummond80}
P.~D. Drummond and C.~W. Gardiner.
\newblock Generalised {P}-representations in quantum optics.
\newblock \emph{Journal of Physics A: Mathematical and General}, 13\penalty0
  (7):\penalty0 2353, 1980.
\newblock \doi{https://doi.org/10.1088/0305-4470/13/7/018}.

\bibitem[Drummond and Mortimer(1991)]{Drummond91}
P.~D. Drummond and I.~K. Mortimer.
\newblock Computer simulations of multiplicative stochastic differential
  equations.
\newblock \emph{Journal of Computational Physics}, 93\penalty0 (1):\penalty0
  144--170, 1991.
\newblock \doi{https://doi.org/10.1016/0021-9991(91)90077-X}.

\bibitem[Drummond and Walls(1980)]{Drummond80b}
P.~D. Drummond and D.~F. Walls.
\newblock Quantum theory of optical bistability. {I}. nonlinear polarisability
  model.
\newblock \emph{Journal of Physics A: Mathematical and General}, 13\penalty0
  (2):\penalty0 725--741, 1980.
\newblock \doi{https://doi.org/10.1088/0305-4470/13/2/034}.

\bibitem[Drummond et~al.(2016)Drummond, Opanchuk, Rosales-Z\'arate, Reid, and
  Forrester]{Drummond16}
P.~D. Drummond, B.~Opanchuk, L.~Rosales-Z\'arate, M.~D. Reid, and P.~J.
  Forrester.
\newblock Scaling of boson sampling experiments.
\newblock \emph{Phys. Rev. A}, 94:\penalty0 042339, 2016.
\newblock \doi{https://doi.org/10.1103/PhysRevA.94.042339}.

\bibitem[Drummond and Opanchuk(2020)]{Drummond20}
Peter~D. Drummond and Bogdan Opanchuk.
\newblock Initial states for quantum field simulations in phase space.
\newblock \emph{Phys. Rev. Research}, 2:\penalty0 033304, 2020.
\newblock \doi{https://doi.org/10.1103/PhysRevResearch.2.033304}.

\bibitem[Else et~al.(2016)Else, Bauer, and Nayak]{Else16}
Dominic~V. Else, Bela Bauer, and Chetan Nayak.
\newblock Floquet time crystals.
\newblock \emph{Phys. Rev. Lett.}, 117:\penalty0 090402, 2016.
\newblock \doi{https://doi.org/10.1103/PhysRevLett.117.090402}.

\bibitem[Fan et~al.(2017)Fan, Zhang, Shen, and Zhai]{Fan17}
Ruihua Fan, Pengfei Zhang, Huitao Shen, and Hui Zhai.
\newblock Out-of-time-order correlation for many-body localization.
\newblock \emph{Science Bulletin}, 62\penalty0 (10):\penalty0 707--711, 2017.
\newblock \doi{https://doi.org/10.1016/j.scib.2017.04.011}.

\bibitem[Feit et~al.(1982)Feit, Fleck, and Steiger]{Feit82}
M.~D. Feit, J.~A. Fleck, and A.~Steiger.
\newblock Solution of the {Schrödinger} equation by a spectral method.
\newblock \emph{Journal of Computational Physics}, 47\penalty0 (3):\penalty0
  412--433, 1982.
\newblock \doi{http://dx.doi.org/10.1016/0021-9991(82)90091-2}.

\bibitem[Ferrie(2011)]{Ferrie11}
Christopher Ferrie.
\newblock Quasi-probability representations of quantum theory with applications
  to quantum information science.
\newblock \emph{Reports on Progress in Physics}, 74\penalty0 (11):\penalty0
  116001, 2011.
\newblock \doi{https://doi.org/10.1088/0034-4885/74/11/116001}.

\bibitem[Finazzi et~al.(2015)Finazzi, Le~Boit\'e, Storme, Baksic, and
  Ciuti]{Finazzi15}
S.~Finazzi, A.~Le~Boit\'e, F.~Storme, A.~Baksic, and C.~Ciuti.
\newblock Corner-space renormalization method for driven-dissipative
  two-dimensional correlated systems.
\newblock \emph{Phys. Rev. Lett.}, 115:\penalty0 080604, 2015.
\newblock \doi{https://doi.org/10.1103/PhysRevLett.115.080604}.

\bibitem[Foini and Kurchan(2019)]{Foini19}
Laura Foini and Jorge Kurchan.
\newblock Eigenstate thermalization hypothesis and out of time order
  correlators.
\newblock \emph{Phys. Rev. E}, 99:\penalty0 042139, 2019.
\newblock \doi{https://doi.org/10.1103/PhysRevE.99.042139}.

\bibitem[Frigo and Johnson(2005)]{FFTW05}
Matteo Frigo and Steven~G. Johnson.
\newblock The design and implementation of {FFTW3}.
\newblock \emph{Proceedings of the IEEE}, 93\penalty0 (2):\penalty0 216--231,
  2005.
\newblock \doi{https://doi.org/10.1109/JPROC.2004.840301}.

\bibitem[Gardiner(1991)]{QuantumNoise}
C.~W. Gardiner.
\newblock \emph{Quantum Noise}.
\newblock Springer-Verlag, Berlin, 1991.
\newblock ISBN 9783662096444, 9783662096420.

\bibitem[G{\"a}rttner et~al.(2017)G{\"a}rttner, Bohnet, Safavi-Naini, Wall,
  Bollinger, and Rey]{Garttner17}
Martin G{\"a}rttner, Justin~G. Bohnet, Arghavan Safavi-Naini, Michael~L. Wall,
  John~J. Bollinger, and Ana~Maria Rey.
\newblock Measuring out-of-time-order correlations and multiple quantum spectra
  in a trapped-ion quantum magnet.
\newblock \emph{Nature Physics}, 13\penalty0 (8):\penalty0 781--786, 2017.
\newblock \doi{https://doi.org/10.1038/nphys4119}.

\bibitem[G\"arttner et~al.(2018)G\"arttner, Hauke, and Rey]{Garttner18}
Martin G\"arttner, Philipp Hauke, and Ana~Maria Rey.
\newblock Relating out-of-time-order correlations to entanglement via
  multiple-quantum coherences.
\newblock \emph{Phys. Rev. Lett.}, 120:\penalty0 040402, 2018.
\newblock \doi{https://doi.org/10.1103/PhysRevLett.120.040402}.

\bibitem[Gehrke et~al.(2012)Gehrke, Sperling, and Vogel]{Gehrke12}
C.~Gehrke, J.~Sperling, and W.~Vogel.
\newblock Quantification of nonclassicality.
\newblock \emph{Phys. Rev. A}, 86:\penalty0 052118, 2012.
\newblock \doi{https://doi.org/10.1103/PhysRevA.86.052118}.

\bibitem[Gilchrist et~al.(1997)Gilchrist, Gardiner, and Drummond]{Gilchrist97}
A.~Gilchrist, C.~W. Gardiner, and P.~D. Drummond.
\newblock Positive {P} representation: Application and validity.
\newblock \emph{Phys. Rev. A}, 55:\penalty0 3014--3032, 1997.
\newblock \doi{https://doi.org/10.1103/PhysRevA.55.3014}.

\bibitem[Glauber(1963{\natexlab{a}})]{Glauber63}
Roy~J. Glauber.
\newblock Coherent and incoherent states of the radiation field.
\newblock \emph{Phys. Rev.}, 131:\penalty0 2766--2788, 1963{\natexlab{a}}.
\newblock \doi{https://doi.org/10.1103/PhysRev.131.2766}.

\bibitem[Glauber(1963{\natexlab{b}})]{Glauber63a}
Roy~J. Glauber.
\newblock The quantum theory of optical coherence.
\newblock \emph{Phys. Rev.}, 130:\penalty0 2529--2539, 1963{\natexlab{b}}.
\newblock \doi{https://doi.org/10.1103/PhysRev.130.2529}.

\bibitem[Goblot et~al.(2019)Goblot, Rauer, Vicentini, Le~Boit\'e, Galopin,
  Lema\^{\i}tre, Le~Gratiet, Harouri, Sagnes, Ravets, Ciuti, Amo, and
  Bloch]{Goblot19}
V.~Goblot, B.~Rauer, F.~Vicentini, A.~Le~Boit\'e, E.~Galopin, A.~Lema\^{\i}tre,
  L.~Le~Gratiet, A.~Harouri, I.~Sagnes, S.~Ravets, C.~Ciuti, A.~Amo, and
  J.~Bloch.
\newblock Nonlinear polariton fluids in a flatband reveal discrete gap
  solitons.
\newblock \emph{Phys. Rev. Lett.}, 123:\penalty0 113901, 2019.
\newblock \doi{https://doi.org/10.1103/PhysRevLett.123.113901}.

\bibitem[He and Lu(2017)]{He17}
Rong-Qiang He and Zhong-Yi Lu.
\newblock Characterizing many-body localization by out-of-time-ordered
  correlation.
\newblock \emph{Phys. Rev. B}, 95:\penalty0 054201, 2017.
\newblock \doi{https://doi.org/10.1103/PhysRevB.95.054201}.

\bibitem[Hoffmann et~al.(2008)Hoffmann, Corney, and Drummond]{Hoffmann08}
Scott~E. Hoffmann, Joel~F. Corney, and Peter~D. Drummond.
\newblock Hybrid phase-space simulation method for interacting {Bose} fields.
\newblock \emph{Phys. Rev. A}, 78:\penalty0 013622, 2008.
\newblock \doi{https://doi.org/10.1103/PhysRevA.78.013622}.

\bibitem[Houck et~al.(2012)Houck, Türeci, and Koch]{Houck12}
A.~A. Houck, H.~E. Türeci, and J.~Koch.
\newblock On-chip quantum simulation with superconducting circuits.
\newblock \emph{Nature Physics}, 8:\penalty0 292--299, 2012.
\newblock \doi{https://doi.org/10.1038/nphys2251}.

\bibitem[Huber et~al.(2021)Huber, Kirton, and Rabl]{HUber20b}
Julian Huber, Peter Kirton, and Peter Rabl.
\newblock {Phase-Space Methods for Simulating the Dissipative Many-Body
  Dynamics of Collective Spin Systems}.
\newblock \emph{SciPost Phys.}, 10:\penalty0 45, 2021.
\newblock URL \url{https://scipost.org/10.21468/SciPostPhys.10.2.045}.

\bibitem[Hush et~al.(2013)Hush, Szigeti, Carvalho, and Hope]{Hush13}
M.~R. Hush, S.~S. Szigeti, A.~R.~R. Carvalho, and J.~J. Hope.
\newblock {Controlling spontaneous-emission noise in measurement-based feedback
  cooling of a {Bose-Einstein} condensate}.
\newblock \emph{New J. Phys.}, 15\penalty0 (11):\penalty0 113060, 2013.
\newblock \doi{https://doi.org/10.1088/1367-2630/15/11/113060}.

\bibitem[Husimi(1940)]{Husimi40}
K.~Husimi.
\newblock Some formal properties of the density matrix.
\newblock \emph{Proc. Phys. Math. Soc. Jpn.}, 22:\penalty0 264--314, 1940.
\newblock \doi{https://doi.org/10.11429/ppmsj1919.22.4_264}.

\bibitem[Ikeda and Watanabe(1988)]{Ikeda88}
Nobuyuki Ikeda and Shinzo Watanabe.
\newblock \emph{Stochastic Differential Equations and Diffusion Processes},
  volume~24 of \emph{North-Holland Mathematical Library}.
\newblock North Holland, 2nd edition, 1988.
\newblock ISBN 0444861726, 9780444861726.

\bibitem[Javanainen and Ruostekoski(2006)]{Javanainen06}
Juha Javanainen and Janne Ruostekoski.
\newblock Symbolic calculation in development of algorithms: split-step methods
  for the {Gross–Pitaevskii} equation.
\newblock \emph{Journal of Physics A: Mathematical and General}, 39\penalty0
  (12):\penalty0 L179, 2006.
\newblock \doi{https://doi.org/10.1088/0305-4470/39/12/L02}.

\bibitem[Ji et~al.(2015)Ji, Gladilin, and Wouters]{Kal15}
Kai Ji, Vladimir~N. Gladilin, and Michiel Wouters.
\newblock Temporal coherence of one-dimensional nonequilibrium quantum fluids.
\newblock \emph{Phys. Rev. B}, 91:\penalty0 045301, 2015.
\newblock \doi{https://doi.org/10.1103/PhysRevB.91.045301}.

\bibitem[Jumarie(1999)]{Jumarie99}
Guy Jumarie.
\newblock Complex-valued {Wiener} measure: An approach via random walk in the
  complex plane.
\newblock \emph{Statistics and Probability Letters}, 42\penalty0 (1):\penalty0
  61--67, 1999.
\newblock \doi{https://doi.org/10.1016/S0167-7152(98)00194-1}.

\bibitem[Jumarie(2005)]{Jumarie05}
Guy Jumarie.
\newblock On the representation of fractional brownian motion as an integral
  with respect to {$(dt)^a$}.
\newblock \emph{Applied Mathematics Letters}, 18\penalty0 (7):\penalty0
  739--748, 2005.
\newblock \doi{https://doi.org/10.1016/j.aml.2004.05.014}.

\bibitem[Kelley and Kleiner(1964)]{Kelley64}
P.~L. Kelley and W.~H. Kleiner.
\newblock Theory of electromagnetic field measurement and photoelectron
  counting.
\newblock \emph{Phys. Rev.}, 136:\penalty0 A316--A334, 1964.
\newblock \doi{https://doi.org/10.1103/PhysRev.136.A316}.

\bibitem[Kheruntsyan et~al.(2012)Kheruntsyan, Jaskula, Deuar, Bonneau,
  Partridge, Ruaudel, Lopes, Boiron, and Westbrook]{Kheruntsyan12}
K.~V. Kheruntsyan, J.-C. Jaskula, P.~Deuar, M.~Bonneau, G.~B. Partridge,
  J.~Ruaudel, R.~Lopes, D.~Boiron, and C.~I. Westbrook.
\newblock Violation of the {Cauchy-Schwarz} inequality with matter waves.
\newblock \emph{Phys. Rev. Lett.}, 108:\penalty0 260401, 2012.
\newblock \doi{https://doi.org/10.1103/PhysRevLett.108.260401}.

\bibitem[Kiesel et~al.(2008)Kiesel, Vogel, Parigi, Zavatta, and
  Bellini]{Kiesel08}
T.~Kiesel, W.~Vogel, V.~Parigi, A.~Zavatta, and M.~Bellini.
\newblock Experimental determination of a nonclassical {Glauber-Sudarshan $P$}
  function.
\newblock \emph{Phys. Rev. A}, 78:\penalty0 021804, 2008.
\newblock \doi{https://doi.org/10.1103/PhysRevA.78.021804}.

\bibitem[Kiesewetter et~al.(2014)Kiesewetter, He, Drummond, and
  Reid]{Kiesewetter14}
S.~Kiesewetter, Q.~Y. He, P.~D. Drummond, and M.~D. Reid.
\newblock Scalable quantum simulation of pulsed entanglement and
  {Einstein-Podolsky-Rosen} steering in optomechanics.
\newblock \emph{Phys. Rev. A}, 90:\penalty0 043805, 2014.
\newblock \doi{https://doi.org/10.1103/PhysRevA.90.043805}.

\bibitem[Kinsler and Drummond(1991)]{Kinsler91}
P.~Kinsler and P.~D. Drummond.
\newblock Quantum dynamics of the parametric oscillator.
\newblock \emph{Phys. Rev. A}, 43:\penalty0 6194--6208, 1991.
\newblock \doi{https://doi.org/10.1103/PhysRevA.43.6194}.

\bibitem[Klobas et~al.(2020)Klobas, Vanicat, Garrahan, and Prosen]{Klobas20}
Katja Klobas, Matthieu Vanicat, Juan~P Garrahan, and Toma{\v{z}} Prosen.
\newblock Matrix product state of multi-time correlations.
\newblock \emph{Journal of Physics A: Mathematical and Theoretical},
  53\penalty0 (33):\penalty0 335001, 2020.
\newblock \doi{https://doi.org/10.1088/1751-8121/ab8c62}.

\bibitem[Kloeden and Platen(1992)]{Kloeden92}
Peter~E. Kloeden and Eckhard Platen.
\newblock \emph{Numerical solution of stochastic differential equations}.
\newblock Stochastic Modelling and Applied Probability. Springer-verlag, Berlin
  Heidelberg, 1992.
\newblock ISBN 978-3-540-54062-5.
\newblock \doi{https://doi.org/10.1007/978-3-662-12616-5}.

\bibitem[Korbicz et~al.(2005)Korbicz, Cirac, Wehr, and Lewenstein]{Korbicz05}
J.~K. Korbicz, J.~I. Cirac, Jan Wehr, and M.~Lewenstein.
\newblock Hilbert's 17th problem and the quantumness of states.
\newblock \emph{Phys. Rev. Lett.}, 94:\penalty0 153601, 2005.
\newblock \doi{https://doi.org/10.1103/PhysRevLett.94.153601}.

\bibitem[Krumm et~al.(2016)Krumm, Sperling, and Vogel]{Krumm16}
F.~Krumm, J.~Sperling, and W.~Vogel.
\newblock Multitime correlation functions in nonclassical stochastic processes.
\newblock \emph{Phys. Rev. A}, 93:\penalty0 063843, 2016.
\newblock \doi{https://doi.org/10.1103/PhysRevA.93.063843}.

\bibitem[Kubo et~al.(1985)Kubo, Toda, and Hashitsume]{Kubo85}
Ryogo Kubo, Morikazu Toda, and Natsuki Hashitsume.
\newblock \emph{Statistical Physics {II}}.
\newblock Springer-Verlag, Berlin, 1985.
\newblock ISBN 978-3-540-53833-2.
\newblock \doi{https://doi.org/10.1007/978-3-642-58244-8}.

\bibitem[Lamprecht et~al.(2002)Lamprecht, Olsen, Drummond, and
  Ritsch]{Lamprecht02}
C.~Lamprecht, M.~K. Olsen, P.~D. Drummond, and H.~Ritsch.
\newblock Positive-{P} and {Wigner} representations for quantum-optical systems
  with nonorthogonal modes.
\newblock \emph{Phys. Rev. A}, 65:\penalty0 053813, 2002.
\newblock \doi{https://doi.org/10.1103/PhysRevA.65.053813}.

\bibitem[Lax(1968)]{Lax68}
Melvin Lax.
\newblock Quantum noise. {XI}. multitime correspondence between quantum and
  classical stochastic processes.
\newblock \emph{Phys. Rev.}, 172:\penalty0 350--361, 1968.
\newblock \doi{https://doi.org/10.1103/PhysRev.172.350}.

\bibitem[Lee(1995)]{Lee95}
Hai-Woong Lee.
\newblock Theory and application of the quantum phase-space distribution
  functions.
\newblock \emph{Physics Reports}, 259\penalty0 (3):\penalty0 147--211, 1995.
\newblock \doi{https://doi.org/10.1016/0370-1573(95)00007-4}.

\bibitem[Lewis-Swan and Kheruntsyan(2014)]{Lewis-Swan14}
R.~J. Lewis-Swan and K.~V. Kheruntsyan.
\newblock Proposal for demonstrating the {Hong–Ou–Mandel} effect with
  matter waves.
\newblock \emph{Nature Commun.}, 5:\penalty0 3752, 2014.
\newblock \doi{https://doi.org/10.1038/ncomms4752}.

\bibitem[Lewis-Swan and Kheruntsyan(2015)]{Lewis-Swan15}
R.~J. Lewis-Swan and K.~V. Kheruntsyan.
\newblock Proposal for a motional-state {Bell} inequality test with ultracold
  atoms.
\newblock \emph{Phys. Rev. A}, 91:\penalty0 052114, 2015.
\newblock \doi{https://doi.org/10.1103/PhysRevA.91.052114}.

\bibitem[Liew and Savona(2010)]{Liew10}
T.~C.~H. Liew and V.~Savona.
\newblock Single photons from coupled quantum modes.
\newblock \emph{Phys. Rev. Lett.}, 104:\penalty0 183601, 2010.
\newblock \doi{https://doi.org/10.1103/PhysRevLett.104.183601}.

\bibitem[Läuchli and Kollath(2008)]{Lauchli08}
Andreas~M Läuchli and Corinna Kollath.
\newblock Spreading of correlations and entanglement after a quench in the
  one-dimensional {Bose-Hubbard} model.
\newblock \emph{Journal of Statistical Mechanics: Theory and Experiment},
  2008\penalty0 (05):\penalty0 P05018, 2008.
\newblock \doi{https://doi.org/10.1088/1742-5468/2008/05/p05018}.

\bibitem[Maldacena et~al.(2016)Maldacena, Shenker, and Stanford]{Maldacena16}
Juan Maldacena, Stephen~H. Shenker, and Douglas Stanford.
\newblock A bound on chaos.
\newblock \emph{Journal of High Energy Physics}, 2016\penalty0 (8):\penalty0
  106, 2016.
\newblock \doi{https://doi.org/10.1007/JHEP08(2016)106}.

\bibitem[Mandel(1966)]{Mandel66}
L.~Mandel.
\newblock Antinormally ordered correlations and quantum counters.
\newblock \emph{Phys. Rev.}, 152:\penalty0 438--451, 1966.
\newblock \doi{https://doi.org/10.1103/PhysRev.152.438}.

\bibitem[Mandt et~al.(2015)Mandt, Sadri, Houck, and Türeci]{Mandt15}
Stephan Mandt, Darius Sadri, Andrew~A Houck, and Hakan~E Türeci.
\newblock Stochastic differential equations for quantum dynamics of spin-boson
  networks.
\newblock \emph{New Journal of Physics}, 17\penalty0 (5):\penalty0 053018,
  2015.
\newblock \doi{https://doi.org/10.1088/1367-2630/17/5/053018}.

\bibitem[Mathey et~al.(2014)Mathey, Clark, and Mathey]{Mathey14}
Amy~C. Mathey, Charles~W. Clark, and L.~Mathey.
\newblock Decay of a superfluid current of ultracold atoms in a toroidal trap.
\newblock \emph{Phys. Rev. A}, 90:\penalty0 023604, 2014.
\newblock \doi{https://doi.org/10.1103/PhysRevA.90.023604}.

\bibitem[Midgley et~al.(2009)Midgley, W\"uster, Olsen, Davis, and
  Kheruntsyan]{Midgley09}
S.~L.~W. Midgley, S.~W\"uster, M.~K. Olsen, M.~J. Davis, and K.~V. Kheruntsyan.
\newblock Comparative study of dynamical simulation methods for the
  dissociation of molecular {Bose-Einstein} condensates.
\newblock \emph{Phys. Rev. A}, 79:\penalty0 053632, 2009.
\newblock \doi{https://doi.org/10.1103/PhysRevA.79.053632}.

\bibitem[Mocza{\l}a-Dusanowska et~al.(2019)Mocza{\l}a-Dusanowska, Dusanowski,
  Gerhardt, He, Reindl, Rastelli, Trotta, Gregersen, H\"ofling, and
  Schneider]{Moczala-Dusanowska19}
Magdalena Mocza{\l}a-Dusanowska, {\L}ukasz Dusanowski, Stefan Gerhardt, Yu~Ming
  He, Marcus Reindl, Armando Rastelli, Rinaldo Trotta, Niels Gregersen, Sven
  H\"ofling, and Christian Schneider.
\newblock Strain-tunable single-photon source based on a quantum
  dot–micropillar system.
\newblock \emph{ACS Photonics}, 6\penalty0 (8):\penalty0 2025--2031, 2019.
\newblock \doi{https://doi.org/10.1021/acsphotonics.9b00481}.

\bibitem[Moreva et~al.(2017)Moreva, Gramegna, Brida, Maccone, and
  Genovese]{Moreva17}
Ekaterina Moreva, Marco Gramegna, Giorgio Brida, Lorenzo Maccone, and Marco
  Genovese.
\newblock Quantum time: Experimental multitime correlations.
\newblock \emph{Phys. Rev. D}, 96:\penalty0 102005, 2017.
\newblock \doi{https://doi.org/10.1103/PhysRevD.96.102005}.

\bibitem[Moyal(1949)]{Moyal49}
J.~E. Moyal.
\newblock Quantum mechanics as a statistical theory.
\newblock \emph{Mathematical Proceedings of the Cambridge Philosophical
  Society}, 45\penalty0 (01):\penalty0 99--124, 1949.
\newblock \doi{https://doi.org/10.1017/S0305004100000487}.

\bibitem[Ng and S\o{}rensen(2011)]{Ng11}
Ray Ng and Erik~S. S\o{}rensen.
\newblock Exact real-time dynamics of quantum spin systems using the
  positive-{P} representation.
\newblock \emph{J. Phys. A}, 44:\penalty0 065305, 2011.
\newblock \doi{https://doi.org/10.1088/1751-8113/44/6/065305}.

\bibitem[Ng et~al.(2013)Ng, S\o{}rensen, and Deuar]{Ng13}
Ray Ng, Erik~S. S\o{}rensen, and Piotr Deuar.
\newblock Simulation of the dynamics of many-body quantum spin systems using
  phase-space techniques.
\newblock \emph{Phys. Rev. B}, 88:\penalty0 144304, 2013.
\newblock \doi{https://doi.org/10.1103/PhysRevB.88.144304}.

\bibitem[Norrie et~al.(2005)Norrie, Ballagh, and Gardiner]{Norrie05}
A.~A. Norrie, R.~J. Ballagh, and C.~W. Gardiner.
\newblock Quantum turbulence in condensate collisions: An application of the
  classical field method.
\newblock \emph{Phys. Rev. Lett.}, 94:\penalty0 040401, 2005.
\newblock \doi{https://doi.org/10.1103/PhysRevLett.94.040401}.

\bibitem[Olsen et~al.(2002)Olsen, Plimak, and Fleischhauer]{Olsen02}
M.~K. Olsen, L.~I. Plimak, and M.~Fleischhauer.
\newblock Quantum-theoretical treatments of three-photon processes.
\newblock \emph{Phys. Rev. A}, 65:\penalty0 053806, 2002.
\newblock \doi{https://doi.org/10.1103/PhysRevA.65.053806}.

\bibitem[Olsen et~al.(2004)Olsen, Melo, Dechoum, and Khoury]{Olsen04}
M.~K. Olsen, A.~B. Melo, K.~Dechoum, and A.~Z. Khoury.
\newblock Quantum phase-space analysis of the pendular cavity.
\newblock \emph{Phys. Rev. A}, 70:\penalty0 043815, 2004.
\newblock \doi{https://doi.org/10.1103/PhysRevA.70.043815}.

\bibitem[Opanchuk et~al.(2013)Opanchuk, Polkinghorne, Fialko, Brand, and
  Drummond]{Opanchuk13}
Bogdan Opanchuk, Rodney Polkinghorne, Oleksandr Fialko, Joachim Brand, and
  Peter~D. Drummond.
\newblock Quantum simulations of the early universe.
\newblock \emph{Annalen der Physik}, 525\penalty0 (10-11):\penalty0 866--876,
  2013.
\newblock \doi{https://doi.org/10.1002/andp.201300113}.

\bibitem[Opanchuk et~al.(2018)Opanchuk, Rosales-Z\'arate, Reid, and
  Drummond]{Opanchuk18}
Bogdan Opanchuk, Laura Rosales-Z\'arate, Margaret~D. Reid, and Peter~D.
  Drummond.
\newblock Simulating and assessing boson sampling experiments with phase-space
  representations.
\newblock \emph{Phys. Rev. A}, 97:\penalty0 042304, 2018.
\newblock \doi{https://doi.org/10.1103/PhysRevA.97.042304}.

\bibitem[Opanchuk et~al.(2019)Opanchuk, Rosales-Z\'{a}rate, Reid, and
  Drummond]{Opanchuk19}
Bogdan Opanchuk, Laura Rosales-Z\'{a}rate, Margaret~D. Reid, and Peter~D.
  Drummond.
\newblock Robustness of quantum {Fourier} transform interferometry.
\newblock \emph{Opt. Lett.}, 44\penalty0 (2):\penalty0 343--346, 2019.
\newblock \doi{https://doi.org/10.1364/OL.44.000343}.

\bibitem[Pietraszewicz et~al.(2019)Pietraszewicz, Stobi\ifmmode~\acute{n}\else
  \'{n}\fi{}ska, and Deuar]{stob}
J.~Pietraszewicz, M.~Stobi\ifmmode~\acute{n}\else \'{n}\fi{}ska, and P.~Deuar.
\newblock Correlation evolution in dilute {Bose-Einstein} condensates after
  quantum quenches.
\newblock \emph{Phys. Rev. A}, 99:\penalty0 023620, 2019.
\newblock \doi{https://doi.org/10.1103/PhysRevA.99.023620}.

\bibitem[Plimak and Olsen(2014)]{Plimak14}
L.~I. Plimak and M.~K. Olsen.
\newblock Quantum-field-theoretical approach to phase–space techniques:
  Symmetric {Wick} theorem and multitime {Wigner} representation.
\newblock \emph{Annals of Physics}, 351:\penalty0 593 -- 619, 2014.
\newblock \doi{https://doi.org/10.1016/j.aop.2014.09.010}.

\bibitem[Plimak et~al.(2001)Plimak, Olsen, Fleischhauer, and Collett]{Plimak01}
L.~I. Plimak, M.~K. Olsen, M.~Fleischhauer, and M.~J. Collett.
\newblock Beyond the {Fokker-Planck} equation: Stochastic simulation of
  complete {Wigner} representation for the optical parametric oscillator.
\newblock \emph{Europhysics Letters ({EPL})}, 56\penalty0 (3):\penalty0
  372--378, 2001.
\newblock \doi{https://doi.org/10.1209/epl/i2001-00529-8}.

\bibitem[Plimak et~al.(2003)Plimak, Fleischhauer, Olsen, and Collett]{Plimak03}
L.~I. Plimak, M.~Fleischhauer, M.~K. Olsen, and M.~J. Collett.
\newblock Quantum-field-theoretical approach to phase-space techniques:
  Generalizing the positive-{P} representation.
\newblock \emph{Phys. Rev. A}, 67:\penalty0 013812, 2003.
\newblock \doi{https://doi.org/10.1103/PhysRevA.67.013812}.

\bibitem[Polkovnikov(2010)]{Polkovnikov10}
Anatoli Polkovnikov.
\newblock Phase space representation of quantum dynamics.
\newblock \emph{Annals of Physics}, 325\penalty0 (8):\penalty0 1790 -- 1852,
  2010.
\newblock \doi{http://dx.doi.org/10.1016/j.aop.2010.02.006}.

\bibitem[Ringbauer et~al.(2018)Ringbauer, Costa, Goggin, White, and
  Alessandro]{Ringbauer18}
Martin Ringbauer, Fabio Costa, Michael~E. Goggin, Andrew~G. White, and Fedrizzi
  Alessandro.
\newblock Multi-time quantum correlations with no spatial analog.
\newblock \emph{NPJ Quantum Information}, 4:\penalty0 37, 2018.
\newblock \doi{https://doi.org/10.1038/s41534-018-0086-y}.

\bibitem[Ross et~al.(2021)Ross, Deuar, Shin, Thomas, Henson, Hodgman, and
  Truscott]{Ross21}
J.~A. Ross, P.~Deuar, D.~K. Shin, K.~F. Thomas, B.~M. Henson, S.~S. Hodgman,
  and A.~G Truscott.
\newblock Survival of the quantum depletion of a condensate after release from
  a harmonic trap in theory and experiment, 2021.
\newblock URL \url{https://arxiv.org/abs/2103.15283}.
\newblock arXiv:2103.15283.

\bibitem[Saito and Matsumoto(2008)]{SFMT}
Mutsuo Saito and Makoto Matsumoto.
\newblock Simd-oriented fast {Mersenne} twister: a 128-bit pseudorandom number
  generator.
\newblock In Alexander Keller, Stefan Heinrich, and Harald Niederreiter,
  editors, \emph{{Monte Carlo} and Quasi-{Monte Carlo} Methods 2006}, pages
  607--622, Berlin, Heidelberg, 2008. Springer Berlin Heidelberg.
\newblock ISBN 978-3-540-74496-2.
\newblock \doi{https://doi.org/10.1007/978-3-540-74496-2_36}.

\bibitem[Schmidt and Koch(2013)]{Schmidt13}
Sebastian Schmidt and Jens Koch.
\newblock Circuit qed lattices: Towards quantum simulation with superconducting
  circuits.
\newblock \emph{Annalen der Physik}, 525\penalty0 (6):\penalty0 395--412, 2013.
\newblock \doi{https://doi.org/10.1002/andp.201200261}.

\bibitem[Schneider et~al.(2016)Schneider, Winkler, Fraser, Kamp, Yamamoto,
  Ostrovskaya, and Höfling]{Schneider16}
C.~Schneider, K.~Winkler, M.~D. Fraser, M.~Kamp, Y.~Yamamoto, E.~A.
  Ostrovskaya, and S.~Höfling.
\newblock Exciton-polariton trapping and potential landscape engineering.
\newblock \emph{Reports on Progress in Physics}, 80\penalty0 (1):\penalty0
  016503, 2016.
\newblock \doi{https://doi.org/10.1088/0034-4885/80/1/016503}.

\bibitem[Shen et~al.(2017)Shen, Zhang, Fan, and Zhai]{Shen17}
Huitao Shen, Pengfei Zhang, Ruihua Fan, and Hui Zhai.
\newblock Out-of-time-order correlation at a quantum phase transition.
\newblock \emph{Phys. Rev. B}, 96:\penalty0 054503, 2017.
\newblock \doi{https://doi.org/10.1103/PhysRevB.96.054503}.

\bibitem[Sinatra et~al.(2002)Sinatra, Lobo, and Castin]{Sinatra02}
Alice Sinatra, Carlos Lobo, and Yvan Castin.
\newblock The truncated {{Wigner}} method for {Bose}-condensed gases: limits of
  validity and applications.
\newblock \emph{Journal of Physics B: Atomic, Molecular and Optical Physics},
  35\penalty0 (17):\penalty0 3599, 2002.
\newblock \doi{https://doi.org/10.1088/0953-4075/35/17/301}.

\bibitem[Smith and Gardiner(1989)]{Smith89}
A.~M. Smith and C.~W. Gardiner.
\newblock Simulations of nonlinear quantum damping using the positive {P}
  representation.
\newblock \emph{Phys. Rev. A}, 39:\penalty0 3511--3524, 1989.
\newblock \doi{https://doi.org/10.1103/PhysRevA.39.3511}.

\bibitem[Spekkens(2008)]{Spekkens08}
Robert~W. Spekkens.
\newblock Negativity and contextuality are equivalent notions of
  nonclassicality.
\newblock \emph{Phys. Rev. Lett.}, 101:\penalty0 020401, 2008.
\newblock \doi{https://doi.org/10.1103/PhysRevLett.101.020401}.

\bibitem[Sperling(2016)]{Sperling16}
J.~Sperling.
\newblock Characterizing maximally singular phase-space distributions.
\newblock \emph{Phys. Rev. A}, 94:\penalty0 013814, 2016.
\newblock \doi{https://doi.org/10.1103/PhysRevA.94.013814}.

\bibitem[Sperling and Vogel(2020)]{Sperling20}
J~Sperling and W~Vogel.
\newblock Quasiprobability distributions for quantum-optical coherence and
  beyond.
\newblock \emph{Physica Scripta}, 95\penalty0 (3):\penalty0 034007, 2020.
\newblock \doi{https://doi.org/10.1088/1402-4896/ab5501}.

\bibitem[Spohn(1980)]{Spohn80}
Herbert Spohn.
\newblock Kinetic equations from hamiltonian dynamics: Markovian limits.
\newblock \emph{Rev. Mod. Phys.}, 52:\penalty0 569--615, 1980.
\newblock \doi{https://doi.org/10.1103/RevModPhys.52.569}.

\bibitem[Steel et~al.(1998)Steel, Olsen, Plimak, Drummond, Tan, Collett, Walls,
  and Graham]{Steel98}
M.~J. Steel, M.~K. Olsen, L.~I. Plimak, P.~D. Drummond, S.~M. Tan, M.~J.
  Collett, D.~F. Walls, and R.~Graham.
\newblock Dynamical quantum noise in trapped {Bose-Einstein} condensates.
\newblock \emph{Phys. Rev. A}, 58:\penalty0 4824--4835, 1998.
\newblock \doi{https://doi.org/10.1103/PhysRevA.58.4824}.

\bibitem[Sudarshan(1963)]{Sudarshan63}
E.~C.~G. Sudarshan.
\newblock {Equivalence of Semiclassical and Quantum Mechanical Descriptions of
  Statistical Light Beams}.
\newblock \emph{Phys. Rev. Lett.}, 10:\penalty0 277--279, 1963.
\newblock \doi{https://doi.org/10.1103/PhysRevLett.10.277}.

\bibitem[Swingle et~al.(2016)Swingle, Bentsen, Schleier-Smith, and
  Hayden]{Swingle16}
Brian Swingle, Gregory Bentsen, Monika Schleier-Smith, and Patrick Hayden.
\newblock Measuring the scrambling of quantum information.
\newblock \emph{Phys. Rev. A}, 94:\penalty0 040302, 2016.
\newblock \doi{https://doi.org/10.1103/PhysRevA.94.040302}.

\bibitem[\'Swis{\l}ocki and Deuar(2016)]{Swislocki16}
Tomasz \'Swis{\l}ocki and Piotr Deuar.
\newblock Quantum fluctuation effects on the quench dynamics of thermal
  quasicondensates.
\newblock \emph{Journal of Physics B: Atomic, Molecular and Optical Physics},
  49\penalty0 (14):\penalty0 145303, 2016.
\newblock \doi{https://doi.org/10.1088/0953-4075/49/14/145303}.

\bibitem[Syrwid et~al.(2017)Syrwid, Zakrzewski, and Sacha]{Syrwid17}
Andrzej Syrwid, Jakub Zakrzewski, and Krzysztof Sacha.
\newblock Time crystal behavior of excited eigenstates.
\newblock \emph{Phys. Rev. Lett.}, 119:\penalty0 250602, 2017.
\newblock \doi{https://doi.org/10.1103/PhysRevLett.119.250602}.

\bibitem[Thapliyal et~al.(2015)Thapliyal, Banerjee, Pathak, Omkar, and
  Ravishankar]{Thapliyal15}
Kishore Thapliyal, Subhashish Banerjee, Anirban Pathak, S.~Omkar, and
  V.~Ravishankar.
\newblock Quasiprobability distributions in open quantum systems: Spin-qubit
  systems.
\newblock \emph{Annals of Physics}, 362:\penalty0 261--286, 2015.
\newblock \doi{https://doi.org/10.1016/j.aop.2015.07.029}.

\bibitem[Tsukiji et~al.(2016)Tsukiji, Iida, Kunihiro, Ohnishi, and
  Takahashi]{Tsukiji16}
Hidekazu Tsukiji, Hideaki Iida, Teiji Kunihiro, Akira Ohnishi, and Toru~T.
  Takahashi.
\newblock Entropy production from chaoticity in {Yang-Mills} field theory with
  use of the {Husimi} function.
\newblock \emph{Phys. Rev. D}, 94:\penalty0 091502, 2016.
\newblock \doi{https://doi.org/10.1103/PhysRevD.94.091502}.

\bibitem[Veitch et~al.(2012)Veitch, Ferrie, Gross, and Emerson]{Veitch12}
Victor Veitch, Christopher Ferrie, David Gross, and Joseph Emerson.
\newblock Negative quasi-probability as a resource for quantum computation.
\newblock \emph{New Journal of Physics}, 14\penalty0 (11):\penalty0 113011,
  2012.
\newblock \doi{https://doi.org/10.1088/1367-2630/14/11/113011}.

\bibitem[Vogel(2008)]{Vogel08}
Werner Vogel.
\newblock Nonclassical correlation properties of radiation fields.
\newblock \emph{Phys. Rev. Lett.}, 100:\penalty0 013605, 2008.
\newblock \doi{https://doi.org/10.1103/PhysRevLett.100.013605}.

\bibitem[Werner and Drummond(1997)]{Werner97}
M.~J. Werner and P.~D. Drummond.
\newblock Robust algorithms for solving stochastic partial differential
  equations.
\newblock \emph{Journal of Computational Physics}, 132\penalty0 (2):\penalty0
  312 -- 326, 1997.
\newblock \doi{https://doi.org/10.1006/jcph.1996.5638}.

\bibitem[Wigner(1932)]{Wigner32}
E.~Wigner.
\newblock {On the Quantum Correction For Thermodynamic Equilibrium}.
\newblock \emph{Phys. Rev.}, 40:\penalty0 749--759, 1932.
\newblock \doi{https://doi.org/10.1103/PhysRev.40.749}.

\bibitem[Wilczek(2012)]{Wilczek12}
Frank Wilczek.
\newblock Quantum time crystals.
\newblock \emph{Phys. Rev. Lett.}, 109:\penalty0 160401, 2012.
\newblock \doi{https://doi.org/10.1103/PhysRevLett.109.160401}.

\bibitem[Wouters and Savona(2009)]{Wouters09}
Michiel Wouters and Vincenzo Savona.
\newblock Stochastic classical field model for polariton condensates.
\newblock \emph{Phys. Rev. B}, 79:\penalty0 165302, 2009.
\newblock \doi{https://doi.org/10.1103/PhysRevB.79.165302}.

\bibitem[W\"uster et~al.(2017)W\"uster, Corney, Rost, and Deuar]{Wuster17}
S.~W\"uster, J.~F. Corney, J.~M. Rost, and P.~Deuar.
\newblock Quantum dynamics of long-range interacting systems using the
  positive-{$P$} and gauge-{$P$} representations.
\newblock \emph{Phys. Rev. E}, 96:\penalty0 013309, 2017.
\newblock \doi{https://doi.org/10.1103/PhysRevE.96.013309}.

\bibitem[Zhang et~al.(2017)Zhang, Hess, Kyprianidis, Becker, Lee, Smith,
  Pagano, Potirniche, Potter, Vishwanath, Yao, and Monroe]{Zhang17}
J.~Zhang, P.~W. Hess, A.~Kyprianidis, P.~Becker, A.~Lee, J.~Smith, G.~Pagano,
  I.~D. Potirniche, A.~C. Potter, A.~Vishwanath, N.~Y. Yao, and C.~Monroe.
\newblock Observation of a discrete time crystal.
\newblock \emph{Nature}, 543\penalty0 (7644):\penalty0 217, 2017.
\newblock \doi{https://doi.org/10.1038/nature21413}.

\end{thebibliography}

\end{document}